\documentclass[useAMS,usenatbib,usegraphicx]{mn2e}
\usepackage{rotating}
\usepackage{pdflscape}
\bibpunct {(} {)} {;} {a} {} {,}

\title[Recovering galaxy properties from SED-fitting II]{Recovering galaxy stellar population properties from broad-band spectral energy distribution fitting II. The case with unknown redshift}
\author[Pforr, Maraston \& Tonini]{Janine~Pforr$^{1,2}$\thanks{email: pforr@noao.edu}, Claudia~Maraston$^2$, Chiara~Tonini$^{2,3}$\\
$^1$ NOAO, 950 N. Cherry Ave., Tucson, AZ, 85719, USA\\
$^2$ Institute of Cosmology and Gravitation, University of Portsmouth, Dennis Sciama Building, Burnaby Road, Portsmouth, PO1 3FX, UK\\
$^3$ Centre for Astrophysics and Supercomputing, Swinburne University of Technology, Hawthorn, Victoria 3122, Australia}
\begin{document}

\newcommand{\hs}{\emph{HyperZspec}}
\newcommand{\hz}{\emph{HyperZ}}
\newcommand{\q}{\emph{Q}}

\maketitle

\begin{abstract}
In a recent work (Paper I) we explored the dependence of galaxy stellar population properties derived from broad-band spectral energy distribution fitting on the fitting parameters, e.g. star formation histories (SFHs), age grid, metallicity, initial mass function (IMF), dust reddening and reddening law, filter setup and wavelength coverage. In this paper we consider the case that also redshift is a free parameter in the fit and study whether one can obtain reasonable estimates of photometric redshifts and stellar population properties at once. As in Paper I we use mock star-forming as well as passive galaxies placed at various redshifts (0.5 to 3) as test particles. Mock star-forming galaxies are extracted from a semi-analytical galaxy formation model. We show that for high-redshift star-forming galaxies photometric redshifts, stellar masses and reddening can be determined simultaneously when using a broad wavelength coverage (including the Lyman and the 4000 \AA \space break) and a wide template setup in the fit. Masses are similarly well recovered (median $\sim0.2$ dex) as at fixed redshift. For old galaxies with little recent star formation (which are at lower redshift in the simulation) masses are better recovered than in the fixed redshift case, such that the median recovered stellar mass improves by up to $0.3$ dex (at fixed IMF) whereas the uncertainty in the redshift accuracy increases by only $\sim0.05$. However, a failure in redshift recovery also means a failure in mass recovery. As at fixed redshift mismatches in SFH and degeneracies between age, dust and now also redshift cause underestimated ages, overestimated reddening and underestimated masses. Stellar masses are best determined at low redshift without reddening in the fit. Masses are then underestimated by only $\sim0.1$ dex whereas redshifts are similarly well recovered. Not surprisingly, the recovery of properties is substantially better for passive galaxies, for which e.g. the mass is recovered only slightly worse than at fixed redshift (underestimated by $\sim0.02$ dex instead of $\sim0.01$, at fixed IMF) using a setup including metallicity effects. In all cases, the recovery of physical parameters is crucially dependent on the wavelength coverage adopted in the fitting because the redshift recovery depends on the wavelength coverage. As is well known redshifts are best recovered for a wavelength coverage including the Lyman and 4000 \AA \space break. As in Paper I, all effects from changing templates, the wavelength coverage and filters are quantified and scaling relations for the  transformation of stellar masses obtained using different fitting parameters, including stellar population models, are provided.
\end{abstract}
\begin{keywords}
galaxies: general -- galaxies: evolution -- galaxies: formation -- galaxies: fundamental parameters
\end{keywords}


\section{Introduction}
The robust derivation of galaxy physical properties is a crucial requirement for galaxy formation and evolution studies. 
Fitting stellar population models to data in order to derive the galaxy properties can be approached in several ways. In particular, we shall focus here on the fitting of the broad-band spectral energy distribution (SED), which is widely adopted in the literature, especially for high-redshift galaxies. In a previous paper (hereafter Paper I) we analysed the robustness of the derivation of galaxy physical properties when redshift was measured spectroscopically or known and used as a prior in the fit. However, many surveys  aiming to probe galaxy formation and evolution, e.g. CANDELS \citep{Grogin2011,Koekemoer2011}, COMBO-17 \citep{Wolf2001}, COSMOS \citep{Scoville2007}, DEEP2 \citep{Davis2003}, GMASS \citep{GMASS}, GOODS \citep{GOODS}, MUNICS \citep{MUNICS}, SDSS \citep{SDSS}, SERVS \citep{SERVS}, and surveys primarily undertaken for cosmological studies, e.g. DES (http://www.darkenergysurvey.org/), rely on photometrically derived redshifts. In particular, this is the case when probing the high-redshift Universe for which high signal to noise spectra are difficult to obtain for large numbers of galaxies. As is well known, redshifts are derived photometrically through the identification of strong spectral features, such as the Lyman-break at 912 \AA \space and the 4000 \AA -break, and the overall spectral shape, the SED. Two methods are  commonly used in the literature: 1) \textit{template fitting} and 2) \textit{empirical training sets}. Template fitting compares model SEDs to the data \citep{Sawicki1998} and the SED with the best fit determines the redshift \citep[][]{Baum1962}. Empirical training set techniques rely on a large set of galaxies with known colours and spectroscopic redshifts from which a relation between observed colour and redshift is inferred and then used to obtain redshifts for galaxies for which only colour information is available \citep{Connolly1995,Brunner1997,Wang1998}. A recent review of the method of SED-fitting can be found in \citet{Walcher2011}. A variety of codes for the determination of photometric redshifts are publicly available to carry out this task using various methods (e.g., \citealt{Benitez2000}, \hz \space by \citealt*{Bol2000}, ANNZ by \citealt*{Firth2003}, IMPZ by \citealt{Babbedge2004}, LePhare by \citealt{Ilbert2006}, ZEBRA by \citealt{Feldmann2006}, EAZY by \citealt{Brammer2008}, GalMC by \citealt*{Acquaviva2011}).
In this paper we use \hz \space \citep{Bol2000} consistently with Paper I. The performance of \hz\space with respect to the determination of photometric redshifts has been extensively studied in \citet{Bol2000} and \citet{Hildebrandt2010}. Here we use it as a tool in combination with the \citet[][hereafter M05]{M05} stellar population models to understand how well photometric redshifts as well as stellar population properties can be determined at once as a function of fitting parameters.\\ 
\noindent As in Paper I, we complement the literature work by using the M05 stellar population models and also a wider redshift range and galaxy type.\\
\noindent The methodology is as in Paper I. We use mock galaxies with known physical properties - age, metallicity, reddening, mass and star formation rate. These are then treated like observed data for which we carry out SED-fitting with several different setups. Input properties and those derived from the fitting are compared in order to  understand the robustness of the derived properties. For simulating star-forming galaxies we use models from the semi-analytic galaxy formation code GalICS \citep[Galaxies in Cosmological Simulations, ][]{Galics} as updated in \citet{Tonini2009, Tonini2010} based on the M05 stellar population model. Passive galaxies are simple stellar populations. As key results, we provide the average uncertainties of stellar properties that are associated with the various assumptions in the fitting and scaling relations to unify different works.\\
The paper is organised as follows. We recap the mock galaxy samples in Section \ref{mocks}. The method of SED-fitting and the different fitting setups are addressed in Section \ref{sed}. We present and discuss our results in Section \ref{results}. Our results are compared with the literature in Section \ref{litcompzfree} and scaling relations that allow the homogenising of data sets are provided in Section \ref{scalerelzfree}. Section \ref{summzfree} gives a summary of our work.\\
Throughout the paper we use a standard cosmology of $H_0=71.9$ km/s/Mpc, $\Omega_{\Lambda}=0.742$ and $\Omega_{M}=0.258$ and Vega magnitudes.


\section{Mock galaxy samples}\label{mocks}

We use the same two sets of mock galaxies - passive and star-forming - as in Paper I. Mock star-forming galaxies are extracted from the semi-analytic galaxy formation model GalICS \citep{Galics} and are based on M05 stellar population models \citep{Tonini2009, Tonini2010}.  As in \citet{Tonini2010} observer's frame spectra are obtained by redshifting both, reddened and unreddened, spectra to the redshift corresponding to the simulation time step. The spectra are then filtered with a chosen set of filters which provides us with the observer's frame broad-band magnitudes. These are scattered with Gaussian errors with 3 $\sigma$ errorbars choosing a value of $\sigma=0.1$. The adopted errorbars in the fitting reflect the typical errorbars of the COSMOS survey \citep{Capak2007} for $z\leq1$ and those of the GOODS star-forming objects \citep{Gia2004} for $z\geq2$. The errors of the first are $\sim0.05$ mag for optical and IRAC bands and $\sim0.3$ mag for near-IR bands, for the latter survey errors are typically $\sim$ 0.1 mag. In cases where the photometric error in a band is $<0.05$ mag we apply a minimum photometric error of $0.05$ mag for that band in the fitting. At each redshift  ($z=0.5$, $1$, $2$, $3$) a sample of $100$ mock galaxies is used.\\
The catalogues used in this paper are identical to the ones used in Paper I allowing us to directly compare the results.\\
The stellar population properties of the mock star-forming galaxies were shown in Paper I (their Fig. 1-5, see a summary in their table 1) to which we refer the reader for more details. Here, we summarise the main properties of the mock star-forming galaxies: 
\begin{enumerate}
\item mass-weighted ages of galaxies at z$\leq$ 1 are around a few Gyr, those at redshift $\geq2$ are mostly younger than 1 Gyr
\item mass-weighted metallicities become metal richer over time, yet half-solar metallicity is hardly reached at z$\sim$0 
\item stellar masses range from $\log M^*\approx 8 - 11.5 \log M_{\odot}$
\item every galaxy contains a fraction of young stars at every redshift since star formation never stops entirely in the semi-analytic model
\item star formation rates reach $100 M_{\odot}$/yr at $z\geq2$, at $z=0.5$ they are generally lower than $10 M_{\odot}$/yr
\item the star formation histories of most high-redshift galaxies are predominantly rising, most low-redshift galaxies exhibit bimodal star formation histories in the sense that SFR decreases after an initial rise  
\item dust reddening is applied following \citet{Tonini2010} as $E(B-V) = 0.33 \cdot (Log(SFR)-2) + 1/3$  and $E(B-V) = 0$ for SFRs less than $10\,M_{\odot}/yr$ \citep{Daddi2007a,M10} using the \citet{Calzetti} extinction curve.
\end{enumerate}
Our single burst populations which evolve passively span the same redshift range as mock star-forming galaxies and their properties can be summarized as:
\begin{enumerate}
\item varying ages depending on redshift, e.g. $1\leq$ age $\leq7$ Gyr at $z=0.5$, $1\leq$ age $\leq1.5$ Gyr at $z=3$
\item solar metallicity
\item $10^{10.5}\leq M^*\leq10^{12}\,M_{\odot}$ in steps of $0.25$ dex
\item dust free
\end{enumerate}
Both mock galaxy samples are based on a Salpeter IMF \citep{Salp}.


\section{Spectral Energy Distribution Fitting}\label{sed}
We perform the SED-fitting (comparison between theoretical template spectra and observed broad-band magnitudes) in the same way as described in \citet{M06} and in Paper I. We use the \hz-code of \citet{Bol2000} adapted for our needs (i.e. with an inclusion of a minimum age parameter, M. Bolzonella, private communication) which is based on a $\chi^2$-minimisation\footnote{$\chi^2=\sum_{i=1}^{N_{filters}}{[\frac{F_{obs,i}-b\times F_{temp,i}}{\sigma_i}]}^2$ with $F_{obs,i}$ and $F_{temp,i}$ as observed and template fluxes in filter i, respectively. The photometric uncertainty is given by $\sigma_i$, the normalisation factor between template and observed fluxes is $b$. The reduced $\chi^2_{\nu}$ is defined as $\frac{\chi^2}{\nu}$ where $\nu$ describes the number of degrees of freedom.}. For each template the $\chi^2_{\nu}$ is computed. Templates span a wide range in parameters such as star formation mode, age and metallicity. The best-fit solution is the combination of parameters with the minimum $\chi^2$. \\
Galaxy internal reddening is implemented in \hz \space through various empirically-derived laws: Milky Way by \citet{Allen76}, Milky Way by \citet{Seaton79}, Large Magellanic Cloud by \citet{Fitz86}, Small Magellanic Cloud by \citet{Prevot84}, Calzetti's law for local starburst galaxies \citep{Calzetti} and no reddening. Unless stated otherwise (e.g. Section \ref{rlzf}) for all fits presented in the results that include reddening ('the reddened case') the best fit is chosen among all reddening options in \hz.\\ 
At each redshift we constrain galaxy ages to be younger than the age of the Universe.\\ 
Galaxy age, star formation law, metallicity, dust reddening and reddening law are determined from the best-fit solution where age is describes the time elapsed since the onset of star formation and thus equals the age of the oldest stellar population present in the template. Stellar masses and star formation rates are then calculated by our own separate code\footnote{While \hz \space provides the normalisation factor $b$ between template and observed galaxy, the composite stellar population templates once processed through the BC03 \textit{csp\_galaxev} pipeline are not normalised to one solar mass at each time step and an additional normalisation depending on age has to be applied. This is not done internally in the \hz-code that we are using, hence we employ our own mass calculation code.} developed in \citet{Daddi05} and M06 using the normalisation between template and observed SED.\\
We draw our set of SED templates from the simple and composite stellar population models of M05.\\
The variety of template and filter setups investigated in this paper are summarizes in the following sections.


\subsection{Model Template Setups}\label{difftemp}
We employ the same template setups as in Paper I, namely i) a wide template setup and ii) an only-$\tau$ template setup.\\
The wide template setup consists of 32 types of theoretical spectra covering a wide range of star formation histories (SSPs, exponentially-declining with star formation time scale $\tau=$0.1, 0.3, and 1 Gyr, truncated SFR with truncation of star formation after 0.1, 0.3 and 1 Gyr and constant SFR) and metallicities ($\frac{1}{5}\,Z_{\odot}$, $\frac{1}{2}\,Z_{\odot}$, $Z_{\odot}$ and $2\,Z_{\odot}$) adopted from \citet{M06}. We repeat the fitting with with a mono-metallicity wide setup to study the effect of metallicity.\\
For $\tau$-type models we adopt the template setup of \citet{S05} containing solar metallicity $\tau$-models for values of $\tau=0.01,0.05,0.1,0.2,0.5,1,2$ and 5 Gyr and constant star formation.

\noindent Both template setups consist of theoretical spectra and are based on Salpeter IMF and M05 stellar populations for consistency with the mock galaxies. For each a full age grid (221 ages) is considered. As in Paper I we explore different resolutions of the template age grid. The default is an age grid of 221 values\footnote{Minimum and maximum age are $100,000$ yrs and $20$ Gyr, respectively.} and we test the effect of rebinning it to 51 values with the \hz-internal option. Furthermore, we obtain results for an age grid restricted to ages older than $0.1$ Gyr which is a commonly used practice in the literature \citep[e.g.][]{Bol2009,Wuyts2009} to reduce overshining as shown in M10 and Paper I.

\noindent Stellar population model effects are investigated by adopting BC03 models for the wide setup. Results for this are summarised in Appendix \ref{bc03results}. The reverse case - mock galaxies based on PEGASE \citep{Pegase}\footnote{PEGASE and BC03 models are quite similar due to similar input physics.} models and fitted with M05 models - is shown in Appendix \ref{pegase}.

\noindent The role of the IMF is addressed by changing the IMF in the wide template setup from its default (Salpeter IMF) to Chabrier, Kroupa and a top-heavy IMF (with slope $x\sim0$).  As pointed out in Paper I the IMF influences the overall SED shape of a galaxy due to the varying number of different-mass stars for each IMF. Our test will show how derived parameters change when a mismatching IMF is used in the fitting templates for Salpeter IMF mock galaxies. All our conclusions with regard to the IMF will be based on this case only. The outcome might be different when the mock galaxies are based on an IMF other than Salpeter.\\
The semi-analytic model was calibrated for a Salpeter IMF, which plays a role mainly in
the prescriptions for the supernovae feedback and the metallicity
evolution. An exploration of the systematics related
to the IMF of the simulated galaxies is beyond the scope of this paper.

\subsection{Wavelength range included in the fit} \label{filters}
In Paper I we addressed the importance of the wavelength coverage adopted in the fit. We found that a broad wavelength coverage (from rest-frame UV to rest-frame near-IR) leads to the most robust results in agreement with the findings of M06 and \citet{Kannappan}.\\
The wavelength coverage is of even greater importance when redshift needs to be determined because the fit relies on the correct identification of the position of important spectral features from which the redshift can be inferred when compared to their rest-frame position in the spectrum.\\
\noindent We carry out a comprehensive test of various filter setups and their performance in recovering the galaxy physical properties to answer these questions. For this we use the same filter setups as in Paper I which are: 
\begin{enumerate}
\item UBVRI\footnote{Johnson-Cousin system} JHK IRAC (3.6, 4.5, 5.8 $\mu$m)\footnote{For z=3 star-forming objects also the IRAC 4 band at 8$\mu$m was included.}, UBVRI JHK, UBVRI JH, UBVRI J, UBVRI
\item UBVRI IRAC
\item BVRI JHK IRAC, VRI JHK IRAC, RI JHK IRAC
\item u' g' r' i' z' (SDSS) [z=0.5]
\item UBVRI \emph{Y} JHK IRAC
\item BRIK
\end{enumerate}

\subsection{Photometric uncertainties}\label{photunc}
The effect of photometric uncertainties is studied using two different catalogues consisting of the same 100 unreddened galaxies at $z=0.5$, one containing randomised magnitudes\footnote{Magnitudes are randomised using a Gaussian and three $\sigma$ photometric errors where $\sigma\sim0.1$.}, the other the original magnitudes. For both the stellar population properties are derived with the same method, i.e. we use the wide setup and exclude reddening in the SED-fitting. The absence of reddening allows us to explore the effect of photometric uncertainties without the influence of the age-dust degeneracy.\\ 
In the case of the mock passive galaxies we fit the randomised magnitudes with the simple stellar population that was used to simulate them and without dust reddening. This ensures an exact match in metallicity and star formation history and deviations of age and stellar mass from the true values are due to photometric uncertainties.


\section{Results}\label{results}

In this section we compare the results derived from SED-fitting to the true values for both samples of mock galaxies. As in Paper I we compare the results depending on template setup (star formation history, metallicity, age grid, IMF, stellar population model\footnote{see Appendix \ref{bc03results}}), wavelength coverage, and reddening law. Differences to the results in Paper I are caused by the additional degree of freedom in the fitting due to the unknown redshift. Throughout this section we compare our results to those obtained with known redshift (Paper I).\\

\noindent As in Paper I we start by leaving dust reddening out of the procedure (both in the fitting and the mock galaxies, referred to as 'unreddened case' or 'case without reddening') for understanding the contributions of multiple stellar populations in the galaxies and their degeneracies with redshift.  We then repeat the exercise including dust reddening. For the mock star-forming galaxies this means that reddening is included in both the mocks and the fitting. The mock passive galaxies are always dust free and we investigate the results for excluding and including reddening in the fitting. We provide an overview over all fits for redshifts $0.5$, $1$, $2$ and $3$ in Table \ref{overfits}. We quantify which template and filter sets provide the closest match between input and fitted properties by defining the quality estimator \emph{Q} as in Paper I:
\[
Q=\sqrt{\sum_{i}{(\Delta)^2}}\,\,\rmn{\space with}\,\, \Delta=x_{i,fitting}-x_{i,input},
\]
where i is the number of objects and \emph{x} represents one of the following: redshift, age, metallicity, reddening, stellar mass or SFR. \q=0 and $\Delta=0$ is the ideal case.\\
We showed in M10 and Paper I that models with the smallest $\chi^2_{\nu}$ do not necessarily coincide with the best physical solution. For this reason we abstain from showing any comparisons of $\chi^2_{\nu}$ but focus instead on the effects of different template setups and wavelength coverage on the recovered redshifts, ages, SFHs and stellar masses. We do not show results for metallicity and reddening when redshift is a free parameter, as these are very similar to those at fixed redshift reported in Paper I. Additionally to Paper I, we also have the opportunity to see how the derivation of other properties, such as age and stellar mass, depends on the derived redshift. As a reference we show the result for the wide setup in each figure. Results are summarised for redshift 0.5 and 2 in quantitative form in Tables \ref{sfoverres}1 to \ref{poverres2}4.\\

\subsection{Star-forming galaxies}

\subsubsection{Photometric redshift}\label{zresults}
\begin{figure*}
\centering\includegraphics[width=124mm]{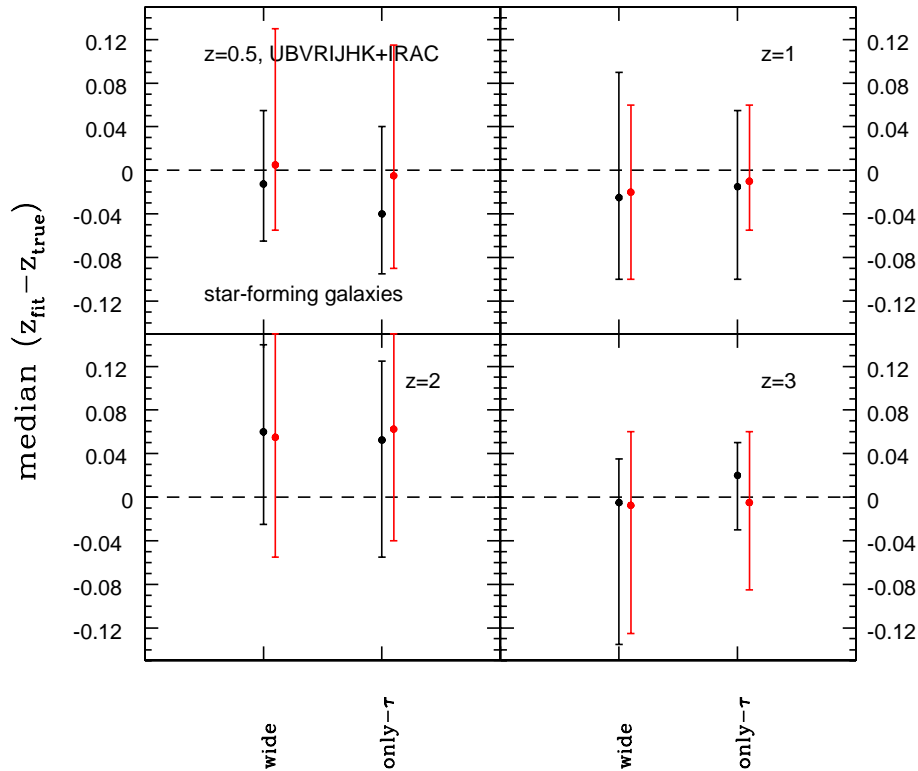}
\caption[Redshift recovery for mock star-forming galaxies as function of template setup]{\label{zhistdtzf} Median redshift recovery for mock star-forming galaxies as a function of template setup, namely wide and only-$\tau$ using a broad wavelength coverage (UBVRIJHK+IRAC). Redshift increases from top left to bottom right. Black symbols refer to the case without reddening, red to the reddened case. Errorbars are 68\% confidence levels.}
\end{figure*}
\begin{figure*}
\centering\includegraphics[width=144mm]{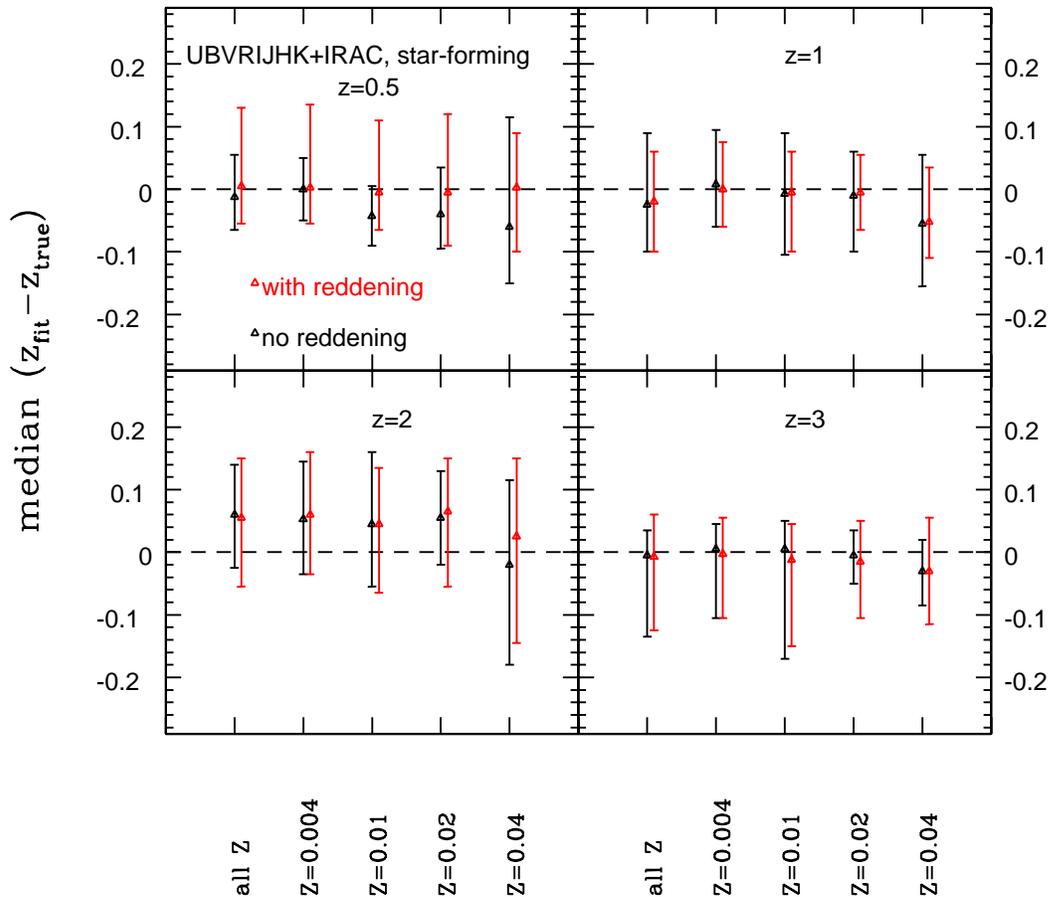}
\caption[Median redshift recovery for mock star-forming galaxies as function of metallicity]{\label{zhistdmetalzf} Median redshift recovery for mock star-forming galaxies as a function of metallicity (increasing from left to right in each panel) using a broad wavelength coverage (UBVRIJHK+IRAC). Black symbols refer to the case without reddening, red to the reddened case. Errorbars are 68\% confidence levels.}
\end{figure*}
\begin{figure*}
\centering\includegraphics[width=144mm]{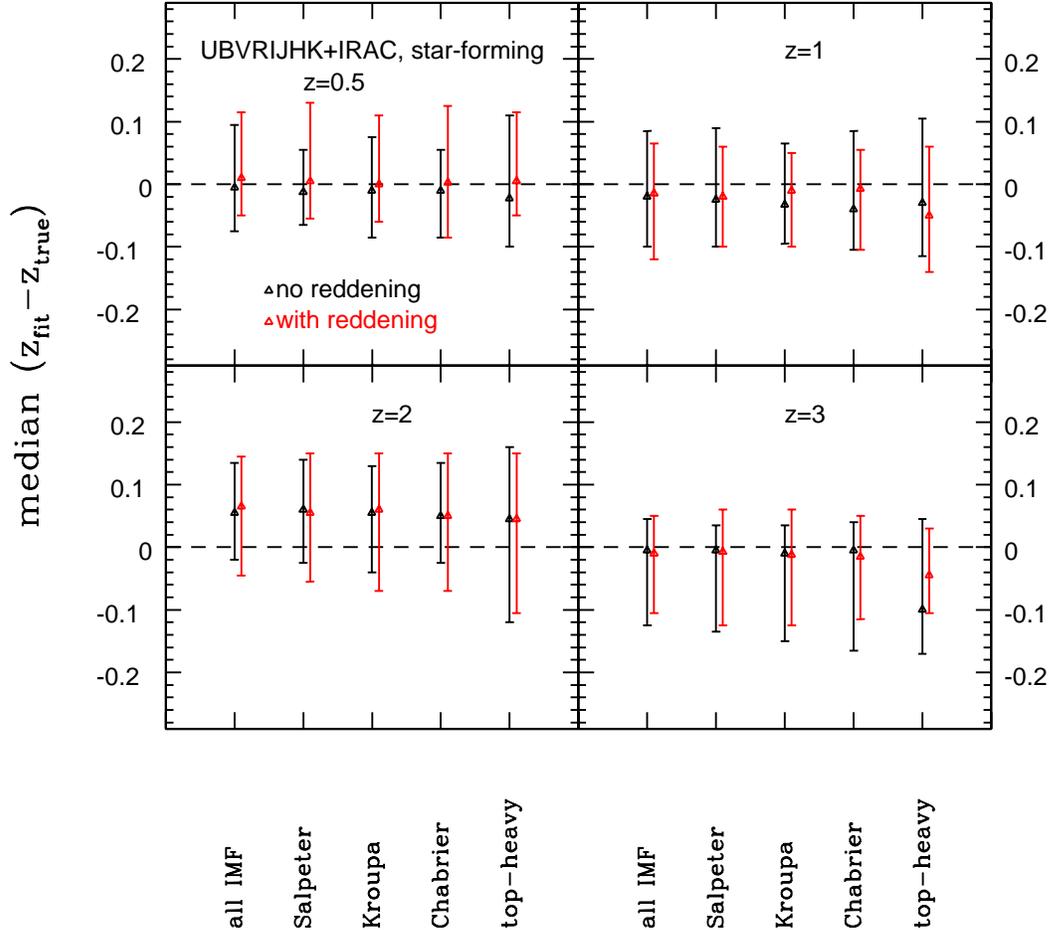}
\caption[Median redshift recovery for mock star-forming galaxies as function of IMF]{\label{zhistdIMFzf} Median redshift recovery for mock star-forming galaxies as a function of IMF using a broad wavelength coverage (UBVRIJHK+IRAC) and a wide setup. Symbols as in Fig. \ref{zhistdmetalzf}.}
\end{figure*}
\begin{figure*}
\centering\includegraphics[width=144mm]{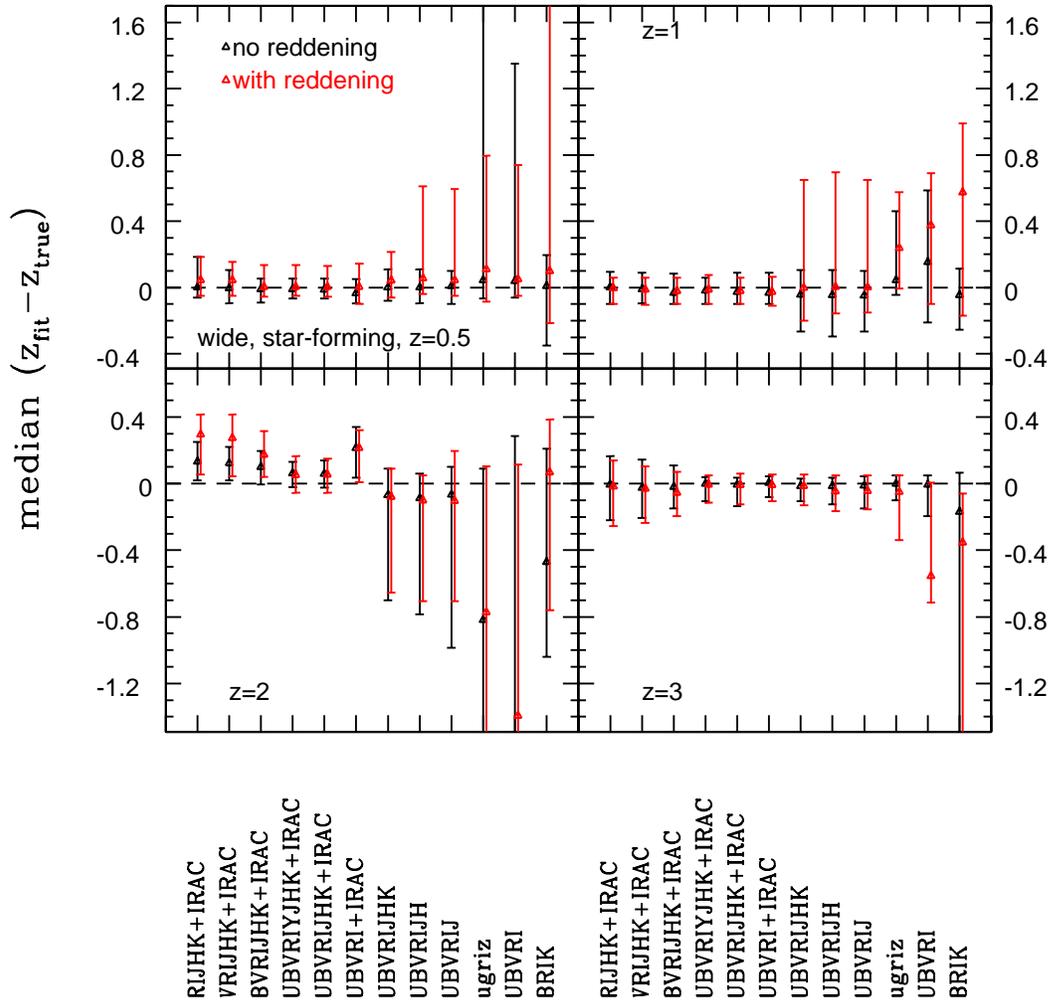}
\caption[Median redshift recovery for the mock star-forming galaxies as function of wavelength coverage]{\label{zhistdfzf} Redshift recovery (median) for mock star-forming galaxies as function of wavelength coverage using a wide template setup. Redshift increases from top left to bottom right. Red symbols refer to the case with reddening, black to the unreddened case. Errorbars reflect 68\% confidence levels.}
\end{figure*}

In this section, we study the redshift recovery for mock star-forming galaxies (Figs. \ref{zhistdtzf}-\ref{zhistdfzf}). First, we show the effect of template setup in Fig. \ref{zhistdtzf}. Overall, the differences in redshift recovery between template setups are small. On average, redshifts are recovered correctly within $\sigma\sim0.1$. The highest redshifts (z$\sim$3) are recovered best. Those for which $z_{true}=2$ are overestimated by a median of 0.06 and show larger scatter (with 68\% confidence levels of size $\sim0.12$) for both template setups. For the only-$\tau$ setup we find more outliers for $z_{true}=2$ objects in the unreddened case. For catastrophic outliers the recovered redshifts are mainly below $1$ and best fits have very short star formation time scales ($\tau\leq0.05$) and very young ages ($<\sim20$ Myr). Indeed, they belong to the youngest objects at this redshift with ages of the oldest stars of 400 Myr and have a rising SFH (compare Fig. 3, panels a) and c) in Paper I). In the wide setup these are fitted with similar ages, but lower metallicity and even shorter star formation times scales (namely SSPs). Neither template setup has options that match the rising SFH of these galaxies. For the only-$\tau$ setup with fixed solar metallicity, the only way left to match the very blue SED best is to choose a lower redshift while for the wide setup other options (bluer SFH in form of SSPs, lower metallicity) are available. Including modest amounts of reddening reduces the outlier fraction and improves the redshift recovery with the only-$\tau$ setup. In particular, an even younger fitted age in combination with a small amount of dust reddening (E(B-V)$\sim 0.1$) and similarly short SFHs as in the unreddened case lead to an acceptable redshift estimate. For the wide setup, differences arising from the inclusion of dust reddening are small. Since the galaxies at $z\geq2$ are intrinsically young and inhabit significant star formation they lack a distinct 4000 \AA \space break. At $z=2$ the adopted filter set does not comprise the Lyman-limit (at 912 \AA \space rest-frame). Hence, two important features aiding the redshift determination are missing, making the process of obtaining redshifts for galaxies in this redshift range very difficult. Galaxies with redshifts around $2$ lie in the so-called redshift desert where the redshift determination is challenging even with spectroscopic methods due to the lack of spectral lines in the optical \citep*[e.g.][]{Renzini2009}. At $z=3$ the Lyman-limit falls into the range of the U filter and thus significantly helps in determining the redshift.\\
\noindent Unlike for the mass recovery, excluding SSPs from the fit does not help to improve the redshift recovery for old galaxies with little star formation (low redshift). At high redshift, the scatter is somewhat smaller.\\
\noindent From these results we conclude that a wide setup is best for the redshift determination at low redshift. At high redshift, the only-$\tau$ setup (including reddening) is better and also more economic due to a smaller number of fitting templates.

\noindent In the next step we investigate the effect of the template metallicity on the redshift determination by using the wide setup in mono-metallicity form. Overall, metallicity effects are small (Fig. \ref{zhistdmetalzf}). In principle, redshifts should be best determined with the lowest metallicity setup because the metallicities of mock star-forming galaxies are predominantly sub-solar (compare Fig. 2 in Paper I). For low redshift galaxies, redshifts are indeed recovered best with the lowest metallicity setup in terms of median offset and scatter. However, at redshift $3$ the best median offset and smallest scatter is achieved with a solar metallicity setup, highlighting the presence of degeneracies in the fitting.\footnote{Note though that when considering the average offset with its $1~\sigma$ uncertainty the redshift recovery is best with the lowest metallicity setup.} The redshift recovery is worst for $z=2$ objects in terms of an offset and in particular for the highest metallicity setup in terms of scatter. Despite having a small effect overall, including reddening impacts the most at low redshift where it helps to correct the observed median offset between true and recovered redshift in the unreddened case for the highest metallicity setups but also increases the scatter. However, redshifts are well recovered at $z=0.5$ with reddening in spite of large offsets for the stellar mass (Section \ref{massresults}).

\noindent Fig. \ref{zhistdIMFzf} illustrates the dependence of redshift determination on the IMF of the templates. Slightly better results are achieved with an IMF in the templates that is the same as in the mocks (Salpeter in this case)\footnote{When considering the average offsets with their $1~\sigma$ uncertainties instead of the median and 68\% confidence levels, Salpeter IMF templates also perform slightly better at most redshift and reddening cases.}. In the unreddened case, at redshifts $0.5$ and $3$ using Salpeter IMF templates results in somewhat smaller scatter and at redshift $1$ the median offset is marginally smaller compared to the other IMF templates. Redshifts are similarly well recovered with Chabrier and Kroupa IMF templates. Templates with a top-heavy IMF perform worst by resulting in the largest offsets and/or largest scatter. The right IMF cannot be identified by the fit as already concluded in Paper I. About 30\% of the objects at each redshift are associated with a top-heavy IMF instead of Salpeter. The effect of adding reddening is again small. For the top-heavy IMF redshift estimates at $z=3$ are improved such that redshifts are less underestimated by choosing larger amounts of dust reddening. This is clearly due to the significantly bluer SED of the top-heavy IMF at young ages. For Chabrier and Kroupa IMF at $z=1$ the offset between recovered and true redshift is improved and scatter is slightly smaller when reddening is involved.

\noindent Finally, in Fig. \ref{zhistdfzf} we show the dependence of the redshift recovery on the wavelength coverage adopted in the fitting. The largest effects are seen for filter setups that do not comprise the rest-frame near-IR and the red optical resulting in larger scatter and overestimated redshifts at $z=0.5$ and $1$. Clearly the old age of the objects is not sufficiently constrained in the fit and gets compensated with a higher redshift. At $z=2$ and $3$, redshifts are underestimated (by a median of up to $\sim1.5$ at for a UBVRI filter setup at $z=2$ without reddening) because the sole use of  optical filter bands excludes the 4000 \AA \space break. At redshift $3$ at least the Lyman-limit is covered by the filters. Additionally, the galaxies' young ages and restricted age range in the fitting helps the redshift recovery. Omitting blue filter bands in the fitting has little effect at low redshift because the 4000 \AA \space break is still included (in R band at $z=0.5$ and I band at $z=1$). At redshift $2$ redshifts are overestimated by a median of up to 0.13 (unreddened case) with these filter setups because the rest-frame wavelength covered by these filters contains information about the Lyman-absorption between Lyman-$\alpha$ (1216 \AA) and the Lyman-limit (912 \AA) although missing the Lyman-limit itself. For redshift $3$ objects, only the scatter increases. Furthermore, a filter setup missing out on the near-IR filter bands JHK, means excluding the 4000 \AA \space break at $z=2$. For $z=3$ this is less harmful because the Lyman-limit is still covered. Including reddening does not alter these trends but the scatter increases. At redshift $1$ redshifts are overestimated by a median of up to 0.6 for the most restricted wavelength coverages.
\begin{figure*}
\centering\includegraphics[width=144mm]{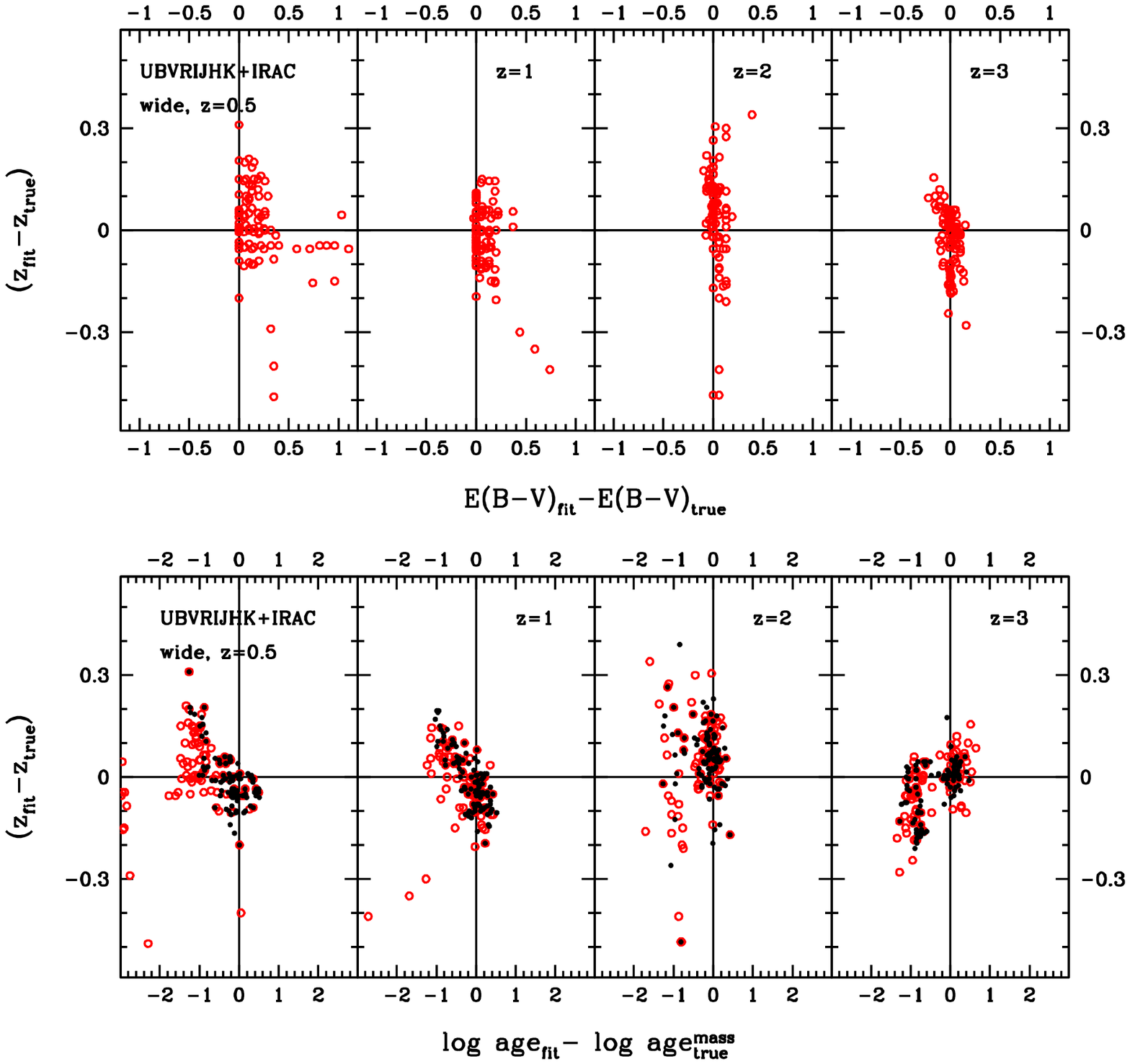}
\caption[Redshift recovery for mock star-forming galaxies as function of reddening and age]{\label{zebvagezf} Redshift recovery for mock star-forming galaxies as a function of difference between derived and true reddening (top row) and between derived and mass-weighted age (bottom row). Redshift increases from left to right. Red symbols refer to the case with reddening, black to the unreddened case. Horizontal and vertical solid lines are plotted for reference.}
\end{figure*}

\noindent In Paper I we demonstrated that the lack of knowledge of the true SFH as well as degeneracies between age, dust and metallicity are the largest obstacles against obtaining robust stellar population properties from the fitting. In particular, no star formation, older ages, larger amounts of dust and higher metallicity make the SED redder compared to star-forming SFHs, younger ages, small amounts or no dust and low metallicity. Here, we have an additional degeneracy due to the unconstrained redshift. The redshift recovery as a function of the difference between derived and true values of reddening and mass-weighted age is shown in Fig. \ref{zebvagezf} for the wide setup and the broadest wavelength coverage. There is no correlation  between redshift recovery and difference in reddening. Fig. \ref{zhistdtzf} already revealed that there is little difference in the redshift recovery between the reddened and unreddened case. However, in the unreddened case the redshift recovery is clearly anti-correlated with the difference in age at the lowest redshifts, i.e. for the oldest galaxies. For underestimated ages  redshifts are overestimated, for overestimated ages they are slightly underestimated. At $z=2$ the connection loosens up and at $z=3$ it turns into a weak correlation, such that for underestimated ages (which are also the youngest true ages, compare Fig. 6 in Paper I) redshifts are also underestimated. Here, underestimated ages and redshifts (and low metallicities) try to compensate shortcomings in the star formation history (SSPs) which does not reflect the true star formation history of these objects (i.e. rising star formation, compare Fig. 3 in Paper I). We concluded in Paper I that SSPs should not be used in the fitting due to their negative effect on the stellar mass recovery. For the redshift recovery the exclusion of SSPs has a negligible effect at low redshift and only the scatter is decreased at higher redshift.

\begin{table*}
\caption[Median offset in recovered redshifts for mock star-forming galaxies]{\label{medztabzf}Redshift recovery as median($z_{fit}-z_{true}$) and 68\% confidence levels for mock star-forming galaxies as a function of template setup, IMF and metallicity at each true redshift.}
\begin{center}
\begin{tabular}{@{}lrrrr}\hline
setup & $z_{true}=0.5$ & $z_{true}=1$ & $z_{true}=2$ & $z_{true}=3$ \\\hline
\textbf{no reddening}	& & & & \\
wide Salpeter	&	$-0.0125_{-0.0525}^{+0.0675}$	&	$-0.0250^{+0.1150}_{-0.0750}$	&	$0.0600_{-0.0850}^{+0.0800}$	&	$-0.0050_{-0.1300}^{+0.0400}$	\\\hline
wide Kroupa	&	$-0.0100_{-0.0800}^{+0.0700}$	&	$-0.0400_{-0.0600}^{+0.1250}$	&	$0.0525_{-0.0925}^{+0.0775}$	&	$-0.0100_{-0.1400}^{+0.0500}$	\\\hline
wide Chabrier	&	$-0.0100_{-0.0750}^{+0.0650}$	&	$-0.0400^{+0.1250}_{-0.0650}$	&	$0.0500_{-0.0750}^{+0.0850}$	&	$-0.0050_{-0.1600}^{+0.0450}$	\\\hline
wide top-heavy	&	$-0.0225_{-0.0775}^{+0.1325}$	&	$-0.0300^{+0.1350}_{-0.0850}$	&	$0.0450_{-0.1650}^{+0.1150}$	&	$-0.1000_{-0.0700}^{+0.1450}$	\\\hline
wide best IMF	&	$-0.0050_{-0.0700}^{+0.1000}$	&	$-0.0200_{-0.0800}^{+0.1000}$	&	$0.0550^{+0.0800}_{-0.0750}$	&	$-0.0050_{-0.1200}^{+0.0500}$	\\\hline
wide Z=0.004	&	$0.0000_{-0.0700}^{+0.1000}$		&	$0.0075^{+0.0875}_{-0.0675}$		&	$0.0525_{-0.0875}^{+0.0925}$	&	$0.0050_{-0.1100}^{+0.0400}$		\\\hline
wide Z=0.01	&	$-0.0425_{-0.0475}^{+0.0475}$	&	$-0.0075^{+0.0975}_{-0.0975}$	&	$0.0450_{-0.1000}^{+0.1150}$	&	$0.0050_{-0.1750}^{+0.0450}$		\\\hline
wide Z=0.02	&	$-0.0400_{-0.0550}^{+0.0750}$	&	$-0.0100_{-0.0900}^{+0.0700}$	&	$0.0550_{-0.0750}^{+0.0750}$	&	$-0.0050_{-0.0450}^{+0.0400}$	\\\hline
wide Z=0.04	&	$-0.0600_{-0.0900}^{+0.1750}$	&	$-0.0550_{-0.1000}^{+0.1100}$	&	$-0.0200_{-0.1600}^{+0.1350}$&	$-0.0300^{+0.0500}_{-0.0550}$	\\\hline
only-$\tau$	&	$-0.0400_{-0.0550}^{+0.0800}$	&	$-0.0150_{-0.0850}^{+0.0700}$	&	$0.0525_{-0.1075}^{+0.0725}$	&	$0.0200^{+0.0300}_{-0.0500}$		\\\hline
\textbf{with reddening}	& & & & \\
wide Salpeter	&	$0.0050_{-0.0600}^{+0.1250}$		&	$-0.0200_{-0.0800}^{+0.0800}$	&	$0.0550_{-0.1100}^{+0.0950}$	&	$-0.0075_{-0.1175}^{+0.0675}$	\\\hline
wide Kroupa	&	$0.0050_{-0.0600}^{+0.1100}$		&	$-0.0125_{-0.0875}^{+0.0675}$	&	$0.0600_{-0.1150}^{+0.0850}$	&	$-0.0175_{-0.1075}^{+0.0625}$	\\\hline
wide Chabrier	&	$0.0025_{-0.0875}^{+0.1225}$		&	$-0.0075_{-0.0975}^{+0.0625}$	&	$0.0500_{-0.1200}^{+0.1000}$	&	$-0.0150_{-0.1000}^{+0.0650}$	\\\hline
wide top-heavy	&	$0.0050_{-0.0550}^{+0.1100}$		&	$-0.0500_{-0.0900}^{+0.1100}$	&	$0.0450_{-0.1500}^{+0.1050}$	&	$-0.0450_{-0.0600}^{+0.0750}$	\\\hline
wide best IMF	&	$0.0225_{-0.0725}^{+0.1075}$		&	$-0.0100_{-0.1100}^{+0.0750}$	&	$0.0650_{-0.1050}^{+0.0800}$	&	$-0.0075_{-0.0975}^{+0.0675}$	\\\hline
wide Z=0.004	&	$0.0025_{-0.0575}^{+0.1325}$		&	$0.0000_{-0.0600}^{+0.0750}$		&	$0.0600_{-0.0950}^{+0.1000}$	&	$-0.0025_{-0.1025}^{+0.0575}$	\\\hline
wide Z=0.01	&	$-0.0050_{-0.0600}^{+0.1150}$	&	$-0.0050_{-0.0950}^{+0.0650}$	&	$0.0450_{-0.1100}^{+0.0900}$	&	$-0.0125_{-0.1375}^{+0.0575}$	\\\hline
wide Z=0.02	&	$-0.0050_{-0.0850}^{+0.1250}$	&	$-0.0050_{-0.0600}^{+0.0600}$	&	$0.0650_{-0.1200}^{+0.0850}$	&	$-0.0150_{-0.0900}^{+0.0650}$	\\\hline
wide Z=0.04	&	$0.0025_{-0.1025}^{+0.0875}$		&	$-0.0525_{-0.0575}^{+0.0875}$	&	$0.0250_{-0.1700}^{+0.1250}$	&	$-0.0300_{-0.0850}^{+0.0850}$	\\\hline
only-$\tau$	&	$-0.0050_{-0.0850}^{+0.1200}$	&	$-0.0100^{+0.0700}_{-0.0450}$	&	$0.0625_{-0.1025}^{+0.0875}$	&	$-0.0050_{-0.0800}^{+0.0650}$	\\\hline
\end{tabular}
\end{center}
\end{table*}%

\noindent With respect to the findings of Fig. \ref{zebvagezf} we investigated whether redshifts can be better recovered by using only one particular reddening law in the fit or by using a different age grid. Constraining the age grid by rebinning or applying a minimum age of $0.1$ Gyr does not improve the median redshift estimate. For single objects at low redshift a minimum age constraint helps. Additionally, fitting with a minimum age constraint is more efficient, especially for large data samples. Equivalently, we do not find any reddening law performing better than another or the best fit out of all reddening laws (see also Section 4.4). Hence, using only one reddening law, e.g. a Calzetti law, in the fit is sufficient.

\noindent In summary, we find that redshifts are recovered quite well overall and the redshift recovery depends only weakly on template setup, IMF and metallicity because of compensating effects. Since the inclusion of reddening seems to help the redshift recovery the most economic template setup to use is an only-$\tau$ setup with a minimum age of $0.1$ Gyr and including reddening in the fit (one reddening law is sufficient). Indeed, when comparing photometric and spectroscopic redshifts for real galaxies observed within the Spitzer Extragalactic Representative Volume Survey \citep[SERVS][]{Mauduit2012} we find such a setup to perform best (Pforr et al. in prep.). The knowledge of the right IMF and metallicity helps the accuracy of the photometric redshifts, but only by $\Delta z\sim0.05$. Unsurprisingly, redshifts are best recovered with a broad wavelength coverage which comprises the Lyman-limit and the 4000 \AA \space break. Thus, red optical and near-IR filter bands are crucial in recovering the right redshift of high-redshift objects. The redshifts of $z=2$ objects are recovered worst due to a lack of coverage of the Lyman-limit even with our widest filter setup, those of $z=3$ objects are recovered best since both breaks are covered. Median recovered redshifts with their 68\% confidence levels are summarised in Table \ref{medztabzf}.

\subsubsection{Age}\label{ageresults}
\begin{figure*}
\centering
\includegraphics[width=144mm]{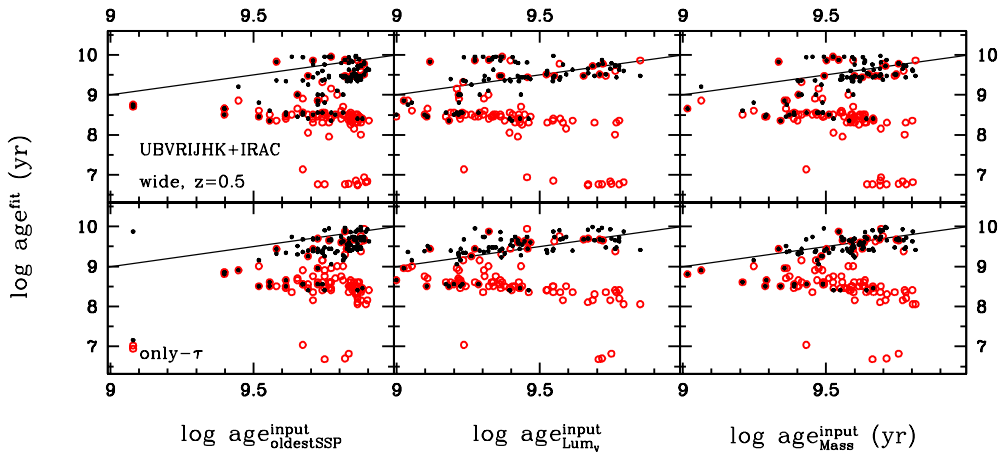}
\caption[Age recovery for mock star-forming galaxies as function of template setup]{\label{agedt05zf} Comparison of derived ages to input ages of mock star-forming galaxies as a function of template setup at redshift $0.5$. Symbols are as in Fig. \ref{zhistdfzf}. \textit{Left:} Age of oldest present SSP. \textit{Middle:} V-band luminosity-weighted ages. \textit{Right:} mass-weighted ages.}
\end{figure*}

In Paper I we established that mass-weighted ages seem to be best recovered in the fit, but that age is overall poorly recovered, especially for the oldest galaxies with little on-going star formation (see Fig. 6 in Paper I). Here, we confirm this result when redshift is a free parameter. In comparison to the fixed redshift case, ages of the oldest galaxies are better recovered (Fig. \ref{agedt05zf}) such that very young ages ($<10^8$ yr)\footnote{As in Paper I these are predominantly fit with highly dust-reddened SSPs.} are less common in the fitting when reddening is included. However, ages are still underestimated in most cases because of the age-dust degeneracy \citep[see e.g.][]{ren06}. Additionally, fitted ages tend to anti-correlate with true ages (luminosity or mass-weighted) for the majority of the oldest objects when reddening is used in the fitting. This is clearly an effect of the added degeneracy with redshift.

\noindent In the unreddened case ages are better recovered compared to the case with reddening and to the fixed redshift case but are still mostly underestimated. The only-$\tau$ setup recovers ages slightly better than the wide setup in the unreddened case. Since differences between the fixed redshift (Paper I) and free redshift case at high redshift are negligible, we do not show it here. Generally, ages at high redshift are better determined because of a smaller age range available in the fitting. Hence, the age-dust degeneracy has little effect and only the youngest obtained ages (below $10^8$ yr) are underestimated. As already discussed in Paper I this is due to two reasons: 1) age-metallicity degeneracy - template metallicities are much higher than mock galaxy metallicities, thus, bluer colours are reproduced by younger ages, and 2) SFH mismatch between templates and mock star-forming galaxies.  

\noindent Mono-metallicity wide template setups help to separate metallicity effects. Similarly to our findings in Paper I, when reddening is switched off and template metallicity is high, sub-solar metallicity galaxies are fitted with a younger age to compensate this mismatch. This trend persists even when reddening is introduced. When redshift is known metallicity effects are partly compensated by the age-dust degeneracy.

\noindent Although IMF effects on the redshift recovery are small, ages are affected more. At fixed redshift ages derived with templates of different IMF differ significantly, especially in the reddened case and at low redshift. Here, ages are much more similar at low redshift and scatter is significantly reduced. The age recovery is otherwise very similar to the fixed redshift case.

\noindent Lastly, we study the wavelength dependence of the derived ages. Results are similar to the ones in Paper I, thus we only summarise them here. In the unreddened case the oldest galaxies ($z=0.5$) are affected very little by a lack in wavelength coverage. Excluding the rest-frame near-IR and red optical filter bands in the fitting focusses the age distribution between $0.1$ and $1$ Gyr. For higher redshift and intrinsically younger objects derived ages become younger as wavelength coverage narrows. The more the wavelength coverage is restricted to the rest-frame UV, the younger the derived ages. Neglecting blue filter bands has a very small effect such that a few objects are rejuvenated. The inclusion of reddening has little impact on the general dependence of derived ages on wavelength coverage. Of course, due to the age-dust degeneracy ages are poorly recovered overall.

\noindent In summary, we find in agreement with Paper I that ages are poorly recovered and underestimated in most cases. This depends on the redshift estimate at low redshift such that ages are underestimated\footnote{in comparison to the mass-weighted age} when redshifts are overestimated and vice versa. At high redshift ages are overestimated when redshift is overestimated and vice versa. However, at low redshift ages are slightly better determined when redshift is a free parameter. The age-metallicity degeneracy has a clearer effect and is not overshadowed by the age-dust degeneracy. Ages derived with different IMFs are more similar to each other than at fixed redshift, scatter is particularly reduced in the reddened case at low redshift. The difficulty of recovering the oldest age remains.

\subsubsection{Stellar Mass}\label{massresults}
\begin{figure*}
\centering\includegraphics[width=144mm]{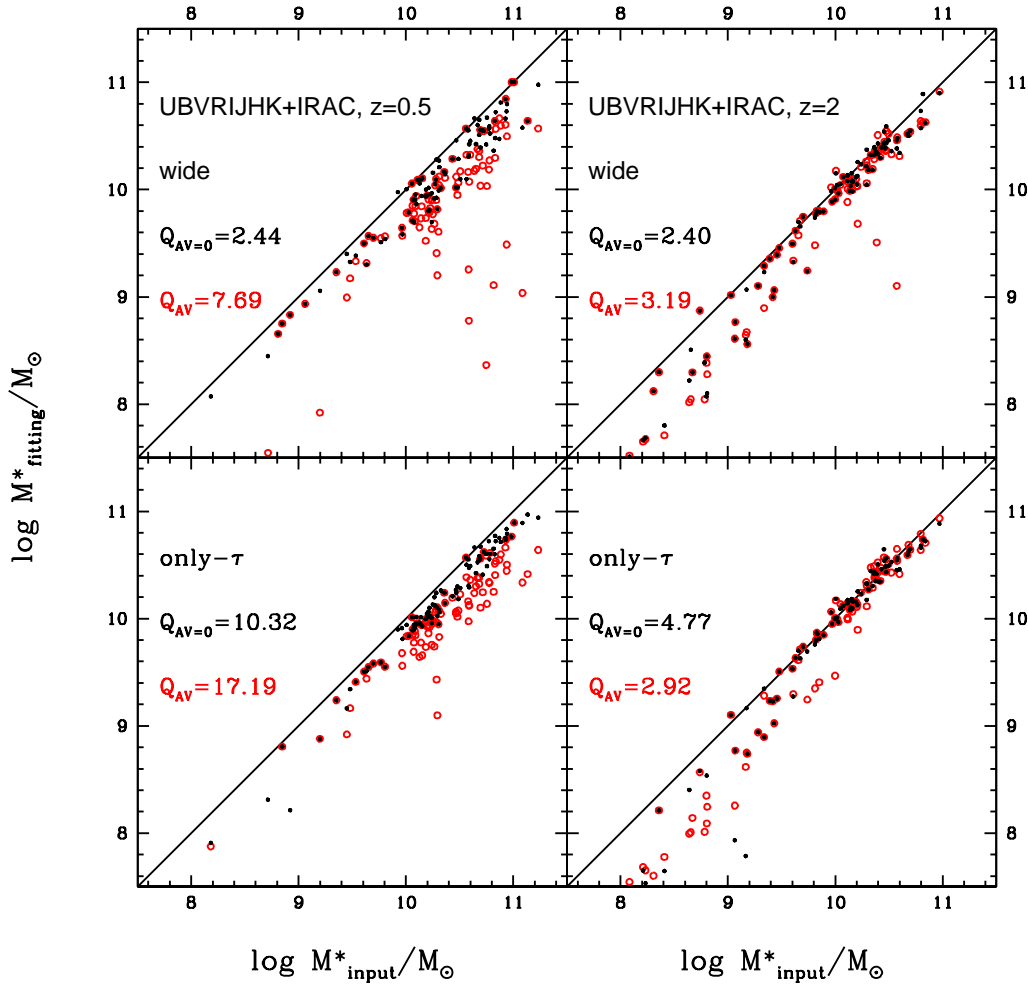}
\caption[Stellar mass recovery for mock star-forming galaxies as a function of template setup]{\label{massdt05zf} Comparison of derived stellar mass to input stellar mass at redshift $0.5$ (left) and $2$ (right) for different template setups (from top to bottom in each panel: wide, only-$\tau$). Red circles refer to cases with reddening, black dots to no reddening. Quality factors are given for the entire mass range for reddened and unreddened cases, respectively.}
\end{figure*}
\begin{figure*}
\centering\includegraphics[width=144mm]{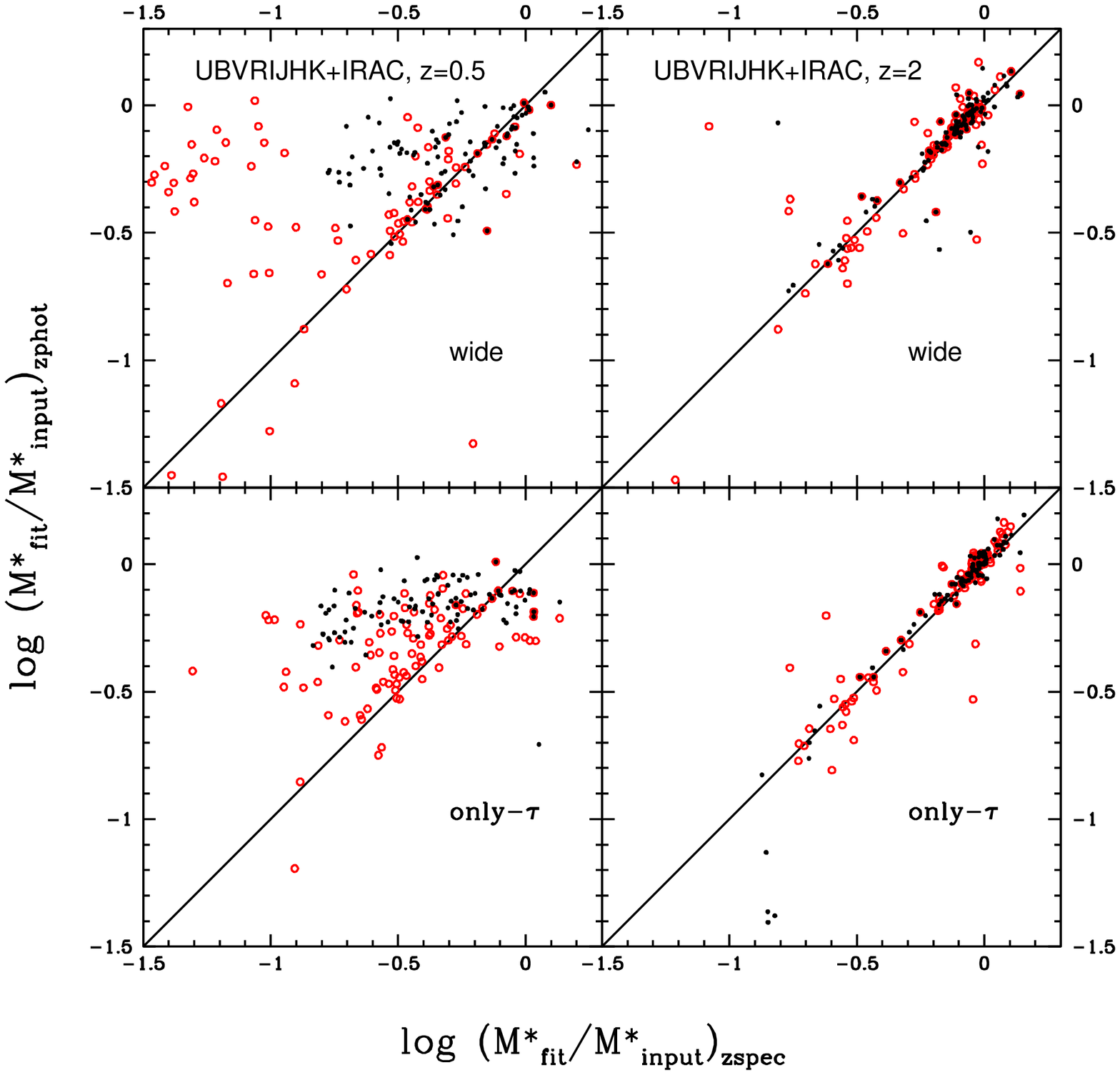}
\caption[Stellar mass recovery for mock star-forming galaxies for fixed and free redshift]{\label{massdtzfzf} Stellar mass recovery at redshift $0.5$ (left) and $2$ (right) as a function of template setup (from top to bottom in each panel: wide, only-$\tau$) when redshift is fixed or a free parameter in the fit. Symbols are as in Fig. \ref{massdt05zf}. The most extreme outliers for which the mass recovery fails when redshift is a free parameter are not shown in this plot.}
\end{figure*}
\noindent We demonstrated in M10 and Paper I that star formation histories are the most important factor for recovering the stellar mass. Hence, the choice of template setup affects the result. At fixed redshift we found that masses are underestimated by a median of $0.6$ dex at low redshift and in the reddened case because of the age-dust degeneracy and SFH mismatch. At high redshift masses are perfectly recovered using inverted-$\tau$ models with high formation redshift and $\tau$'s larger than $0.3$ Gyr (M10 and compare Fig. 11 in Paper I).

\noindent Interestingly, although masses are still underestimated, the mass recovery is on average better at low redshift for the wide setup when redshift is a free parameter in the fitting (Fig. \ref{massdt05zf} and Fig. \ref{massdtzfzf}). In particular, at low redshift this is due to the combination of underestimated redshifts with slightly overestimated ages (see Fig. \ref{zebvagezf} and Section \ref{zresults}) which in turn provides a larger mass estimate. Furthermore, reddening is generally less overestimated in the case of free redshift than at fixed redshift in Paper I. In the unreddened case the improvement on the median mass is $0.1$ dex, in the reddened case $0.3$ dex for one $\sigma$ uncertainties in redshift of $\pm0.06$ and $-0.06$ to $0.12$, respectively (median redshifts are within $\pm\sim0.01$). Overall, differences between the reddened and unreddened case are much smaller than at fixed redshift (Paper I) because redshift compensates for SFH and metallicity mismatch, age-dust degeneracy and overshining - the main reasons why masses are underestimated for the oldest galaxies when redshift is fixed. However, when reddening is included, the mass recovery fails completely for objects for which the redshift is underestimated the most. For these objects a very low redshift combined with a high amount of reddening, mostly very young ages and a single burst SFH results in very low stellar masses. In Paper I solutions with the largest mass underestimation because of the age-dust degeneracy could largely be avoided by 1) excluding SSPs or 2) introducing a minimum age. 
Here, the minimum age constraint is most effective at low redshift (see discussion below). Excluding SSPs has no effect at low redshift, but improves the mass estimate slightly at higher redshift.

\noindent Although masses are generally less underestimated with the only-$\tau$ setup at low redshift because SFHs are a better match, this setup fails for the most metal-poor objects in redshift and mass recovery (both are 0) in the reddened and unreddened case. At high redshift masses are nearly equally as well recovered as in the fixed redshift case for both setups independently of dust reddening as shown in Fig. \ref{massdtzfzf}. While at low redshift the offset in stellar masses is significantly reduced in the case of redshift as a free parameter for most objects, particularly when reddening is used in the fit, differences at high redshift are small.

\noindent In particular, for redshifts above $2$ this means that photometry is enough to accurately determine redshifts and stellar population parameters, such as stellar mass, simultaneously.

\noindent Metallicity effects are of similar size as at fixed redshift implying a 0.2-0.3 dex median difference in mass. The right metallicity provides the best mass recovery. As already found, metallicity effects are small. Hence, it is sufficient and more economic to fit star-forming galaxies with a mono-metallicity setup. 

\noindent It is common practice to introduce a cut in minimum age in the fit in order to improve upon the mass estimate \citep{Wuyts2009,Bol2009} and we demonstrated in Paper I that $age_{min}=0.1$ Gyr works sufficiently well, particularly at low redshift (see Fig. 13 in Paper I). Although the minimum age constraint does not affect the median (or mean) redshift estimate much, the mass estimate still profits from this restriction at low redshift 
such that the mean recovered mass improves from $-0.49\pm0.59$ to $-0.33\pm0.20$ dex (median improves by $\sim 0.01$ dex). At high redshift, the underestimation at the low mass end worsens with this age constraint, the mean recovered redshift is much worse (particularly at $z=3$) but the median recovered redshift is almost unchanged. We find that a simple rebinning of the age grid has a negligible effect at low redshift and can even damage the mass estimation at high redshift, similarly to our findings in Paper I. 

\noindent Differences in stellar mass recovery between using Salpeter, Kroupa and Chabrier IMF templates amount to $\sim0.2$ dex while masses derived with top-heavy IMF templates are very poorly recovered (offsets up to a median difference of 0.9 dex) and show large scatter. 

\begin{figure*}
\centering\includegraphics[width=144mm]{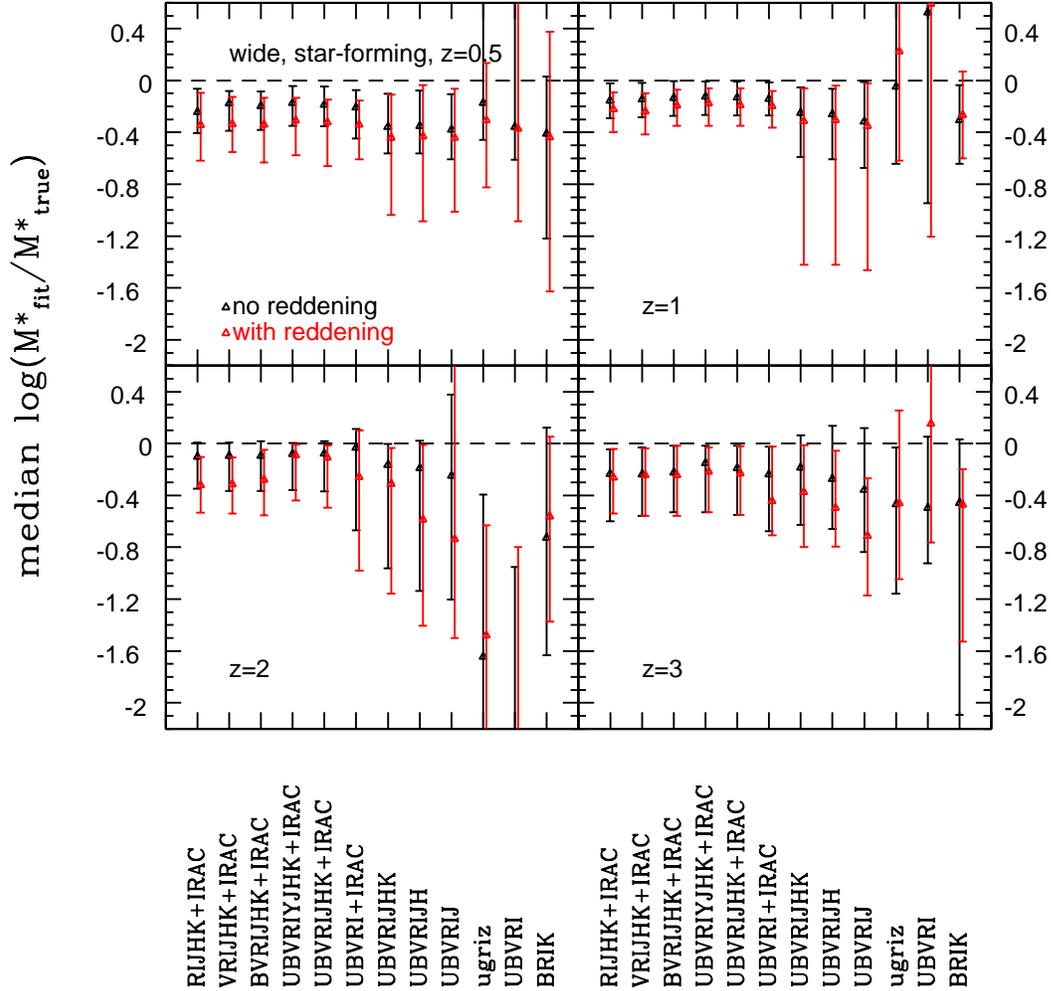}
\caption[Median mass recovery for mock star-forming galaxies as a function of wavelength coverage]{\label{massdf05zf} Median difference between true and recovered stellar mass with 68\% confidence levels as function of wavelength coverage and filter set at redshifts $0.5$, $1$, $2$ and $3$. The filter setup used in the fitting changes from from left to right in each panel. Symbols are as in Fig. \ref{massdt05zf}.}
\end{figure*}

\noindent Finally, we focus on the effect of different filter setups, i.e. wavelength coverage, on the stellar mass recovery for the wide setup (Fig. \ref{massdf05zf}). The mass recovery follows the redshift recovery such that when the redshift is poorly recovered, stellar masses are poorly recovered. For older galaxies with little on-going star formation ($z=0.5-1$) the lack of the rest-frame near-IR (IRAC bands) results in a median mass underestimated by $0.4$ dex and increased scatter. When observed near-IR filter bands are also missing, the scatter increases even further such that now masses are also overestimated. Generally, the scatter is larger when reddening is included. The lack of blue filter bands has only very little effect on the mass recovery because the effect on the redshift recovery is small, too. 
For young galaxies with high SFRs at $z=2$ excluding IRAC and near-IR filter bands in the fit results in a catastrophic failure of the mass recovery. Because the redshift recovery is better at $z=3$, masses are not underestimated as much as those at $z=2$ for similar wavelength coverages. The median recovered mass is underestimated by $\sim 0.3$ dex more than for the broadest wavelength coverage. In the reddened case masses are underestimated by $\sim0.35$ dex at $z=2$.

\noindent At each redshift, masses are best recovered when the wavelength coverage is broad. Especially, the rest-frame near-IR is crucial as already concluded by \citet{Lee2009} and \citet{Kannappan,Bol2009}, M06 and in Paper I at known redshift.

\noindent When redshift is known, we found that masses of older galaxies with little star formation (at $z=0.5$) are recovered worse than when redshift is free but depends only weakly on the wavelength coverage. At fixed, low redshift masses are underestimated by $\sim0.3$ dex in the unreddened case and $\sim0.7$ dex in the reddened case due to the SFH mismatch between template and galaxy and overshining. The scatter is also large. At $z=1$, the main effect is that scatter increases when wavelength coverage decreases. At $z\geq2$ the dependence on wavelength coverage when using the wide setup is similar to the one shown here but less strong. e.g. the median recovered mass is underestimated by one order of magnitude in the reddened case for UBVRI at $z=2$. Masses are better determined at fixed, high redshift when the rest-frame UV is excluded in the fit because the fit is carried out on the mass-sensitive part of the SED while ignoring the contribution of recently formed stars and shortcomings in matching the true SFHs are compensated. When templates that match the real SFHs of these galaxies are used in the fit (inverted-$\tau$ models at $z\geq2$), the stellar mass recovery is nearly independent of the wavelength coverage and mainly scatter increases for narrow wavelength coverages. Besides the SFH mismatch and overshining, here, a large factor is also the recovery of the correct redshift. 

\noindent Excluding SSPs reduces the scatter in the reddened case for fitting setups adopting a narrow wavelength coverage at $z\leq2$ as long as at least 6 filter bands and either near-IR or IRAC bands are added to the optical. At $z=3$ the median mass recovery improves for filter setups excluding the UV. 

\noindent In summary, we find that although stellar masses of star-forming galaxies are still underestimated, they are better determined when redshift is a free parameter in the fit, independently of template setup. This is valid for a redshift recovery within $\sim0.1$ at low redshift. At low redshift, some catastrophic failures occur however, due to catastrophic failures in redshift. Metallicity adds an uncertainty of $\sim0.2$ dex to the mass estimate. The same is true for IMF effects. The mass estimate can be improved with an artificial age constraint in the fit for old galaxies, whereas for young galaxies this does not help. As in the fixed redshift case, masses are best determined using a broad wavelength coverage in the fit because then also redshifts and ages are best determined.

\subsubsection{Star formation history and star formation rate}\label{sfhresults}
\begin{figure*}
\centering\includegraphics[width=144mm]{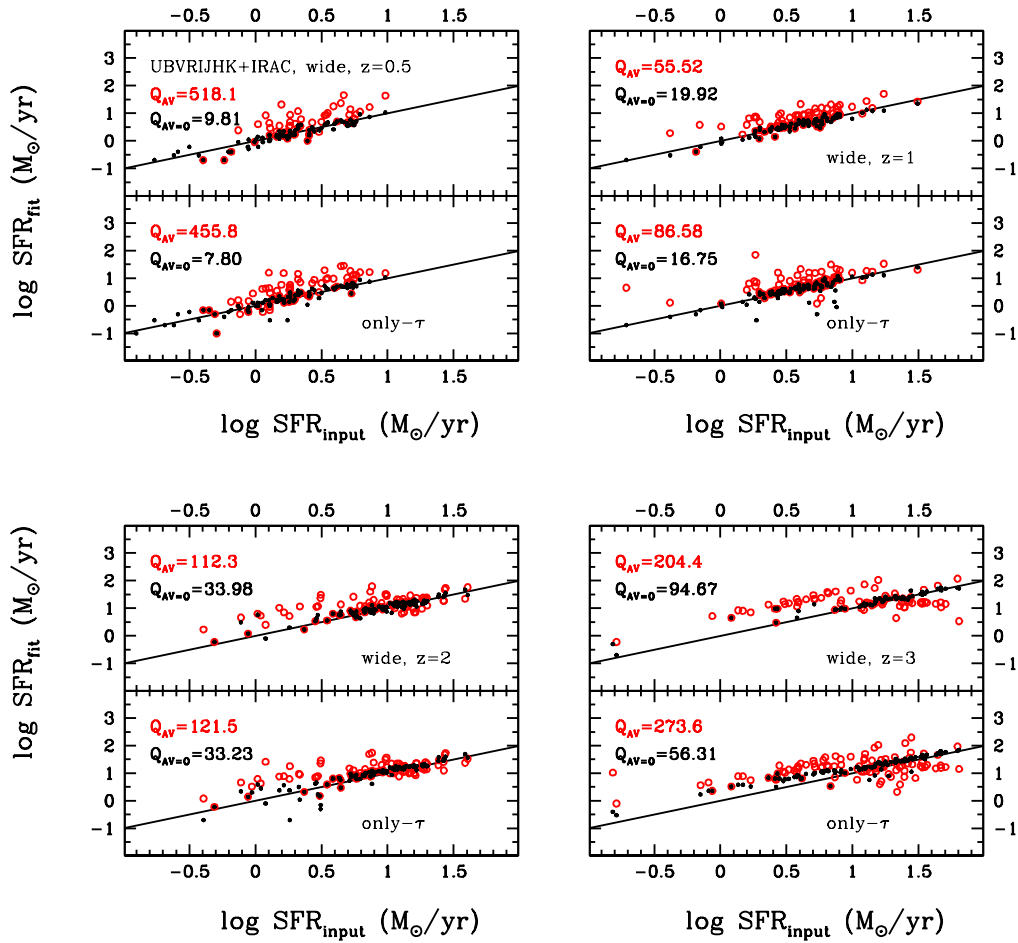}
\caption[Star formation rate recovery for mock star-forming galaxies as a function of template setup]{\label{sfrdt05zf} Comparison between SFR derived from SED-fitting and input SFR as a function of template setup and redshift. The template setups are wide (top in each panel) and only-$\tau$ (bottom in each panel). Symbols are the same as in previous figures. SFR that are zero are not displayed.}
\end{figure*}
\noindent In Fig. \ref{sfrdt05zf} we show the recovery of SFR as a function of template setup. Qualitatively, the SFR recovery is similar to the one at fixed redshift shown in Fig. 18 of Paper I. When redshift is left free, though, SFRs are overestimated less at $z\leq1$ and underestimated more at $z=3$. The mismatch between template and mock galaxy SFH is the biggest driver at redshift 3 where inverted-$\tau$ models (with high formation redshift and $\tau>0.3$) recovered the SFRs best as shown in M10 and Paper I. The effect of the unknown redshift is secondary here.

\noindent Similarly to the mass estimate, SFRs are best recovered when a minimum age constraint is applied in the fitting. 
A simple rebinning of the age grid on the other hand results in higher SFRs at low redshift and in underestimated SFRs at high redshift. These results are very similar to the ones obtained at fixed redshift in Paper I (see their Fig. 19).

\noindent SFRs obtained with Kroupa, Chabrier or top-heavy IMF templates show the same behaviour independently of redshift as a free or fixed parameter (see Paper I). The effect of wavelength coverage is also similar to the fixed redshift case. 

\noindent In summary, when redshift is a free parameter in the fit SFRs are best recovered with the broadest wavelength coverage and a wide template setup. SFR estimates can be improved at low redshift using an age constraint. Using templates with lighter IMFs results in underestimated  SFRs. Reddening causes SFRs to be overestimated instead. These conclusions are similar to the fixed redshift case.\\

\subsection{Passive galaxies}\label{resultspasszf}
\subsubsection{Photometric redshift}\label{zresultspasszf}
\begin{figure*}
\centering\includegraphics[width=124mm]{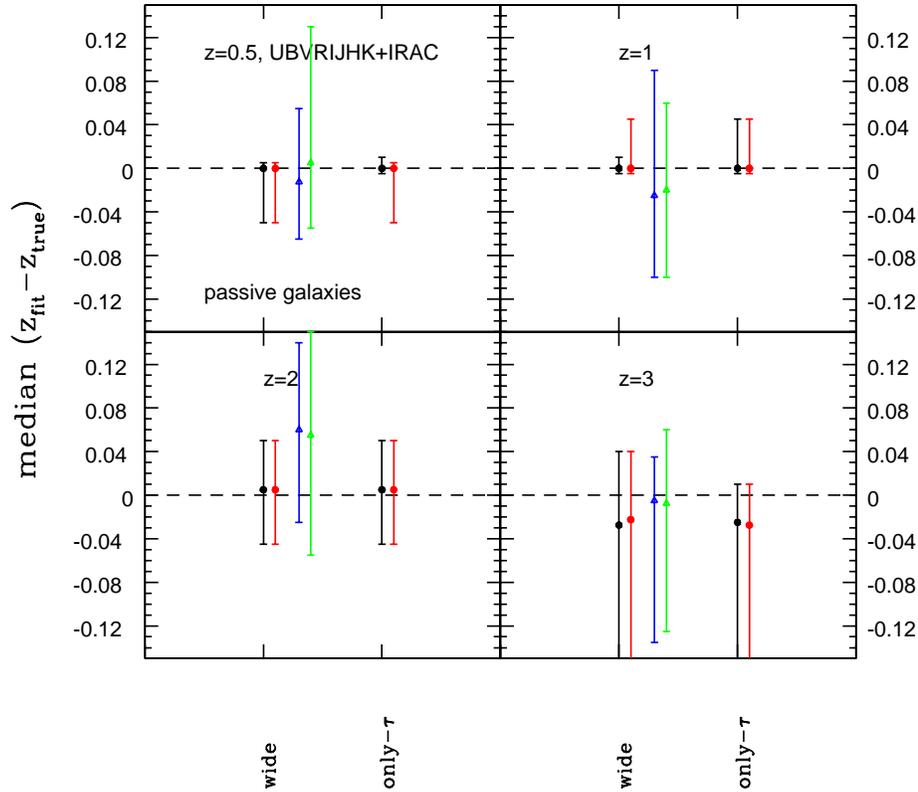}
\caption[Median redshift recovery for mock passive galaxies as a function of template setup]{\label{OLDzhistdtzf} Median redshift recovery for mock passive galaxies as a function of template setup, \textbf{namely wide and only-$\tau$}, using a broad wavelength coverage (UBVRIJHK+IRAC). Black symbols refer to the case without reddening, red ones to the reddened case. The green and blue symbols show the redshift recovery for a wide template setup in the case of mock star-forming galaxies with and without reddening, respectively. Errorbars are 68\% confidence levels.}
\end{figure*}

In Fig. \ref{OLDzhistdtzf} we show the redshift recovery of mock passive galaxies as a function of template setup. Due to their age and passive nature which results in a well defined 4000 \AA \space break we expect the redshift estimation for passive galaxies to be better than those for star-forming galaxies. Overall, redshifts are recovered very well (small or no offsets combined with small scatter), independently of template setup in the unreddened case. Redshifts of $z=3$ objects are recovered slightly worse ($\Delta$ z $\leq0.1$ and larger scatter) than those of lower redshift ones because of the slight mismatch in age\footnote{For ages of $1$ and $1.5$ Gyr the closest matching age of the template age grids are $1.015$ Gyr  and $1.434$ Gyr.} which is compensated by a combination of older age, prolonged SFH and lower redshift. In the reddened case we find a slight increase in scatter at lower redshifts. For the wide setup, the redshift recovery fails for a few objects (5\%) because of degeneracies between age, dust, metallicity and redshift. Instead of being old, these objects are misidentified as very young, very dusty, high metallicity, high redshift single bursts.
\begin{figure*}
\centering\includegraphics[width=144mm]{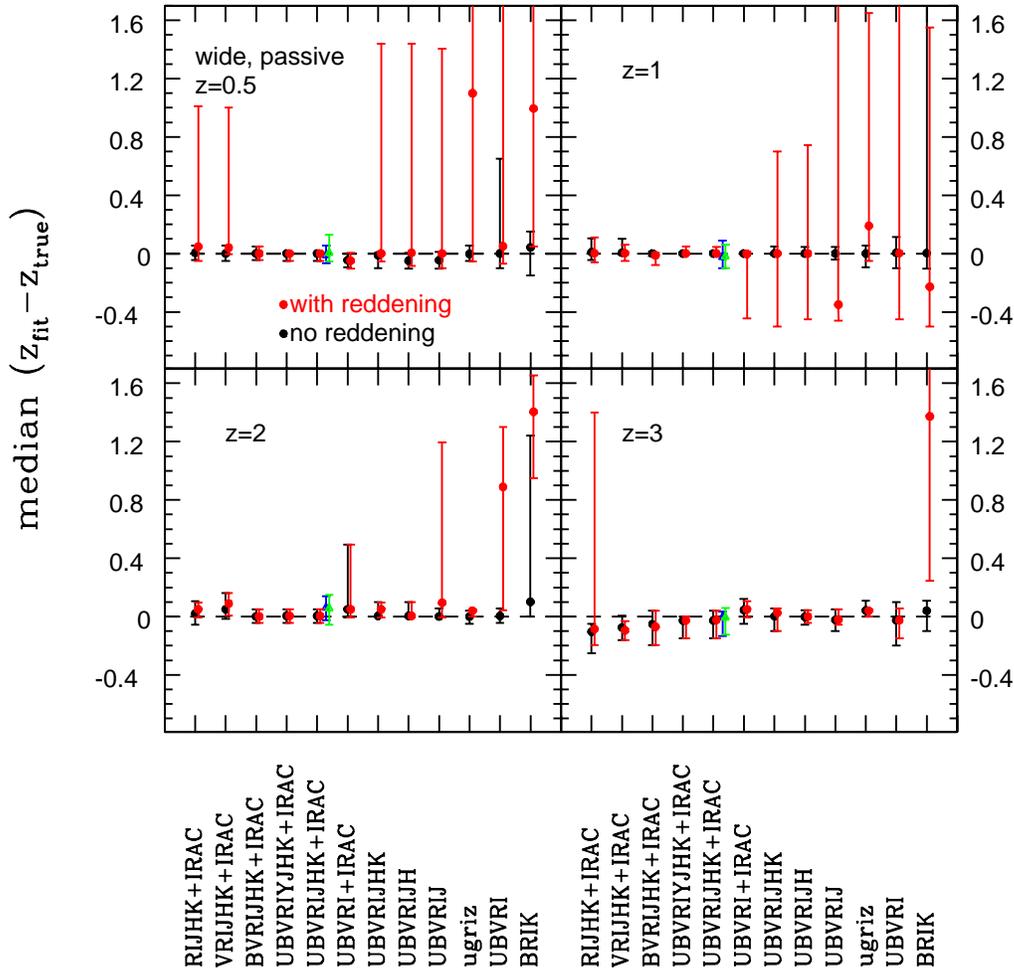}
\caption[Median redshift recovery for mock passive galaxies as a function of wavelength coverage]{\label{OLDzhistdfzf} Median redshift recovery with 68\% confidence levels for mock passive galaxies as a function of wavelength coverage using a wide setup in the fit. Symbols are the same as in Fig. \ref{OLDzhistdtzf}.}
\end{figure*}

\noindent Redshifts are best recovered with templates of the right metallicity. When the template metallicity is slightly lower or higher than the true metallicity, the SFH compensates such that redshifts are still recovered within $\pm0.1$. Redshifts are systematically overestimated for the lowest metallicity templates although lots of objects are fit with the right SFH (namely a SSP). Here, very young ages, large amounts of dust and an overestimated redshift fail in compensating the blue colour required by the metallicity.

\noindent Fig. \ref{zhistdIMFzf} showed that the IMF plays a minor role with respect to the redshift recovery of mock star-forming galaxies. We find the same for the mock passive galaxies and hence abstain from displaying it here. As for the star-forming galaxies, the right IMF cannot be identified for passive galaxies by pure means of minimum $\chi^2_{\nu}$.

\noindent Fig. \ref{OLDzhistdfzf} summarises the effect of wavelength coverage on the redshift determination for passive galaxies. Overall, like for star-forming galaxies, the redshifts of passive galaxies are best determined with a broad wavelength coverage. In the unreddened case, the wavelength dependence is weak, offsets and scatter are very small in most cases. For the most restricted wavelength coverages (mainly for a BRIK filter set) and youngest galaxies scatter increases such that redshifts are overestimated. In the reddened case, the redshift estimate becomes less robust and shows large scatter when the rest-frame near-IR (near-IR and IRAC filter bands) are omitted for the oldest galaxies. Likewise a lack of blue filter bands results in overestimated redshifts at $z=0.5$ because the rest-frame blue and thus the 4000 \AA \space break is not covered anymore. At higher redshift, redshifts are overestimated when the wavelength coverage is most restricted. Additionally, a lack of near-IR filters results in a worse redshift recovery because the 4000 \AA \space break is excluded.

\noindent In summary, the redshift of passive galaxies can be very well determined, independently of SFH and IMF assumed in the template. Metallicity plays an important role in the redshift recovery hence passive galaxies should be fit with a wide range in metallicities. As for star-forming galaxies, redshifts of passive galaxies are best determined when a broad wavelength coverage is used in the fit although if reddening is small this requirement is less crucial. Very narrow wavelength coverages result in overestimated redshifts, especially when reddening is included because of degeneracies between age, dust and redshift.

\subsubsection{Age}\label{ageresultspass}
\begin{figure*}
\centering\includegraphics[width=124mm]{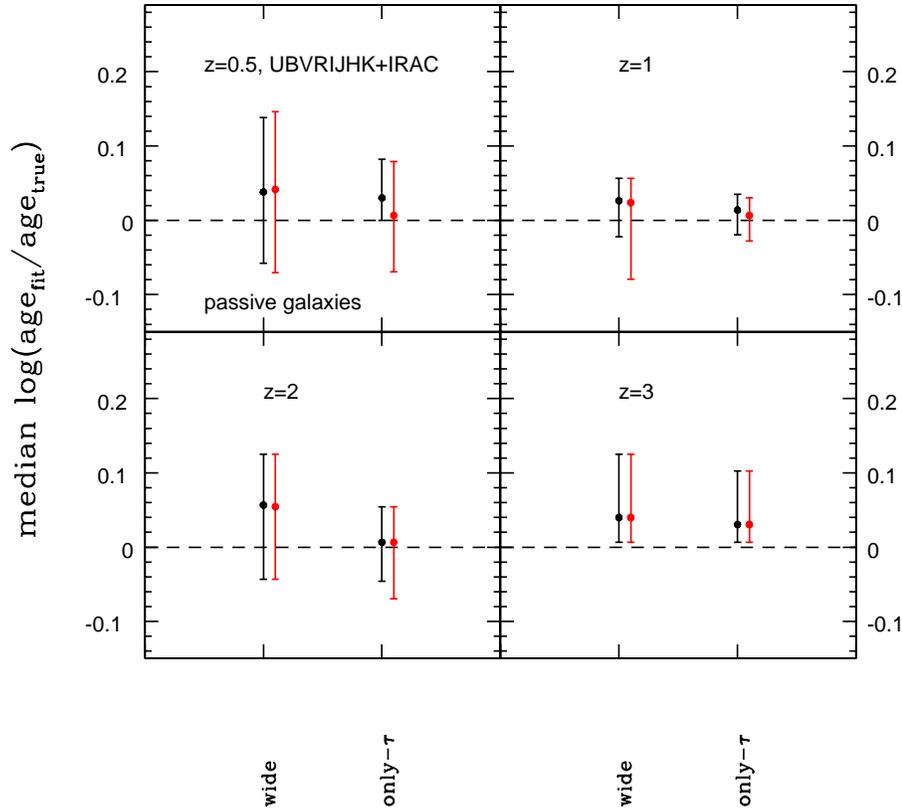}
\caption[Median age recovery for mock passive galaxies as a function of template setup]{\label{oldagedtzf} Recovery of age (median) for mock passive galaxies for different template setups and redshifts. Symbols and errorbars are the same as in previous figures.}
\end{figure*}
The effect of template setup on the derived ages of passive galaxies is similar to the fixed redshift case in Paper I. Fig. \ref{oldagedtzf} compares the performance of the wide and only-$\tau$ template setups\footnote{Results for an only SSP and only solar SSP can be found in Appendix \ref{SSPsetups}.}. Overall, the median recovered ages show very little offset ($\leq0.05$ dex). The wide setup tends to overestimate ages and shows the largest scatter at each redshift due to the wide range in star formation histories and metallicities included in this setup. Ages are still slightly overestimated for the only-$\tau$ setup although it contains only the correct metallicity (solar in our case) because the star formation history mismatch is compensated by older ages although $\tau$-models with short values of $\tau$ closely resemble a SSP. Generally, when redshift is underestimated, ages are overestimated. At $z=3$ both template setups overestimate ages equally. We have seen in Paper I that this is caused by a small mismatch between galaxy age and template age grid\footnote{For ages of 1 and 1.5 Gyr the closest matching age of the template age grids are 1.015 Gyr  and 1.434 Gyr} and photometric uncertainties. This in turn causes underestimated redshifts at $z=3$. The effect of including reddening only increases the scatter by a small amount. Overall, ages are best determined with a template of the right SFH and metallicity (compare Appendix \ref{SSPsetups}) but even setups that inhibit a wider range of SFHs and metallicities recover ages within $\pm0.06$ dex (median offset). This is only slightly larger than at fixed redshift.
\begin{figure*}
\centering\includegraphics[width=144mm]{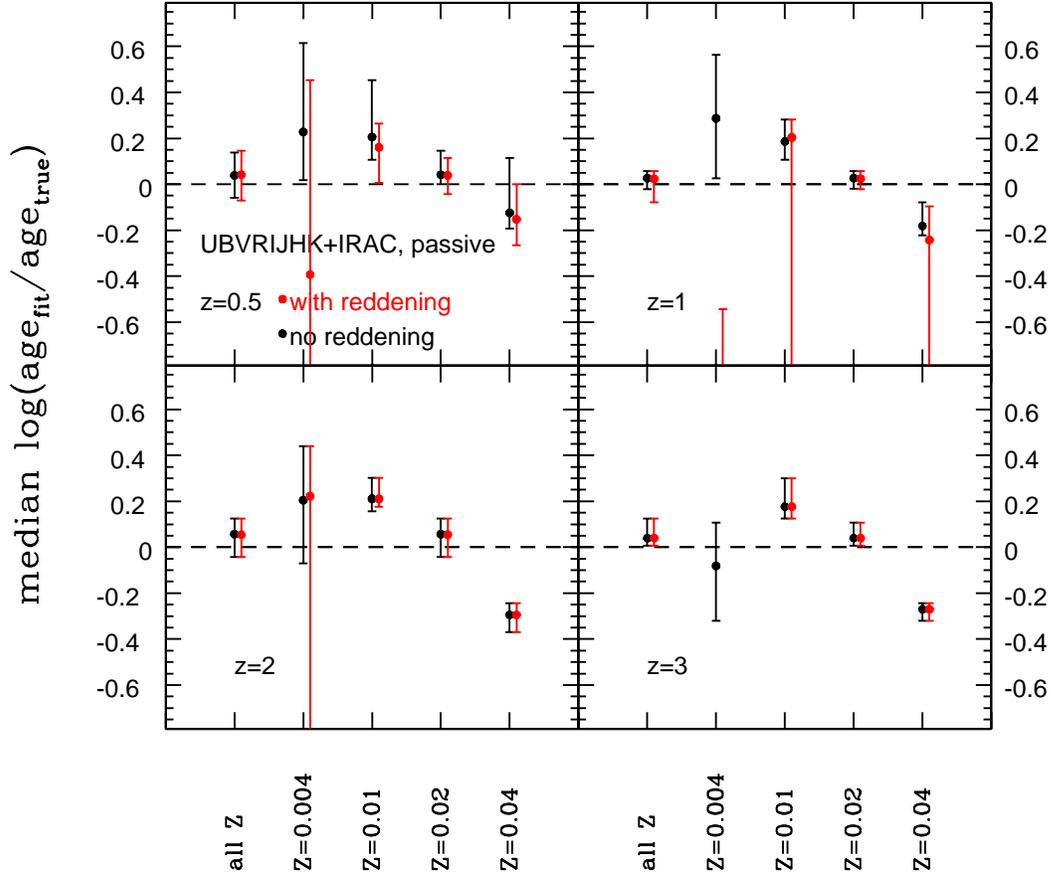}
\caption[Median age recovery for mock passive galaxies as a function of metallicity]{\label{oldagedmetalzf} Median age recovery of mock passive galaxies as a function of redshift and metallicity. Symbols and errorbars are the same as in previous figures. The missing point at $z=1$ for Z=0.004 in the reddened case lies at $-2.55_{-0.60}^{+2.01}$, and at $z=3$ at $-2.28_{-0.35}^{+0.03}$.}
\end{figure*}
\noindent We further investigate metallicity effects on the age recovery in Fig. \ref{oldagedmetalzf} by using mono-metallicity wide setups. Obviously, ages derived with a solar metallicity setup are recovered best. The remaining scatter stems from the SFH mismatch. In the unreddened case, fits with sub-solar metallicity template setups result in overestimated ages because older ages compensate bluer colours due to metallicity. Although most galaxies get assigned the right SFH (a SSP) - meaning the reddest solution is picked by the fit - for the lowest metallicity template setup, ages are still required to be maximally old to compensate for the underestimated metallicity. Yet, this is not effective enough if the true galaxy age is already very old, hence the fit tries to compensate with overestimated redshifts as well. Additionally, as in Paper I, all $\chi_{\nu}^2$ are larger than $2$, most are larger than $10$, and fits are getting worse towards higher redshift. No other parameter in the fit, i.e. age, SFH and redshift, can compensate such a large discrepancy in metallicity. At the highest redshift, ages are underestimated because a combination of longer star formation, younger age and very high redshift gives the best solution ($\chi_{\nu}^2$ are still large). For the highest metallicity template setup, ages are underestimated because of the age-metallicity degeneracy.

\noindent Compared to the case of known redshift, here underestimated (overestimated) redshifts compensate some of the effect. Consequently, ages are underestimated (overestimated) by $\sim0.1$ dex less than in the fixed redshift case. 
The inclusion of reddening affects the age derivation mostly when sub-solar metallicity template setups are used in the fit. For these dust reddening is an additional way to redden galaxy colours. However, because of the age-dust degeneracy galaxies are now identified to be very young (offsets of more than $2$ dex for the lowest metallicity in some cases) and dusty instead of old and dust-free and the scatter is very large (more than $2$ dex). Even though recovered ages are offset when template setups of the wrong metallicity are used in the fit, the age determination works very well when the template setup contains a wide metallicity range because the correct metallicity is identified for most objects and thus ages can be very well recovered.

\noindent Similar to the fixed redshift case, the age recovery is almost independent of the template IMF.
\begin{figure*}
\centering\includegraphics[width=144mm]{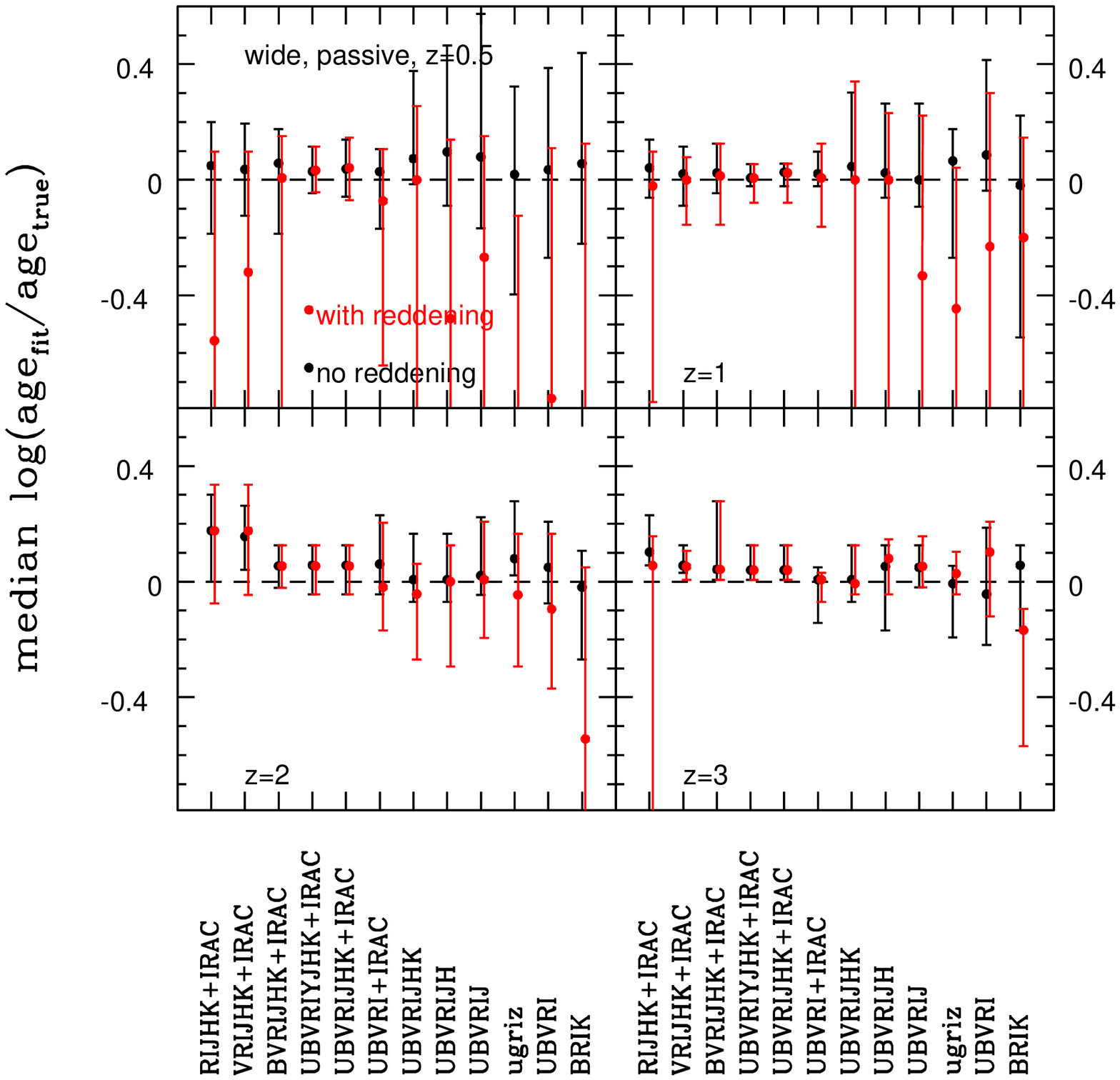}
\caption[Median age recovery for mock passive galaxies as a function of wavelength coverage]{\label{oldagedfzf} Median age recovery of mock passive galaxies as a function of wavelength coverage at redshifts from $0.5$ to $3$. Symbols and errorbars are the same as in previous figures. Missing points at $z=0.5$ lie at $-0.88^{+0.76}_{-1.42}$ for ugriz, at $-2.31^{+2.43}_{-0.72}$ for BRIK and at $z=2$ at $-0.54^{+0.59}_{-0.65}$ for BRIK.}
\end{figure*}

\noindent The dependence of the age recovery on the wavelength coverage shown in Fig. \ref{oldagedfzf} reveals that median ages are recovered within $\sim0.1$ dex for most filter setups in the unreddended case. However, scatter increases with decreasing wavelength coverage. At high redshift setups lacking the bluest filter bands show larger offsets between true and recovered age, ages are overestimated, redshifts are underestimated and star formation histories are mismatched. In comparison to the fixed redshift case, the scatter in the age determination is larger when redshift is an additional free parameter. The age recovery becomes very poor, particularly at low redshift, when reddening is included and the filter coverage lacks rest-frame near-IR and rest-frame red optical filter bands. Fig. \ref{OLDzhistdfzf} illustrated that also redshifts show large offsets and scatter in these cases. This is clearly caused by the degeneracy between age, dust and redshift as the main driver. At low redshift, reddening and redshifts are overestimated, therefore ages have to be underestimated to compensate. The lack of blue filter bands has similar effects. For the youngest galaxies (at redshift $2$) ages are underestimated when the reddest rest-frame filter bands are missing mainly because redshifts are overestimated and show large scatter. When blue rest-frame filter bands are omitted in the fit ages are similarly overestimated as in the unreddened case. At $z=3$, the large scatter in the reddened case for a RIJHK+IRAC filter setup is generated by the degeneracy with redshift which also shows large scatter (see Fig. \ref{OLDzhistdfzf}).
Although we found that a BRIK filter setup recovers ages similarly well as wider wavelength coverages in the case of known redshift, the age recovery fails when redshift is a free parameter because redshifts cannot be constrained reasonably well with these four filter bands alone. Overall, ages are best derived using the full wavelength coverage. Ages of aged and passive galaxies can be better determined from SED-fitting than those of star-forming galaxies because the 4000 \AA \space break is well defined and overshining from young populations is absent.

\noindent In summary, ages are best determined with templates of the right SFH and metallicity but even setups that contain a wider range of SFHs and metallicities recover ages within $\pm0.06$ dex (median). Aged galaxies ($>1$Gyr) should be fit with a wide range in metallicity, as 1) the correct metallicity can be identified in the fit and 2) using the wrong metallicity offsets the recovered ages by $\pm0.2-0.3$ dex in the unreddened case and fails when reddening is included. The choice of IMF has very little influence. The inclusion of reddening in the fit has overall very little effect on the derived ages and mostly increases the scatter. The best age determination is achieved when a broad wavelength coverage is used in the fit. The lack of rest-frame near-IR and red optical filter bands leads to underestimated ages. 

\subsubsection{Stellar Mass}\label{massresultspasszf}
\begin{figure*}
\centering\includegraphics[width=124mm]{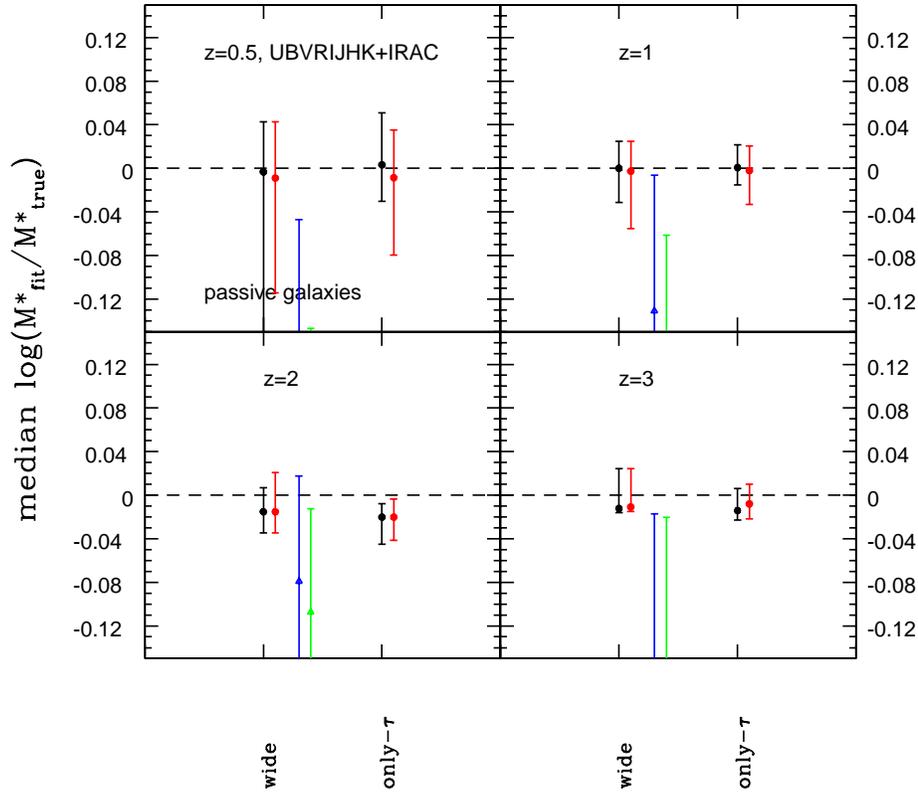}
\caption[Median mass recovery for mock passive galaxies as a function of template setup]{\label{oldmassdtzf} Mass recovery (median) of mock passive galaxies achieved with different template setups as a function of redshift. Symbols and errorbars are the same as in previous figures.}
\end{figure*}
\begin{figure*}
\centering\includegraphics[width=144mm]{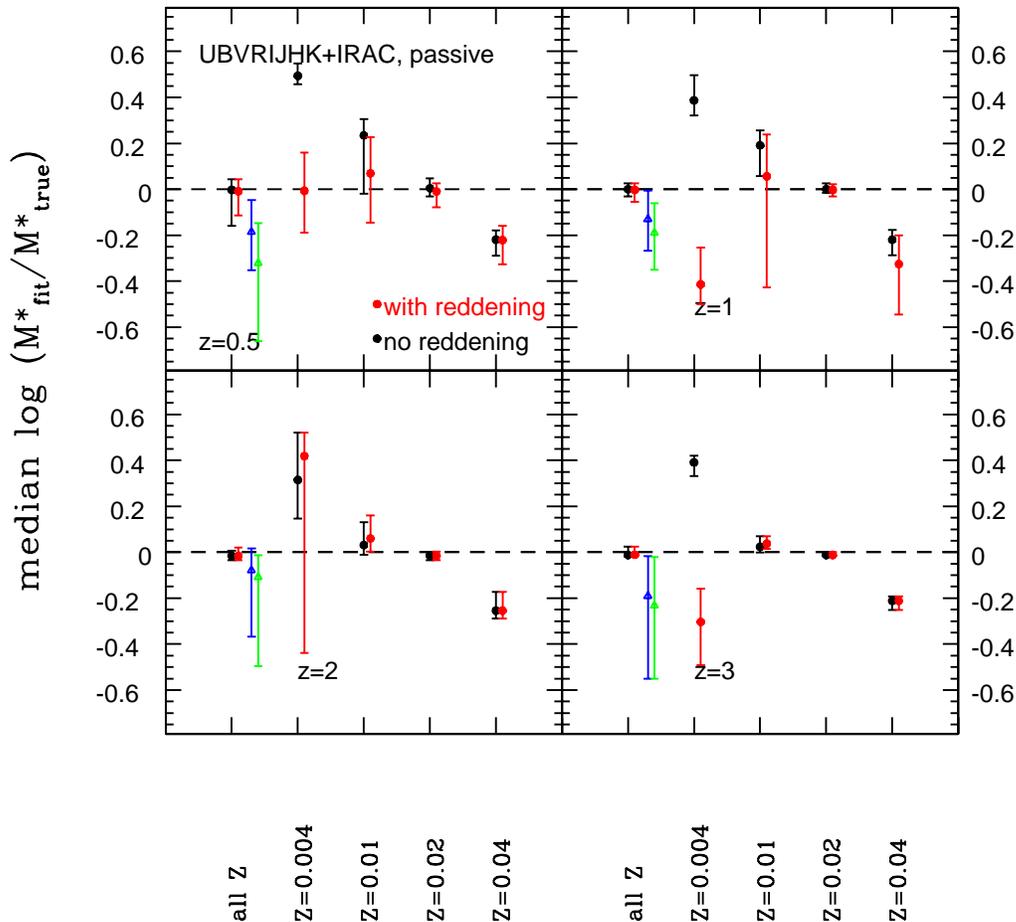}
\caption[Median mass recovery for mock passive galaxies as a function of metallicity]{\label{oldmassdmetalzf} Mass recovery (median) of mock passive galaxies as a function of template metallicity. The input metallicity is solar. Symbols are the same as in Fig. \ref{OLDzhistdtzf}.}
\end{figure*}
\begin{figure*}
\centering\includegraphics[width=144mm]{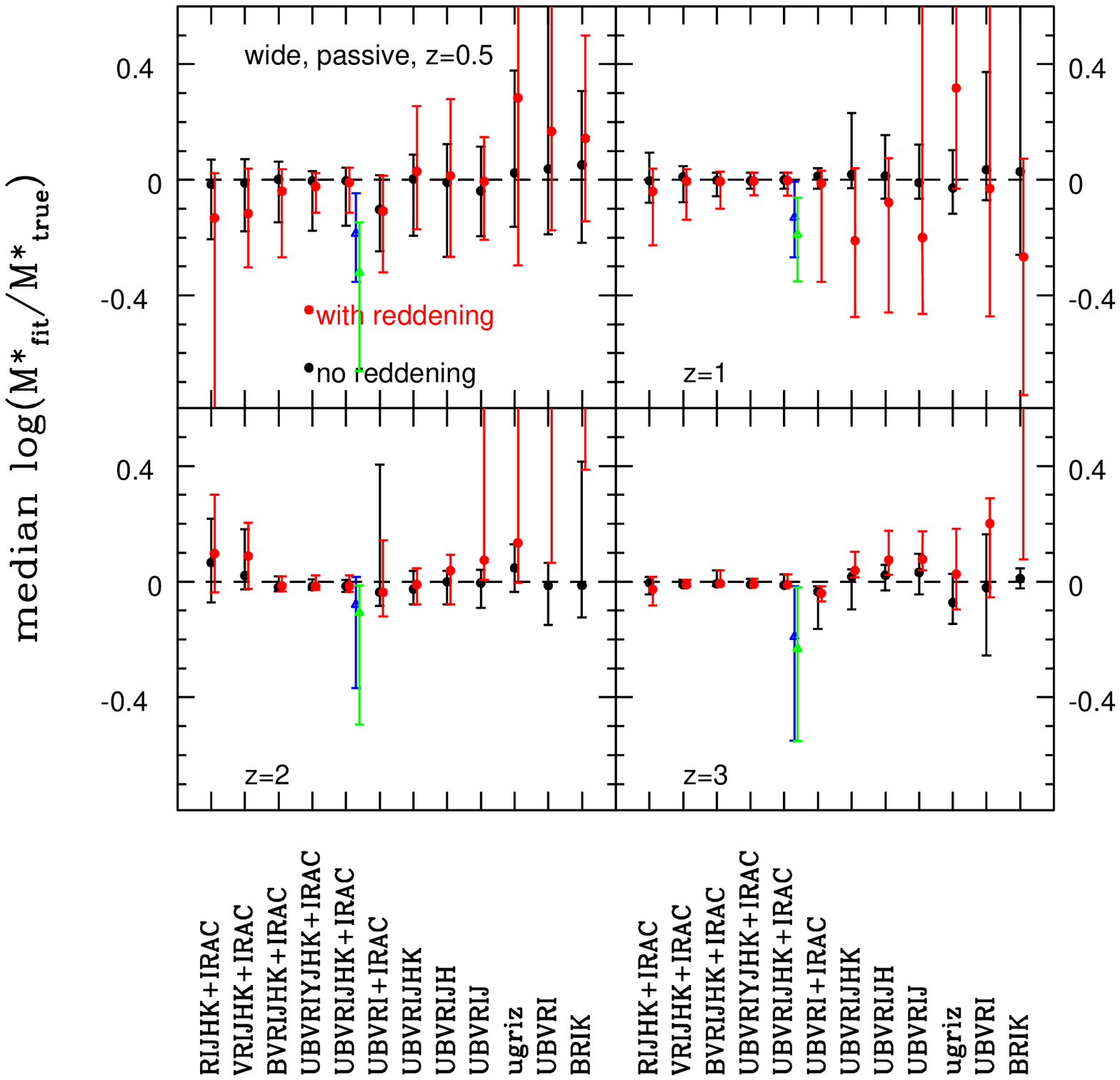}
\caption[Median mass recovery for mock passive galaxies as a function of wavelength coverage]{\label{oldmassdfzf} Mass recovery (median) of mock passive galaxies as a function of wavelength coverage. Symbols are the same as in Fig. \ref{OLDzhistdtzf}. Missing points in the reddened case at $z=2$ are at $0.64_{-0.57}^{+1.56}$ (UBVRI) and at $0.81_{-0.42}^{+0.19}$ (BRIK) and at $z=3$ at $0.78_{-0.71}^{+0.63}$ (BRIK).}
\end{figure*}
\noindent We found in Paper I that stellar masses of passive galaxies can be recovered very well. In this section we investigate if that is still true when redshift is not known. We find that even in this case the median masses of passive galaxies are recovered within $\sim0.02$ dex for both template setups\footnote{This is also true for only SSP setups as shown in Appendix \ref{SSPsetups}. (Fig. \ref{oldmassdtzf}) as at fixed redshift. Scatter is largest for the oldest galaxies (low z) and decreases for younger ones (higher z).} At higher redshift masses tend to be underestimated for all template setups. This is generated by a mismatch in SFH and degeneracies between redshift, SFH and age. Note that also photometric uncertainties and mismatch in age between template age grid and mock galaxy age contribute (see Sections \ref{ageresultspass} and \ref{photuncreszf}). Scatter is smaller than at fixed redshift because redshift compensates. Differences between the reddened and unreddened case are small.

\noindent In the unreddened case the effect of metallicity on the stellar mass recovery is very clear. When metallicity is underestimated masses are overestimated by $\sim0.4$ dex. When metallicity is overestimated masses are underestimated by $\sim0.2$ dex. In comparison to the fixed redshift case for which the mass determination with low metallicity template setups fails at most redshifts, here redshift overcompensates. Overall, masses are best recovered with a solar metallicity template setup because it matches the input metallicity. Since metallicity is recovered in the fit a wide template setup with a broader choice in metallicity recovers masses equally well.

\noindent We established in Paper I that the choice of IMF influences the overall fit only very little. Furthermore, the result for the mass recovery is very similar to the one at fixed redshift (Fig. 30 in Paper I), thus we do no repeat it here. Instead, we summarise the effect into four points:\\
1) Masses are best recovered with a Salpeter IMF template setup.\\
2) The masses obtained using a Kroupa or Chabrier IMF template setup are underestimated at each redshift by $\sim 0.2$ dex. Those obtained with a top-heavy IMF template setup are overestimated by $\sim 1$ dex.\\
3) The correct IMF cannot be identified among all IMFs by means of the minimum $\chi^2_{\nu}$.\\
4) Reddening has very little impact on the previous three points.\\
\noindent We discussed in Paper I that the difference in stellar mass estimates with various IMFs stems from their different mass-to-light ratios ($M^*/L$). As already concluded there and in the previous sections, we again confirm here the notion that, although the choice of IMF is felt in the fit, an identification of the correct IMF based on the minimum $\chi^2_{\nu}$ is impossible.

\noindent We present the dependence of the derived stellar masses on the wavelength coverage in Fig. \ref{oldmassdfzf}. The median recovered stellar mass depends only little on the wavelength coverage in the unreddened case. The biggest effect is an increase in scatter when the wavelength range is restricted because mass is directly correlated with age and redshift, which both have large uncertainties. As for the age and redshift determination the recovery of stellar masses is greatly hampered by the inclusion of reddening due to the many degeneracies. Particularly, a lack in rest-frame near-IR and red optical coverage results in vastly overestimated masses. These filter bands are crucial for the correct redshift and age determination at high redshift because they include the 4000 \AA \space break and contribution of the TP-AGB. At low redshift masses are underestimated when blue filter bands are missing because they cover the rest-frame wavelength range which is most crucial for the redshift and age determination (i.e. the 4000 \AA \space break). At each redshift masses are best recovered when the full wavelength range is used in the fit. Compared to the case of known redshift, masses are more affected by a lack of wavelength coverage and the inclusion of reddening when redshift is a free parameter.

\noindent In summary, stellar masses of passive galaxies can be recovered within $\pm0.02$ dex when a wide setup and a broad wavelength coverage is used even when redshift is a free parameter in the fit. Differences between template setups are small. Metallicity effects can bias the recovered mass by at most $\sim0.5$ dex. Masses are more sensitive to the adopted wavelength coverage and are overestimated when rest-frame near-IR and red optical filter bands are missing. The inclusion of reddening in the fits then generally worsens the mass recovery.

\subsection{Measuring the strength of the latest starburst}\label{burst}
\begin{table}
\caption[Redshifts, ages and masses for galaxies with young starbursts]{\label{bursttab2zf}Derived redshifts, ages and stellar masses for a $10^{11}M_{\odot}$ galaxy at redshift $1$ consisting of a $5$ Gyr old population and either 1 or 10\% of a younger population, respectively. We list characteristics of the burst in the first column, and derived redshifts, ages and masses in the second, third and fourth columns, respectively.}
\begin{center}
\begin{tabular}{@{}lccc}\hline
Burst & derived redshift & derived age & derived $M^*$ \\
no reddening & & & [$M_{\odot}$]\\\hline
1\% 1Gyr	&	0.645	&	3.5 Gyr	&	10.91\\\hline
1\% 100Myr	&	0.965 	&	4.5 Gyr	&	10.83\\\hline
1\% 10Myr	&	1.295	&	3.0 Gyr	&	10.89\\\hline
1\% 1Myr	&	1.240	&	47.5 Myr	&	9.73\\\hline
10\% 1Gyr	&	1.025	&	2.75 Gyr&	10.87\\\hline
10\% 100Myr&	1.125	&	227 Myr	&	10.60\\\hline
10\% 10Myr	&	0.890	&	8.7 Myr	&	9.82\\\hline
10\% 1Myr	&	1.190	&	6.6 Myr	&	10.04\\\hline\hline
Burst & derived redshift & derived age & derived $M^*$ \\
+ reddening & & & [$M_{\odot}$]\\\hline
1\% 1Gyr	&	0.645	&	3.0 Gyr	&  	10.60\\\hline
1\% 100Myr	& 	1.240	&	6.3 Myr	&	9.74\\\hline
1\% 10Myr	&	0.630	&	8.7 Myr	&	8.84\\\hline
1\% 1Myr	& 	1.010	&	13.8 Myr	&	9.31\\\hline
10\% 1Gyr	&	0.605	&	4.75 Gyr	&  	10.70\\\hline
10\% 100Myr	&	0.690	&	7.6 Myr	&	8.98\\\hline
10\% 10Myr	&	0.890	&	8.7 Myr	&	9.82\\\hline
10\% 1Myr	&	1.190	&	6.6 Myr	&   	10.04\\\hline
\end{tabular}
\end{center}
\end{table}%
\noindent The results of M10 and Paper I clearly show that recent star formation influences the fit such that a robust recovery of age, stellar mass, reddening and star formation rate of star-forming galaxies \emph{simultaneously} is very difficult. We further investigated whether the recovered properties then reflect the characteristics of the latest starburst rather than that of the entire galaxy. Here, we carry out the same exercise leaving the redshift free. We use simulated old ($5$ Gyr) and passive galaxies at $z=1$ to which we add a small percentage (1 or 10 \% in mass) of a younger population ($1$ Myr, $10$ Myr, $100$ Myr or $1$ Gyr). The total mass of each simulated galaxy is $10^{11}\,M_{\odot}$. If the burst properties are recovered, then we expect stellar masses to be $10^{9}\,M_{\odot}$ and $10^{10}\,M_{\odot}$ for 1 and 10 \% of young component, respectively. Mass-weighted ages would be $\sim4.5$ (for 10\% young) and $\sim4.95$ Gyr (for 1\% young).\\

\noindent In the fitting we use the broadest wavelength coverage and a wide setup. For simplicity, we stick to solar metallicity for the simulated galaxies and the fitting setup. Results for redshift, age and stellar mass are provided in Table \ref{bursttab2zf}.

\noindent In all cases derived ages and masses are underestimated. Redshifts deviate by up to $0.4$ from the true redshift but because of the many degeneracies there is no systematic trend. Compared to the results obtained at known redshift, ages are similar and much closer to the age of the burst when the burst component accounts for 10\% and is younger than a few $100$ Myr or very young and makes up only 1\%. The mass estimate is improved somewhat when the redshift is overestimated. When reddening is included in the fit ages are very young unless the burst component is close in age to the old component. Due to the age-dust degeneracy ages can even be younger than the age of the burst component. Compared to the fixed redshift case the burst component starts to affect already when the percentage is very low and reddening is included. The combination of underestimated ages with underestimated redshifts results in stellar masses that only reflect the mass of the burst. In some cases the mass is even lower than the burst mass.

\noindent Overall, we confirm that also when redshift is a free parameter in the fit, young burst components can hide a large percentage of the total stellar mass, especially when there is a large gap between the age of the burst and the age of the old component.

\subsection{Reddening laws}\label{rlzf}
\begin{figure}
\centering\includegraphics[width=84mm]{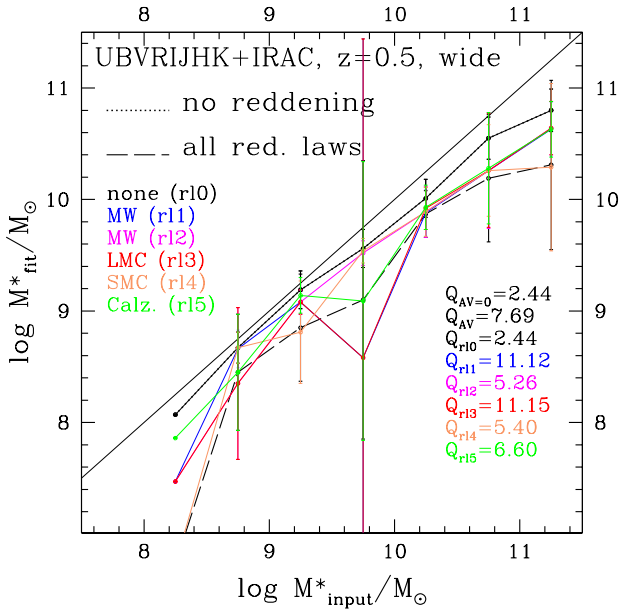}
\caption[Average mass recovery for mock star-forming galaxies as a function of reddening law]{\label{massdtrl05zf} Average mass recovery for different reddening laws at redshift $0.5$. The dotted black line stands for the mass recovery both fit and spectra are free of reddening ($Q_{AV=0}$), the dashed thick black line represents the case that for each object the best fit out of all reddening laws is chosen ($Q_{AV}$). The thin black line shows the case of fitting reddened spectra without reddening ($Q_{rl0}$). Thin coloured lines show the results when only one reddening law is applied in the fit of reddened spectra. We list quality factors $Q$ for the entire mass range.}
\end{figure}
\begin{figure}
\centering\includegraphics[width=84mm]{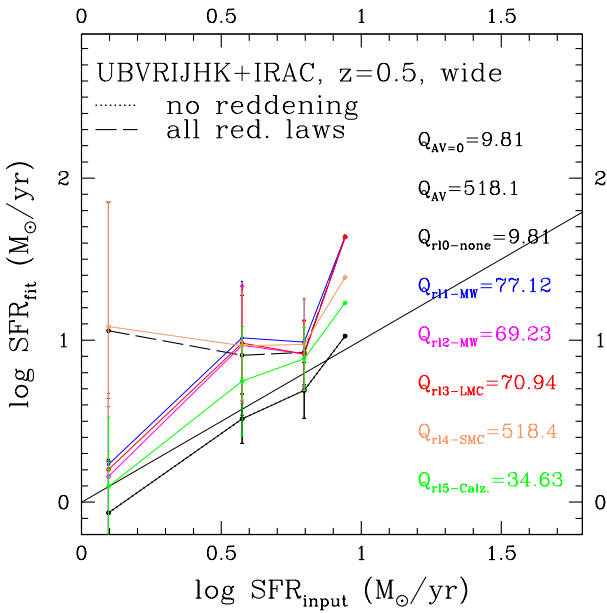}
\caption[Average SFR recovery for the mock star-forming galaxies as a function of reddening law]{\label{sfrdtrl05zf} Average SFR recovery at redshift $0.5$ for different reddening laws used in the fit. Colours and lines are the same as in Fig. \ref{massdtrl05zf}. The data point at the highest SFR at $z=0.5$ is shown for completeness because it consists of only one object. Quality factors $Q$ are listed for the entire SFR range.}
\end{figure}
\noindent In this section we study whether the consideration of only one reddening law in the SED fit, which is the best choice for achieving fast results in case of large numbers of galaxies, penalises the fit results in any way. M06 showed that the highest accuracy for real passive galaxies at redshift 2 is achieved by choosing the best fit among all available reddening options. For the mock star-forming galaxies we consider here one would expect the Calzetti reddening law to outperform all other reddening options because this is the law we assume in the mocks.

\noindent We do not show the redshift recovery as a function of reddening law because results are very similar. Qualitatively, the mass and SFR recovery are the same as in Paper I at fixed redshift. The largest difference in mass recovery between the reddening laws is found at $z=0.5$ (Fig. \ref{massdtrl05zf}) because of the age-dust (and redshift) degeneracy. These galaxies have only little (or no) reddening and the age range available in the fit is largest. Masses are recovered significantly worse with the MW law by Allen and the LMC law. As at \textbf{fixed redshift} the exact type of reddening law is on average less important at high redshift, hence we do not show it again here. In any case, the best mass estimate is achieved when no reddening is used in the fitting independently of the actual amount of reddening present in the galaxies.

\noindent SFRs show a larger sensitivity to the type of reddening law (Fig. \ref{sfrdtrl05zf}). At high redshift the results are very similar to the ones in Paper I, meaning the fitting code is able to identify the right reddening law for most objects. SFRs are underestimated when the fit is performed without reddening. At low redshift, SFRs are generally better recovered than in the fixed redshift case and best recovered when no reddening is used. The Calzetti law performs best when reddening is included. 

\noindent In conclusion, in order to derive the masses and SFRs of old galaxies with little star formation (those at low redshift) reddening should not be used in the fit. For younger, star-forming galaxies (galaxies at high redshift) with dust reddening all reddening laws should be used in the fit, because stellar masses are affected very little by the choice of reddening law but SFRs are estimated best this way since the right law can be picked up when choosing the best fit among all reddening laws.

\subsection{The effect of photometric uncertainties}\label{photuncreszf}
Finally, we compare results obtained using randomised\footnote{Original magnitudes were randomised by scattering them with their three $\sigma$ photometric errors.} and original magnitudes in the fit to study the effect of photometric uncertainties. For simplicity only the unreddened case and mocks at $z=0.5$ are considered. The fitting for star-forming galaxies is carried out with the wide setup, that for passive galaxies with a solar metallicity SSP. Note, that templates do not comprise the exact SFH of the star-forming galaxies.

\begin{figure*}
\centering\includegraphics[width=84mm]{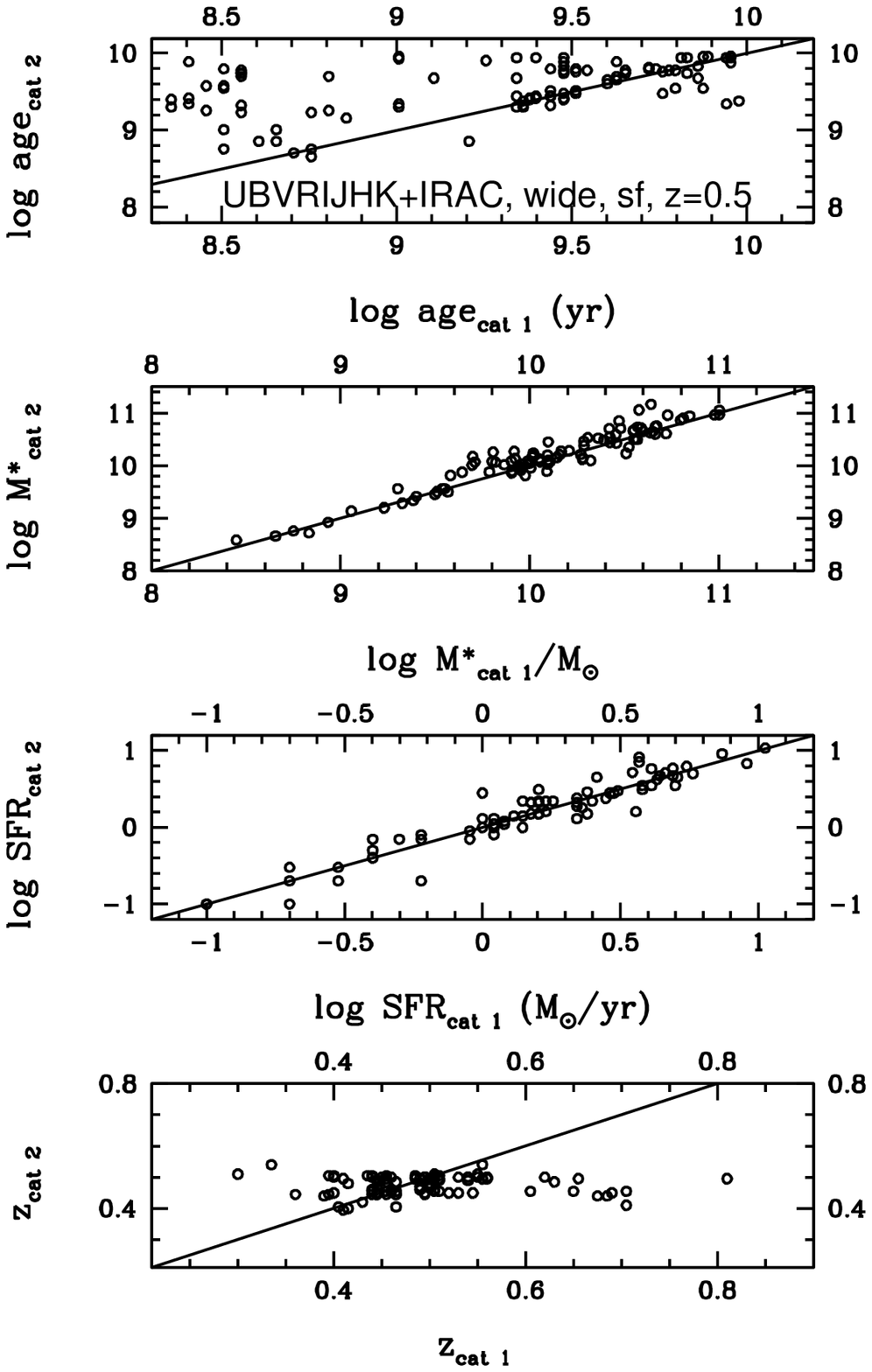}
\includegraphics[width=84mm]{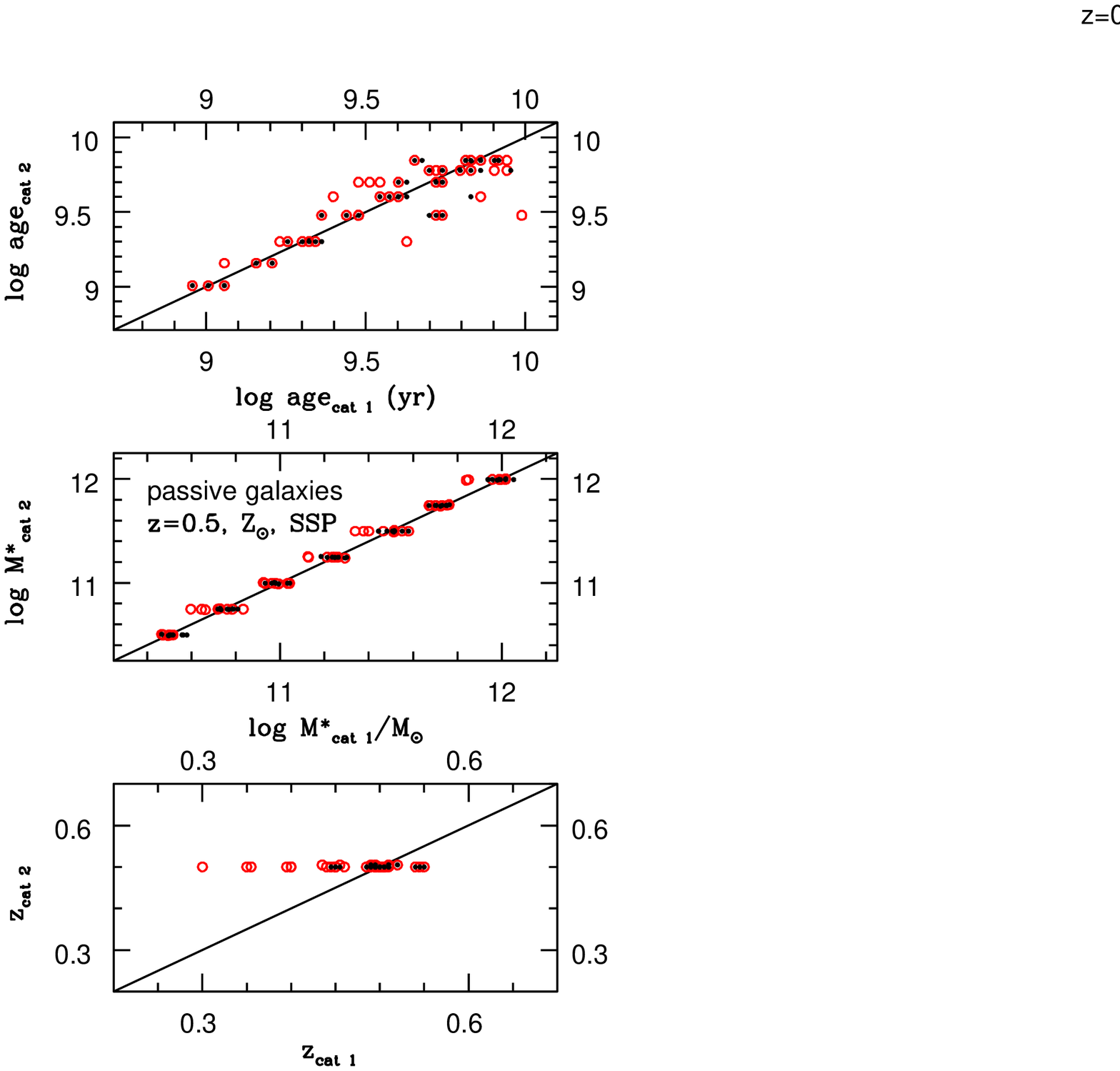}
\caption[Effect of photometric uncertainties on derived properties for mock galaxies when redshift is unknown]{\label{agepumzf} \textit{Left:} The impact of photometric uncertainties for mock star-forming galaxies at $z=0.5$ for a wide template setup and broad wavelength coverage. From top to bottom: derived ages, stellar masses, star formation rates and redshift, respectively. \textit{Right:} Effect for passive galaxies. From top to bottom: derived ages, stellar masses and redshift, respectively.}
\end{figure*}

\noindent First, we compare the derived redshifts of star-forming galaxies (Fig. \ref{agepumzf}, bottom left). The redshift estimate is clearly better when original magnitudes (labelled as $cat 2$ in Fig. \ref{agepumzf}) are used and there are no photometric uncertainties. Redshifts are then determined within $\sim 0.1$. In the randomised case they are only correct within $0.5$. This has the biggest effect on the derived ages (Fig. \ref{agepumzf}, top left). When original magnitudes are used ages are better estimated overall. At fixed redshift this trend was less clear. The maximum deviation is similar to the one in Paper I ($\sim1.6$ dex). 
Despite this large change in age for single objects, masses differ only little; on average by $0.08\pm0.16$ dex similar to the fixed redshift case. 
The mean difference between the derived SFRs is also small and smaller than at fixed redshift.
%
%
\noindent Fig. \ref{agepumzf}, right-hand side, shows that for passive galaxies photometric uncertainties have a small effect in the unreddened case. Redshifts are nearly perfectly recovered for original magnitudes ($cat 2$) and within $0.05$ for randomised magnitudes. Ages on the other hand differ more from each other and also scatter more than at fixed redshift. The largest deviations are found for the oldest galaxies which differ by maximal $\sim 0.2$ dex. Stellar masses differ by $\sim 0.1$ dex at most. The inclusion of reddening as already found worsens all comparisons.\\
As in Paper I, we investigated the effect of changing the size of photometric errorbars on ages, stellar masses and redshift estimates for both passive and star-forming galaxies. We used a wide wavelength coverage and template setup in the fitting for the reddened and unreddened case. For mock star-forming galaxies we studied several cases: 1) scattering original magnitudes with the survey errorbars (from COSMOS and GOODS-S) and fitting with the same, 2) scattering original magnitudes within $\sigma$ obtained from the surveys but fitting with $\sigma=0.1$, 3) scattering original magnitudes with $\sigma=0.1$ and fitting with $\sigma=0.1$, and 4) doubling and tripling photometric errors for the fit for the latter two cases (2) and 3)). For mock passive galaxies we studied cases 2), 3) and 4).\\
For mock star-forming galaxies the median redshift recovery at low redshift and redshift $3$ differs by up to $0.03$ in the unreddened and reddened case. At redshift $2$ the difference can be up to $\sim0.15$, highlighting again that the redshift determination of $z\sim2$ objects is more difficult even with our wide wavelength coverage. Ages can differ a lot more. In the unreddened case the median between mass-weighted and derived age differs by up to $\sim0.2$ dex, in the reddened case by up to $0.4$ dex, but $0.7$ dex for $z=2$ due to the larger uncertainty in redshift. Median offsets for stellar masses are a lot more robust against these variations due to compensating effects and differ maximally by $\leq0.11$ dex in the reddened and unreddened case at each redshift. This is comparable to our findings in Paper I. Variations in median offsets for the mock passive galaxies are very small (for redshift up to $0.05$, for age up to $0.05$, for stellar masses up to $0.08$) and are largest at the lowest redshifts. As in Paper I, doubling or tripling errorbars has a very small effect in any case.

\section{Comparison to results in the literature}\label{litcompzfree}
\noindent Not much work has been devoted to the study of the robustness and accuracy of the derived stellar population parameters from SED-fitting when redshift is a free parameter. On the other hand many studies that involve the derivation of photometric redshifts evaluate the quality of their photometric redshift estimates and calibrate them by using control samples for which spectroscopic redshifts are available \citep[e.g.][to name only a few]{Ilbert2006,Quadri2010,Dahlen2010}. First, we compare our results with \citet{Bol2000} in which they tested their \hz-code. They also studied the dependence of the derived photometric redshifts on various parameters in the fit, such as template setup, photometric uncertainties, wavelength coverage, reddening and reddening law, metallicity and IMF. Then we focus on the results of \citet{Lee2009} and \citet[][W09]{Wuyts2009} for other stellar population parameters when redshift is left free.

\noindent \citet{Bol2000} use a set of simulated galaxies produced from the template set (based on \citealt{BC93} stellar population models) under the assumption of a homogeneous redshift distribution. Ages and spectral types are drawn randomly from the template set. Note that \citet{Bol2000} create their mock galaxies from the  template set that is also used to fit them while our simulated star-forming galaxies stem from a semi-analytic galaxy formation model. Magnitudes in all filter bands are scattered within fixed one $\sigma$ values. These range from $0.05$ to $0.3~$mag (i.e. 5 to 30\% error). They showed when testing their photometric redshift code \hz \space that the determination of photometric redshifts is best for a wide wavelength coverage and small photometric errors. For photometric errors as used for our mock galaxies (i.e. $\sim10$\%) \citet{Bol2000} report a mean photometric accuracy (expressed as $(z_{true}-z_{fit})/N$ under exclusion of catastrophic failures\footnote{Catastrophic failures are defined by \citet{Bol2000} as $|\Delta_z|=|z_{true}-z_{fit}|\geq1$}) of $-0.02$, $-0.02$, $-0.01$ and $0.03$ for redshift bins of $z=0-0.4$, $0.4-1$, $1-2$ and $2-3$, respectively, when using a UBVRIJHK filter set. Photometric redshift accuracies are similar in other studies \citep[e.g.]{Ilbert2006,Wuyts2009}. Our equivalent values but including catastrophic failures are $-0.10$, $-0.14$, $0.27$ and $0.04$ for redshift $z=0.5$, $1$, $2$ and $3$ using a wide setup with reddening\footnote{Median values of $(z_{fit}-z_{true})$ are $0.04$, $-0.003$, $-0.08$ and $-0.015$ for $z=0.5$, $1$, $2$ and $3$ using a wide setup with reddening and a UBVRIJHK filter set.}. Catastrophic failures for $z=1$ amount to 1\% and for $z=2$ to 8\%. At $z=0.5$ and $3$ we do not have catastrophic failures according to the definition of \citet{Bol2000}. Our values for the photometric accuracies are larger than those of \citet{Bol2000} very likely due to the age-dust degeneracy (at low redshift) and SFH mismatch between template and mock galaxies. Additionally, they restrict the reddening to $A_V\leq1~$mag and Calzetti law in their fit while we allow values up to $A_V=3~$mag and use all reddening laws. It is also worth to note that the metallicity of templates and simulated galaxies of \citet{Bol2000} is fixed to the solar value, while both our simulated galaxies and template setup cover a wide range in metallicities. At redshift $3$ however our value for the redshift accuracy is very similar due to the fact that for these objects both the Lyman break and the 4000 \AA \space break are covered by the chosen filter set. This significantly improves the redshift recovery. However, including the rest-frame near-IR improves our mean redshift recovery to $-0.01$, $0.02$, $-0.04$ and $0.03$ for redshift $z=0.5$, $1$, $2$ and $3$ , respectively, using a wide setup and reddening.

\noindent Overall, \citet{Bol2000} conclude that although the redshift determination is most sensitive to the photometric accuracy and wavelength coverage a sufficiently wide range of ages and reddening is still important for deriving accurate photometric redshifts. They find that for high redshift objects reddening is more essential for the redshift accuracy and that the MW and LMC laws are not suitable. Our median recovered redshifts when using different reddening laws (but a full wavelength coverage) differ only very little. Furthermore they conclude that metallicity effects on the redshift recovery are negligible at high-z and that the IMF seems to play a secondary role. Both findings are in agreement with our work. \citet{Bol2000} also showed that the redshifts of old and passive galaxies are determined better than those of younger ones which we confirmed in the previous sections.

\noindent We already compared our results at fixed redshift to the work of \citet{Lee2009} and W09 in Paper I. Let's briefly recap the basis of their works. \citet{Lee2009} address the robustness of stellar population properties derived from SED-fitting of Lyman break galaxies at $3<z\leq5$ (U-, B- and V-dropout galaxies). Mock galaxies are created from a semi-analytic galaxy formation model using BC03 stellar population models with a Chabrier IMF and are selected via the Lyman break criterion. SED-fitting is carried out using BC03 templates of three metallicities (0.2 to 1 $Z_{\odot}$), exponentially declining SFRs ($\tau=0.2 - 15$ Gyr) and Chabrier IMF. Dust extinction is treated with the Calzetti law. When redshift is fixed they find that masses and SFRs tend to be underestimated and mean ages overestimated because of a mismatch in SFH. In the free redshift case, they find that when redshift is underestimated then ages and masses are underestimated and SFRs are overestimated. Compared to their results at fixed redshift stellar masses are underestimated more when redshift is free. While at fixed redshift the means of the intrinsic and recovered mass distribution differ between 19 and 25\%, they differ between 25 and 51\% when redshift is free. We found at fixed redshift a difference of only $\sim15$\% at the highest redshift (using a wide setup, reddening and a minimum age of $0.1$ Gyr). When redshift is free we find the same because the redshift of $z=3$ objects is very well recovered and the available age range in the fitting is small.

\noindent \citet{Lee2009} also report large deviations of mean age and SFR from the true values such that mean SFRs are underestimated and (mass-weighted) ages are overestimated caused by SFH mismatch and overshining. They find that SFRs and ages are better recovered, if only slightly, when redshift is left free. We find that ages and SFRs at $z=3$ are recovered slightly worse because of the added degeneracy with redshift. Recovered redshifts of \citet{Lee2009} are in general slightly underestimated with $(z_{fit}-z_{true})/(1+z_{true})$ ranging from $-0.137$ to $0.043$ for U-dropouts. The ranges are smaller for B and V-dropouts, i.e. they get narrower with increasing redshift. At redshift $3$ we find a median offset in redshift expressed as $(z_{fit}-z_{true})/(1+z_{true})$ of $-0.002$ (and $-0.007$ for the mean) and a maximum deviation of $-0.070$ when using a wide setup and reddening which is slightly smaller than the values of \citet{Lee2009}. In agreement with our findings \citet{Lee2009} also recommend a wide wavelength coverage from the optical to the rest-frame near-IR for the best recovery of the galaxy physical parameters.

\noindent We showed in Paper I and in M10 that when redshift is known using inverted-$\tau$ models with fixed formation redshift ($z_{form}\sim5$) improves the mass and SFR estimate for mock star-forming galaxies at $z=2$ and $3$ and recovers stellar masses perfectly. For the inverted-$\tau$ models the difference between the means of true and recovered stellar mass distributions was less than $1$\%. As explained in Paper I, the use of inverted-$\tau$ models solves both problems identified by \citet{Lee2009} (SFH mismatch and overshining) to be responsible for offsets in recovered stellar population properties.

\noindent Therefore, when redshift is a free parameter, we suggest for high redshift, star-forming objects that redshifts should be determined with the best fit setup first and then used in a separate fit with inverted-$\tau$ models to obtain masses and SFRs.

\noindent This approach is taken by \citet{Wuyts2009}. In the case of unknown redshift they derive redshifts first and use these then in a separate fit to derive stellar population properties, such as ages, reddening, stellar masses and SFRs. Mock galaxies for this exercise are extracted from a hydrodynamical merger simulation. The photometry of their mock galaxies placed at redshifts between $1.5$ and $3$ is based on BC03 models. The SED-fitting is carried out using \hz \space with solar metallicity, Salpeter IMF BC03 templates of three star formation histories: SSP, constant SFR, and a $\tau$-model with $\tau=0.3$ Gyr. Reddening is varied between $A_V$ from 0 to 4 mag but only the Calzetti law is considered. The mock galaxies in W09 are distinguished into disk, merger and spheroid phase of the simulation. Thus the mock star-forming galaxies used in our work compare mostly to galaxies in the disk and merger phase of W09, the mock passive galaxies to galaxies in their spheroid phase. At fixed redshift W09 find reddening, stellar masses and SFRs to be underestimated on average. The properties of spheroidal galaxies are better recovered than those of the disk and merger phase in general.
When redshift needs to be derived in the fitting, they use the template fitting code EAZY \citep{Brammer2008} which fits a linear combination of templates to the data and the redshift is derived by evaluating the probability distribution function. Their median $\Delta z/ (1+z)$ is $0.031$ which is comparable to our findings. The main effect found by W09 on the stellar parameters caused by this uncertainty in redshift is the broadening of the 68\% confidence levels by 1\% in $\Delta\,log\, age$, 7\% in $\Delta\,log\, M^*$, 5\% in $\Delta E(B-V)$ and 1\% in $\Delta\,log\, SFR$. They find offsets in stellar mass and reddening only for the dustiest objects. For these derived stellar masses are larger by $0.05$ dex and reddening smaller by $\sim0.02$ mag. But they point out that these effects are smaller than those caused by SFH mismatches and reddening and might be template setup specific. Compared to W09 the increase in stellar mass estimates that we find are not confined to the dustiest objects but mainly affect the oldest galaxies with little on-going star formation.

\noindent  Finally, \citet{Pozzetti2007} note for similar redshift accuracies an additional $\sim0.2$ dex (at $z<0.4, \sim0.1$ dex at $z=2$) uncertainty in the stellar mass due to photometric redshift accuracies for their sample of K-selected galaxies from the VIMOS VLT Deep Survey \citep{LeFevre2003b}.

\section{Homogenising derived properties via scaling relations}\label{scalerelzfree}
As in Paper I we provide scaling relations for stellar mass that allow one to transform derived masses from one fitting setup to another when redshift is a free parameter in the fit (Tables \ref{scalezfree} and \ref{scalepasszfree}). We use a least squares fit and gain statistical robustness by using the entire merger tree for mock star-forming galaxies. Note that deviations in stellar mass for a single object between fitting setups can be significant and that these relations provide information about statistical differences.

\section{Summary and conclusions}\label{summzfree}
We study the dependence of recovered galaxy stellar population properties (photometric redshifts, ages, star formation rates, stellar masses) through broad-band spectral energy distribution fitting on the various parameters of the fit. Such parameters are star formation history, metallicity, age grid, IMF, reddening law, wavelength coverage and filter setup used in the fitting. This work is complementary to \citet[][Paper I]{Pforr2012} where the redshift was known and used to constrain the fit. We use the same mock galaxies (passive and star-forming) as in Paper I with known properties as test particles. The investigated redshift range is $0.5\leq z\leq3$. The photometry of mock galaxies is based on the stellar population models by \citet{M05} which we also used for the fitting templates. Results for BC03 stellar population models (for mock galaxies and fitting templates) are discussed in the Appendix. Mock galaxies are treated as real observed galaxies such that their observed-frame magnitudes are obtained and SED-fitting with various templates is carried out on them. The properties derived from the fitting are then compared to the true values.

\noindent In Paper I at fixed redshift we found that for mock star-forming galaxies a mismatch in SFH is the main driver for underestimated ages and masses and overestimated reddening and SFRs due to overshining and the age-dust degeneracy. The galaxy physical properties can only  be determined \textit{simultaneously} when SFHs between template and galaxy match. In \citet{Tonini2012a} the effects of uncertainties on the galaxy SFH were analysed in the case of Brightest Cluster Galaxies, where they found a severe mismatch between the SFHs of BCGs in hierarchical models and in toy-models of passively-evolving galaxies (single stellar populations), even if both these kinds of models reproduce the same photometric data. 
The recovery of the galaxy physical properties from SED-fitting in the case of BCGs, and massive ellipticals in general, is greatly affected by the fact that template SFHs cannot take into account the rich merger histories of these objects.\\
For our mock passive galaxies with smooth star formation histories the stellar population properties are much better recovered.

\noindent When redshift is left free in the fit, we can summarise our results for the star-forming galaxies to:\\
- The redshift recovery depends only weakly on the template setup, IMF, metallicity, age grid and reddening law due to compensating effects, thus the most economic option to obtain redshifts is via the use of a mono-metallicity wide setup or an only-$\tau$ setup with a minimum age of $0.1$ Gyr and only one reddening law.\\
- A broad wavelength coverage is even more crucial for the simultaneous and robust determination of galaxy properties (ages, stellar masses, reddening and SFRs) when redshift is a free parameter in the fit because photometric redshifts are best recovered - within $0.01$ to $0.06$ (median $\Delta z$) - when the rest-frame wavelength range comprises important features such as the Lyman-limit or the 4000 \AA \space break.\\
- The mismatch between template and galaxy SFH is still the most important driver for underestimated ages and stellar masses and overestimated reddening and SFRs because of the overshining and degeneracies between age, dust and redshift rather than the unknown redshift.\\
- Stellar masses at low redshift are better recovered when redshift is left free in the fit because redshift compensates. At $z=0.5$ a sacrifice of $\sim0.01$ ($0.05$) in accuracy for the median redshift improves the median recovered stellar mass by $0.1$ dex ($0.3$ dex) in the unreddened (reddened) case. However, the mass recovery fails when the redshift recovery fails completely. \\
- Ages are generally underestimated similar to the fixed redshift case. However, because of compensating effects with redshift, the age recovery is better at low redshift when redshift is a free parameter compared to the case of fixed redshift.\\
- Metallicity effects are minor as in the fixed redshift case.\\
- SFRs are best recovered with a wide setup although the correct SFH is not included. SFRs at low redshift are generally better recovered than at fixed redshift.\\
- Photometric uncertainties account for $0.4$ in redshift uncertainty, masses are affected on average by $0.08\pm0.16$ dex, SFRs are less affected than at fixed redshift.\\
- The effect of the age-dust-redshift degeneracy at low redshift which causes ages and stellar masses to be underestimated and SFRs to be overestimated can be reduced by fitting without reddening and/or introducing an age cap. In this way unrealistically young and dusty solutions are avoided while redshifts are recovered similarly well. \\

\noindent For passive galaxies our results are:\\
- Redshifts of mock passive galaxies are very well determined independently of template setup, SFH and IMF. Again a broad wavelength range guarantees the best redshift recovery.\\
- Stellar masses of passive galaxies can be very well recovered - within $0.02$ dex for a wide setup and wavelength coverage ($0.05$ dex for a wide setup and no reddening but varying wavelength coverage) even when redshift is unknown.\\
- As in the fixed redshift case, metallicity plays an important role and the best option is to use a wide range of metallicities in the fitting. \\
- The age recovery for old passive galaxies is less dependent on the wavelength coverage than for star-forming galaxies but the inclusion of the rest-frame near-IR helps to significantly reduce the scatter. \\
- The effect of photometric uncertainties on redshifts is $0.05$, on ages and masses $0.2$ dex and $0.1$ dex, respectively.\\

\noindent The sensitivity of the SED-fit to the stellar population model was studied in Appendices \ref{bc03results} and \ref{pegase}. In particular, we studied the outcomes of fitting M05 star-forming galaxies with BC03 templates (Appendix \ref{bc03results}) and star-forming galaxies, created by using PEGASE stellar population models \footnote{PEGASE and BC03 models are quite similar due to similar input physics.}\citep{Pegase} in the galaxy formation model, with M05 templates (Appendix \ref{pegase}. Results can be summarised to:\\
- The redshift recovery is significantly worse with BC03 templates for M05 mock star-forming galaxies (Appendix \ref{bc03results}) such that redshift recovery fails in up to $\sim30$\% of the cases. In these cases recovering the stellar mass fails as well (underestimation of more than $2$ dex) and most SFRs are zero. \\
- When PEGASE galaxies are fit with M05 templates interestingly the redshift recovery is comparable to that with BC03 templates.\\
- The redshifts of PEGASE galaxies are recovered (with templates of either stellar population model) worse than those of M05 galaxies with M05 templates and show larger scatter, e.g. at $z=0.5$ the mean $\Delta z=-0.007\pm0.082$ in the unreddened case ($\Delta z=0.011\pm0.115$ in the reddened case) for M05 galaxies fit with M05 templates and $\Delta z=0.015\pm0.133$ ($\Delta z=0.128\pm0.316$) for PEGASE galaxies fit with M05 templates. BC03 templates recover the redshift of PEGASE galaxies at $z=0.5$ similarly well and those of M05 galaxies significantly worse. Thus, it seems to be best to derive redshifts with M05 templates independently of the underlying stellar population model of the mock galaxies.\\ 
- When fitting BC03 models to M05 mock star-forming galaxies ages are older due to the differences in the stellar population model. Consequently, stellar masses are higher and match the input masses better. This better agreement is contrived as it originates from the compensation of the wrong stellar population model.\\
- Equivalent results are provided in the reverse case (fitting of BC03 and M05 templates to PEGASE mock star-forming galaxies, Appendix \ref{pegase}), i.e. masses and ages are underestimated in general. Stellar masses obtained from fits with M05 models are lower than masses derived with BC03 templates and show a clear offset caused by the various differences in the stellar population model (stellar tracks, TP-AGB, etc.) in agreement with findings of e.g. M06. \\

\noindent We conclude that photometric redshifts are recovered very well in almost any case as long as the wavelength coverage is wide. The stellar population parameters of star-forming and passive galaxies can then be reasonably well determined provided one uses the right setup and wavelength coverage. These considerations are important for surveys such as DES for which only photometric redshifts are available. In comparison to the fixed redshift case shown in Paper I masses and SFRs are slightly better determined at low redshift when redshift is left free because of the added degeneracy with redshift. Still the mismatch between observed and assumed SFHs dominates the parameter estimate at low redshift.

\noindent For $z\geq2$ star-forming galaxies good results for redshift, mass and SFR can be obtained simultaneously with a wide setup and no reddening in the fit, too. This means that photometry appears to be sufficient for the robust determination of the stellar population properties of high redshift galaxies. But as shown in Paper I at fixed redshift, the best recovery for stellar masses and SFRs is given when inverted-$\tau$ models are used in the fit. We therefore recommend to obtain redshifts in the most robust way for these galaxies first and to carry out a separate fit at fixed redshift with inverted-$\tau$ models to determine stellar masses and SFRs even better.

\noindent Conclusions about the wavelength coverage remain unchanged from the fixed redshift case, meaning a coverage of UV to near-IR rest-frame wavelengths in the fitting is crucial. Again these effects are quantified and will be useful for the planning of purely photometric surveys and observational proposals.

\noindent As stressed in Paper I a variety of assumptions are made in the literature with regard to fitting methods and parameter setups and thus we confirm what was already concluded by \citet{Lee2009} and in Paper I that one has to be cautious when comparing results from different studies. We also provide scaling relations for the transformation of stellar masses between fitting setups for the free redshift case to ease comparisons between studies that rely on photometric redshifts.

\section*{Acknowledgments}
The authors would like to thank the anonymous referee for a quick report and useful report that enhanced the clarity of the paper. We would also like to thank E. Daddi, D. Thomas, M. Bolzonella, A. Renzini, A. Aragon-Salamanca, C. Conselice, S.  Ellison, M. Dickinson, A. Fontana and A. Grazian for useful discussions and suggestions that improved the paper. We would like to thank A. Cimatti for suggesting the exercise in section \ref{scalerelzfree}. We would also like to thank M. Bolzonella for providing \hs \space and support with the fitting code. JP, CT and CM were supported by the Marie Curie Excellence Team Grant "UniMass", MEXT-CT-2006-042754, P.I. Claudia Maraston. Numerical computations were carried out on the Sciama High Performance Compute (HPC) cluster which is supported by the ICG, SEPNet and the University of Portsmouth. JP is supported by HST program GO-12060, support for which is provided by NASA through a grant from the Space Telescope Science Institute, which is operated by the Association of Universities for Research in Astronomy, Inc., under NASA contract NAS 5-26555.

\bibliographystyle{mn2e}
\bibliography{biblioJaninePhDmnras.bib}

\appendix

\section{Overview over fitting setups}
The explored fitting setups are summarised in Table \ref{overfits}.

\begin{table*}
\begin{minipage}{186mm}
\caption{Overview of performed fits for simulated galaxies at redshift 0.5, 1, 2 and 3. Fits with a modified age grid were only carried out for star-forming galaxies.}\label{overfits}
\centering{
\begin{tabular}{@{}lccccccccccccc}
\hline
Template Setup & \begin{sideways}UBVRI JHK IRAC\end{sideways} & \begin{sideways}UBVRI IRAC\end{sideways} & \begin{sideways}UBVRI JHK\end{sideways} & \begin{sideways}UBVRI JH\end{sideways} & \begin{sideways}UBVRI J\end{sideways} & \begin{sideways}UBVRI\end{sideways} & \begin{sideways}BVRI JHK IRAC\end{sideways} & \begin{sideways}VRI JHK IRAC\end{sideways} & \begin{sideways}RI JHK IRAC\end{sideways} & \begin{sideways}u'g'r'i'z'\end{sideways} & \begin{sideways}UBVRI Y JHK IRAC    \end{sideways}& \begin{sideways}BRIK    \end{sideways}\\\hline\hline
wide setup                                                           & x & x & x & x & x & x & x & x & x & x & x & x\\\hline
wide, age rebin                                                   & x & - & x & - & - & - & - & - & - & - & - & - \\\hline
wide, age $\geq$ 0.1 Gyr                                  & x & - & - & - & - & - & - & - & - & - & - & - \\\hline
wide, $Z=0.004$		                                 & x & - & - & - & - & - & - & - & - & - & - & - \\\hline
wide, $Z=0.01$ 		                                 & x & - & - & - & - & - & - & - & - & - & - & - \\\hline
wide, $Z=0.02$ 		                                 & x & - & - & - & - & - & - & - & - & - & - & - \\\hline
wide, $Z=0.04$ 			                       & x & - & - & - & - & - & - & - & - & - & - & - \\\hline
wide, Kroupa			                                & x & - & - & - & - & - & - & - & - & - & - & - \\\hline
wide, Chabrier			                                & x & - & - & - & - & - & - & - & - & - & - & - \\\hline
wide, top-heavy	                                         & x & - & - & - & - & - & - & - & - & - & - & - \\\hline
only-$\tau$                                                           & x & - & x & - & - & - & - & - & - & - & - & - \\\hline
only SSPs                                                          & x & - & - & - & - & - & - & - & - & - & - & - \\\hline
solar SSP                                                           & x & - & - & - & - & - & - & - & - & - & - & - \\\hline
wide, BC03                                                        & x & - & x & - & - & x & - & - & - & - & - & - \\\hline
\end{tabular}}
\end{minipage}
\end{table*}%

\section{Overview over fitting results}\label{overres}
In Tables B1 - B4 
we list median offsets and 68\% confidence ranges for redshift, age, E(B-V), stellar mass and SFR for a selected number of fitting setups.
\begin{table*}
\vbox to220mm{\vfil Landscape table 1 to go here.
  \caption{}
 \vfil}
 \label{sfoverres}
\end{table*}
\begin{table*}
\vbox to220mm{\vfil Landscape table 2 to go here.
  \caption{}
 \vfil}
 \label{sfoverres2}
\end{table*}
\begin{table*}
\vbox to220mm{\vfil Landscape table 3 to go here.
  \caption{}
 \vfil}
 \label{poverres}
\end{table*}
\begin{table*}
\vbox to220mm{\vfil Landscape table 4 to go here.
 \caption{}
 \vfil}
 \label{poverres2}
\end{table*}

\section{Fitting BC03 templates to M05 galaxies}\label{bc03results}
\begin{figure*}
\centering
\includegraphics[width=124mm]{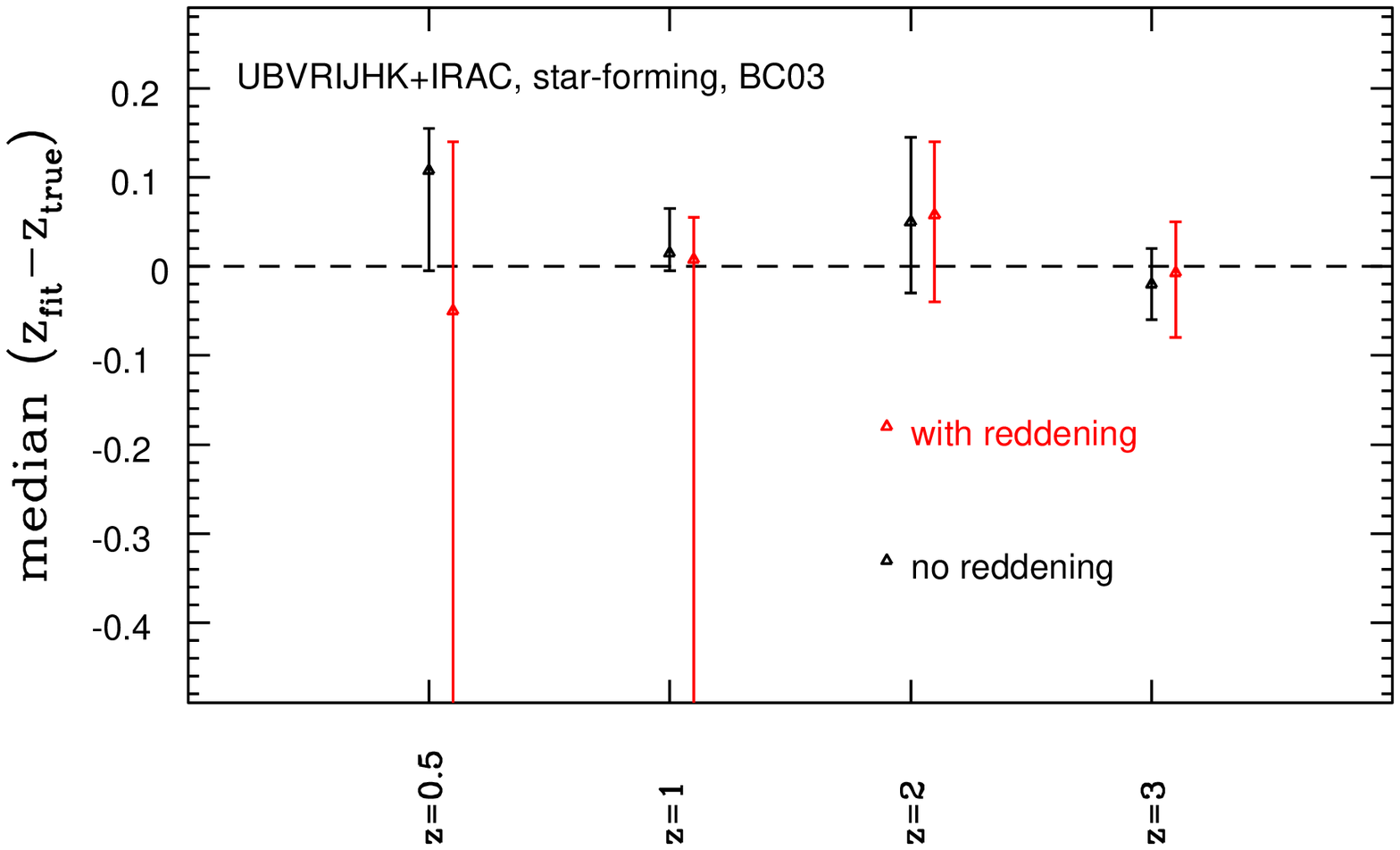}                                                                   
\caption[Redshift recovery for M05 mock star-forming galaxies as function of stellar population models]{\label{Bzhist} Median offset between redshifts derived from SED-fitting with BC03-based templates for mock star-forming galaxies based on M05 photometry and true redshifts. A wide setup and wavelength coverage was used. Black symbols refer to the unreddened case, red ones to the case with reddening. Errorbars are 68\% confidence levels.}
\end{figure*}

\begin{figure}
\centering
\includegraphics[width=84mm]{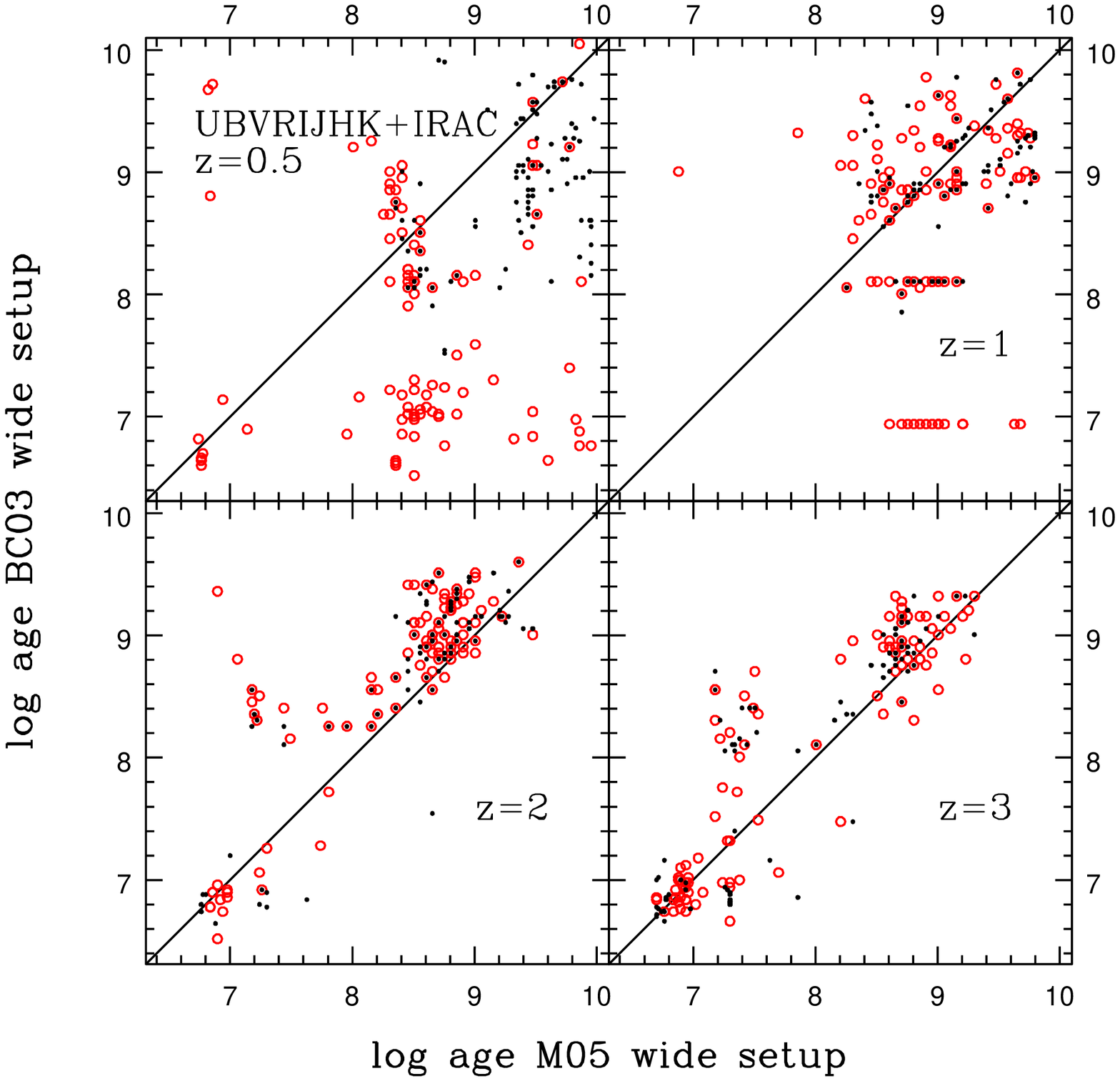}                                                                   
\caption[Ages for M05 mock star-forming galaxies as a function of stellar population model]{\label{Bagezf} SED-fit derived ages when redshift is a free parameter as a function of stellar population synthesis models (M05, BC03) for mock star-forming galaxies based on M05 photometry. The fitting is carried out with a wide setup and wavelength coverage. Black refers to the unreddened case, red to the case including reddening.}
\end{figure}

\begin{figure}
\centering
\includegraphics[width=84mm]{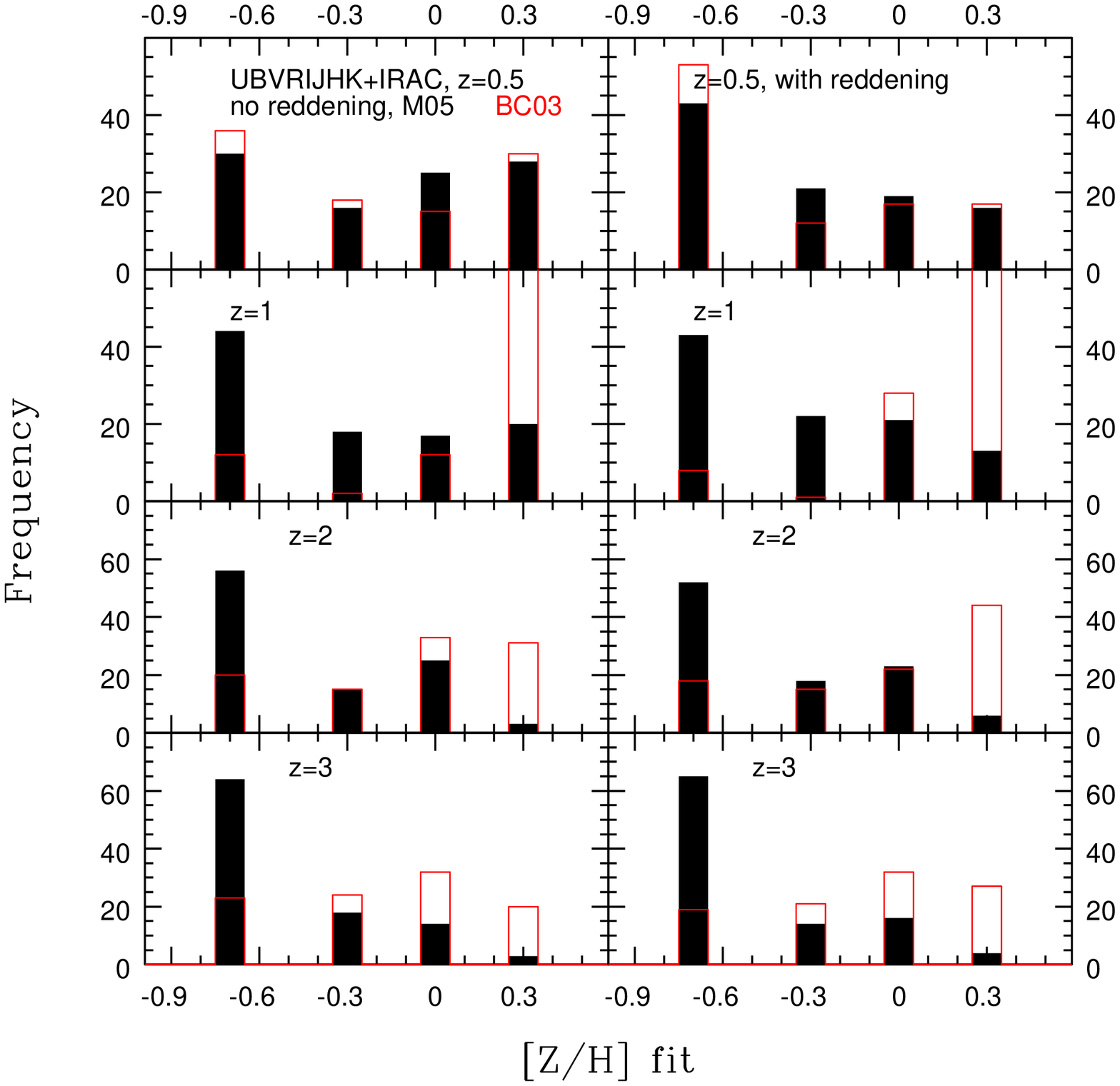}                  
\caption[Same as Fig. \ref{Bagezf} for metallicities]{\label{Bmetalzf} Same as Fig. \ref{Bagezf} for metallicities. Red histograms stand for BC03-based results, black shaded histograms for M05-based results.}
\end{figure}

\begin{figure}
\centering
\includegraphics[width=84mm]{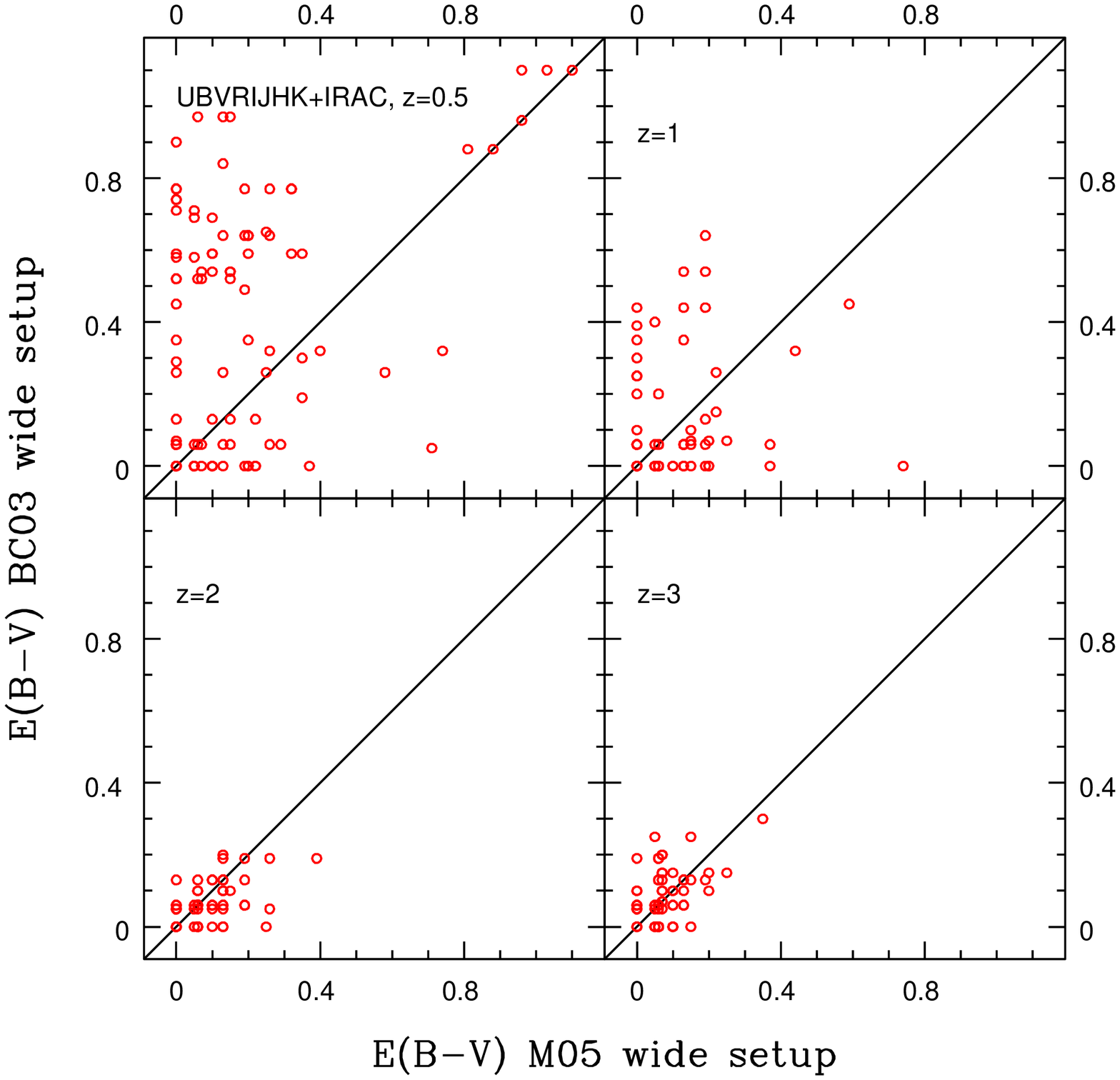}                                                                                                                                  
\caption[Same as Fig. \ref{Bagezf} for $E(B-V)$]{\label{Bebvzf} Same as Fig. \ref{Bagezf} for the comparison of dust attenuation.}
\end{figure}

\begin{figure}
\centering
\includegraphics[width=84mm]{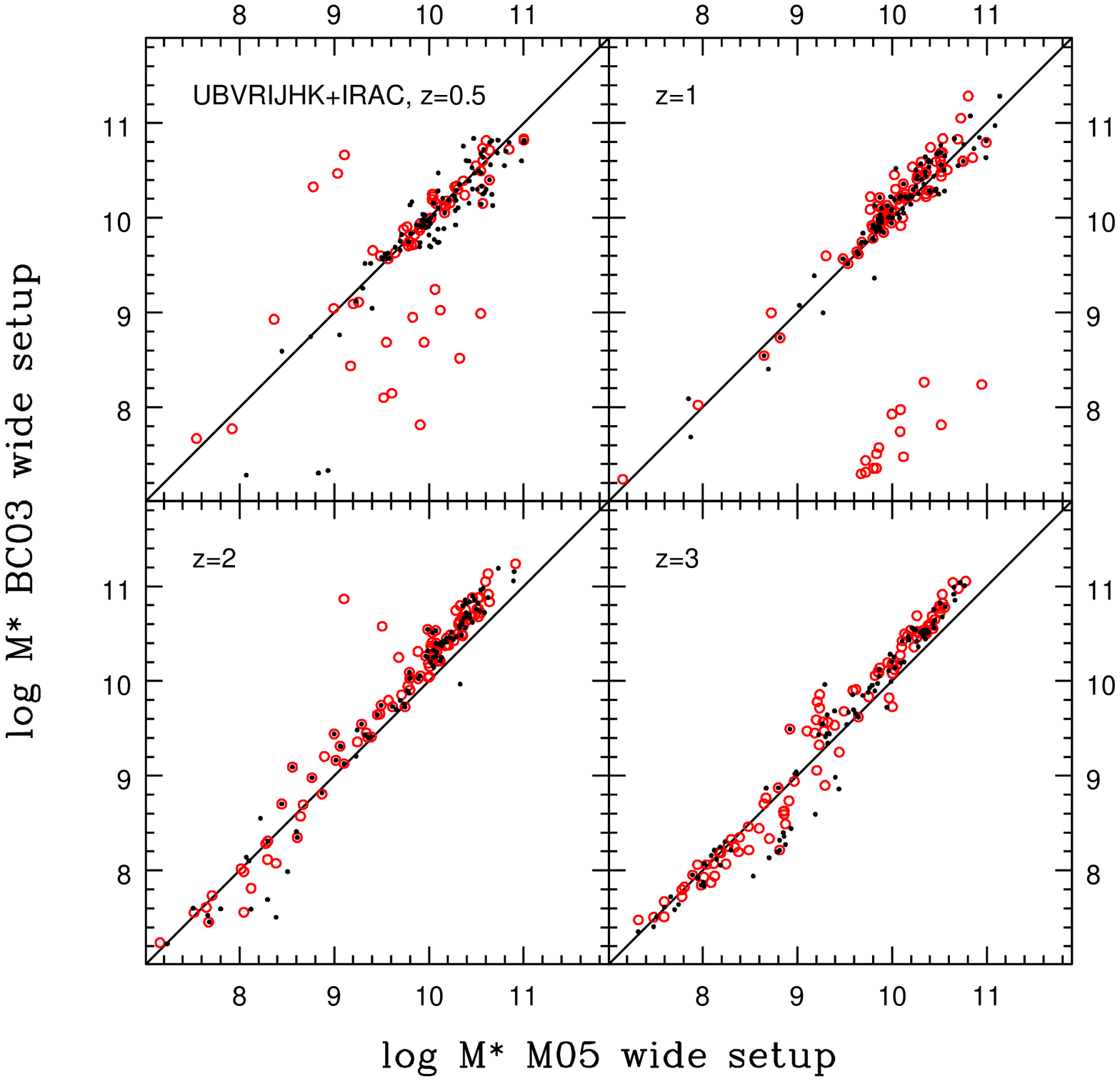}                                                                                                                              
\caption[Same as Fig. \ref{Bagezf} for stellar masses]{\label{Bmasszf} Same as Fig. \ref{Bagezf} but comparing the derived stellar masses. Black refers to the unreddened case, red to the case including reddening.}
\end{figure}

\begin{figure}
\centering
\includegraphics[width=84mm]{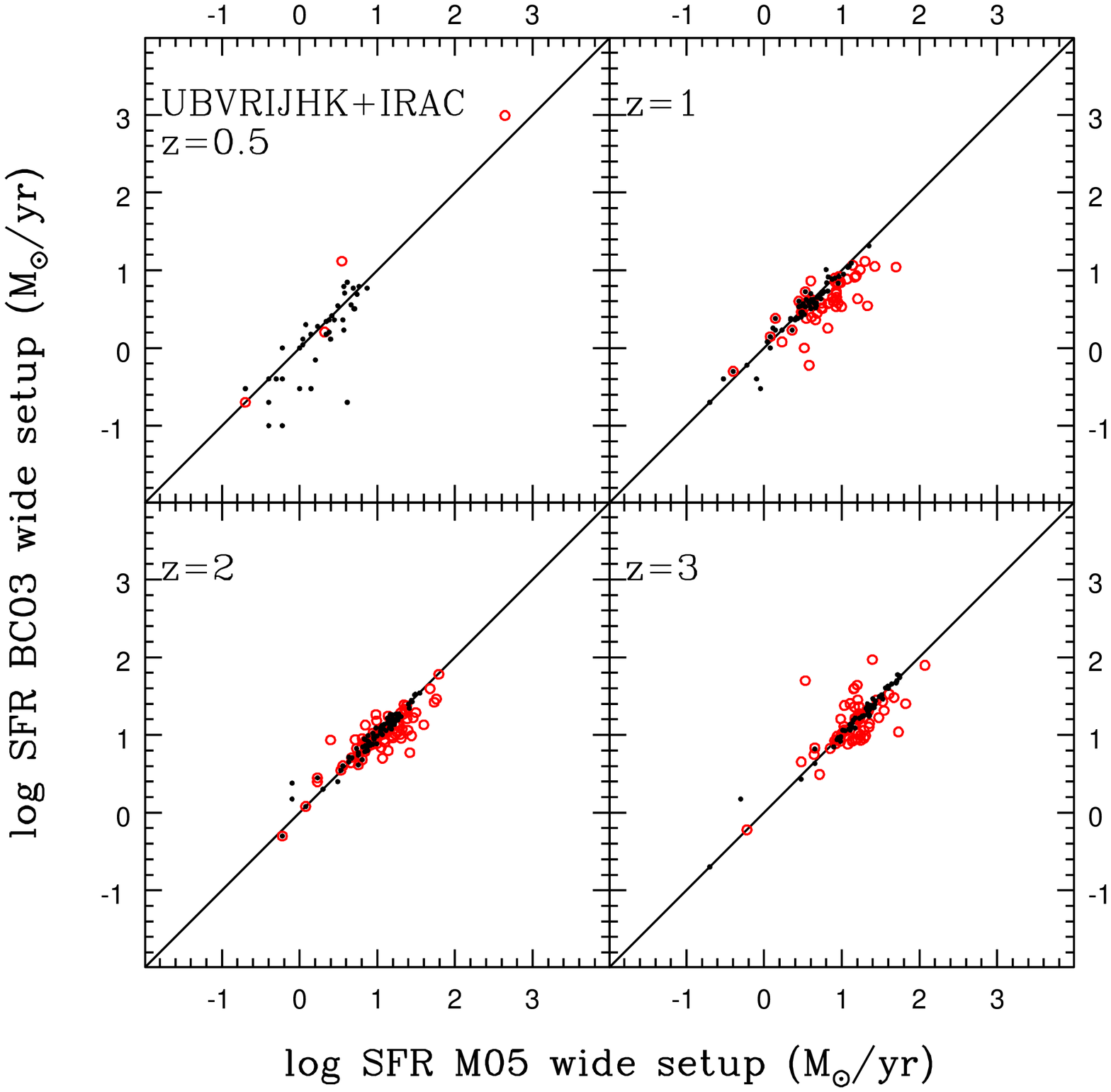}                                                                                      
\caption[Same as Fig. \ref{Bagezf} for SFRs]{\label{Bsfrzf} Same as Fig. \ref{Bagezf} for the comparison of derived star formation rates. Black refers to the unreddened case, red to the case including reddening.}
\end{figure}

Figs. \ref{Bzhist}-\ref{Bsfrzf} show the results for the recovery of redshift, age, metallicity, reddening, stellar mass and SFR of M05 mock galaxies with BC03 templates.

\noindent The redshift recovery is shown in Fig. \ref{Bzhist} which should be compared to Fig. \ref{zhistdtzf}. Redshifts for redshift $2$ and $3$ objects are similarly well recovered as with templates based on the same physical models \textbf{as} the galaxies (M05 in this case, compare Fig. \ref{zhistdtzf}). Offsets and scatter are comparable in both the reddened and unreddened case to those in Fig. \ref{zhistdtzf}. In the unreddened case at redshift $1$ the scatter is smaller than the equivalent value for M05 models. At $z=0.5$ redshifts show are overestimated by a median of $\sim0.1$ and scatter is larger compared to the M05 model fits. When reddening is included, scatter becomes very large at both $z=0.5$ and $1$ because for some objects the redshift recovery fails dramatically due to the age-dust-redshift degeneracy.

\noindent Fig. \ref{Bagezf} shows that for the lowest redshift, ages derived with BC03 templates are on average younger than those derived with M05 templates. Particularly, when reddening is included in the fit, ages from BC03 templates are very young. At $z=1$ this is still the case for a few objects for which also the redshift recovery fails. At fixed redshift ages from BC03 templates were much older. In the unreddened case clearly the degeneracy between age and redshift has a stronger effect when the wrong stellar population model is used. As explained in Paper I BC03 models are redder than M05 models for older ages due to the different stellar tracks. This in combination with a larger redshift (especially at $z=0.5$) requires younger ages from the fit. When reddening is included the degeneracy with dust then drives ages to even younger values and the redshift recovery fails. At higher redshift the differences between ages from BC03 and M05-based templates are similar to the case at fixed redshift, meaning ages from BC03 templates are on average older. We already explained in Paper I that this is a TP-AGB effect and for the youngest ages an effect of the different stellar tracks for the red supergiant phase.

\noindent In Fig. \ref{Bmetalzf} we show the distributions of the best fit metallicities inferred from BC03 and M05 templates. Similar to the fixed redshift case fits with BC03 templates provide on average metal-richer solutions. Again we find the largest differences to the fixed redshift case for the lowest redshifts. Metallicities are also higher on average compared to the fixed redshift case.

\noindent While at fixed redshift the reddening of low redshift galaxies was better estimated with BC03 templates, reddening is even more overestimated when redshift is left free (Fig. \ref{Bebvzf}) because ages and redshifts are underestimated. This again demonstrates the many degeneracies one faces at low redshift where the age range in the fitting is wide and galaxies are mainly old with only little on-going star formation. At high redshift reddening values from M05 and BC03 templates agree better and are comparable to those derived at fixed redshift.

\noindent Masses derived with templates based on BC03 stellar population models are on average higher than M05-based masses at $z\geq2$ which is the same as at fixed redshift albeit with slightly increased scatter (Fig. \ref{Bmasszf}). Again this is clearest for $M^*>10^{9.5}\,M_{\odot}$ where the offset is on average $\sim0.2$ dex. This is a TP-AGB effect. At low redshift on the other hand, masses agree much better in the unreddened case because redshift compensates differences in the stellar population model. Note that compared to the true values masses are still underestimated. However, when reddening is included we find a large offset in stellar mass for the objects for which the redshift recovery fails with BC03 templates. This again is an effect of the age-dust-redshift degeneracy which we pointed out earlier. For 34\% of objects the redshift is falsely identified as being zero (compare also Fig. \ref{Bzhist}), for these objects the mass recovery fails completely. For the remaining outliers (redshift difference $\sim 0.4$) masses are smaller by at least $0.7$ dex. At redshift $1$ they are smaller by at least $\sim2.1$ dex.

\noindent Finally, we show the comparison for the SFRs in Fig. \ref{Bsfrzf}. At high redshift SFRs compare similarly well as in the fixed redshift case. When reddening is included, SFRs derived with BC03 models are on average lower. This is also the case at low redshift. At $z=0.5$ most SFRs obtained with BC03 models are zero in the reddened case because of the failure in redshift and mass recovery. When redshift is fixed, SFRs agreed much better.

\section{SED-fitting results for Pegase semi-analytic galaxies}\label{pegase}
\begin{figure}
\centering\includegraphics[width=84mm]{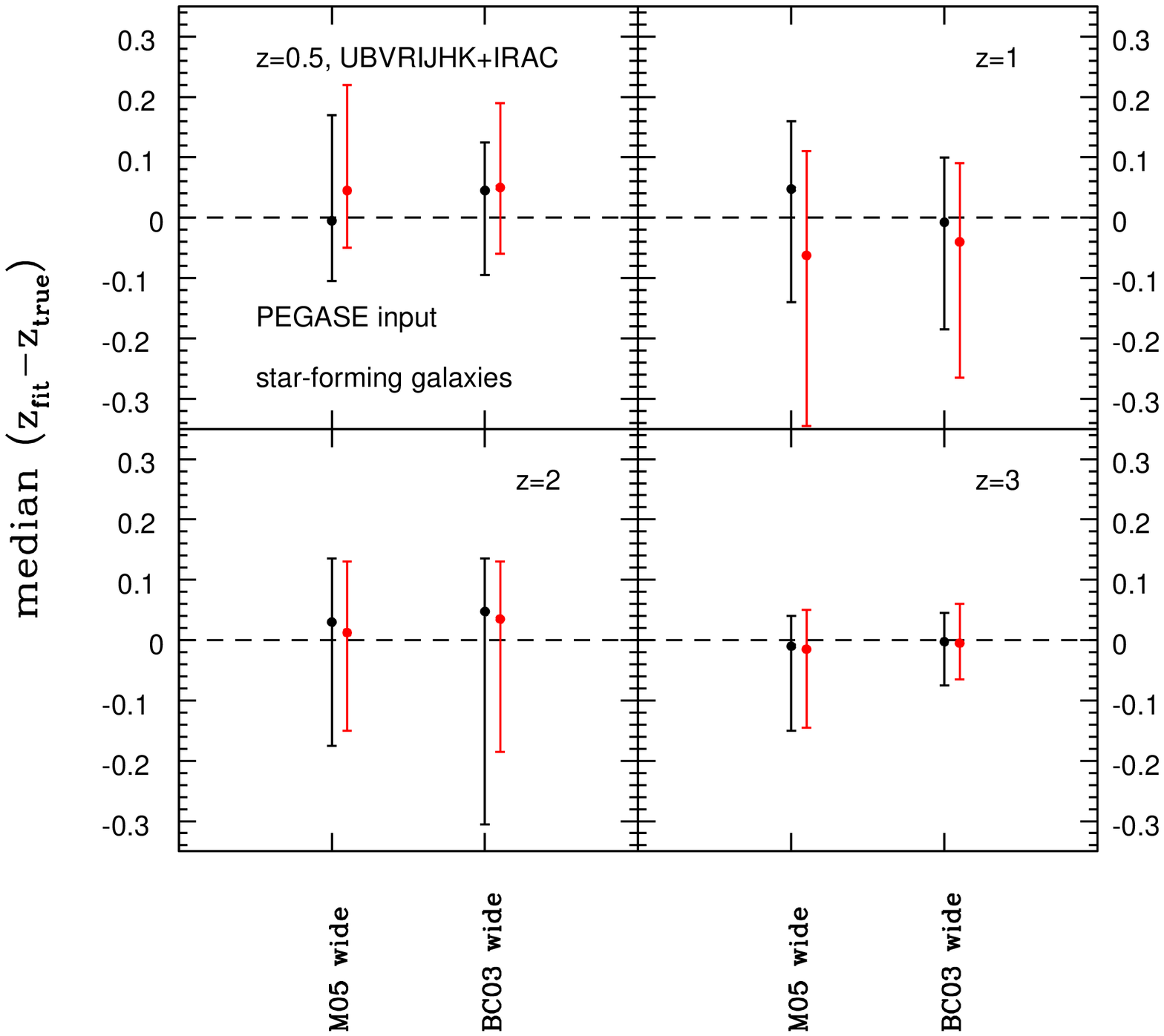}
\caption[Redshift recovery for BC03-type mock star-forming galaxies as a function of stellar population model]{\label{Pzhist1} Median redshift recovery when redshifts are derived from SED-fitting with different stellar population models - M05 (top) and BC03 (bottom) - for mock star-forming galaxies based on PEGASE photometry PEGASE galaxies). A wide setup and wavelength coverage was used. Symbols are the same as in Fig. \ref{Bzhist}. Errorbars are 68\% confidence levels.}
\end{figure}

\begin{figure*}
\centering\includegraphics[width=144mm]{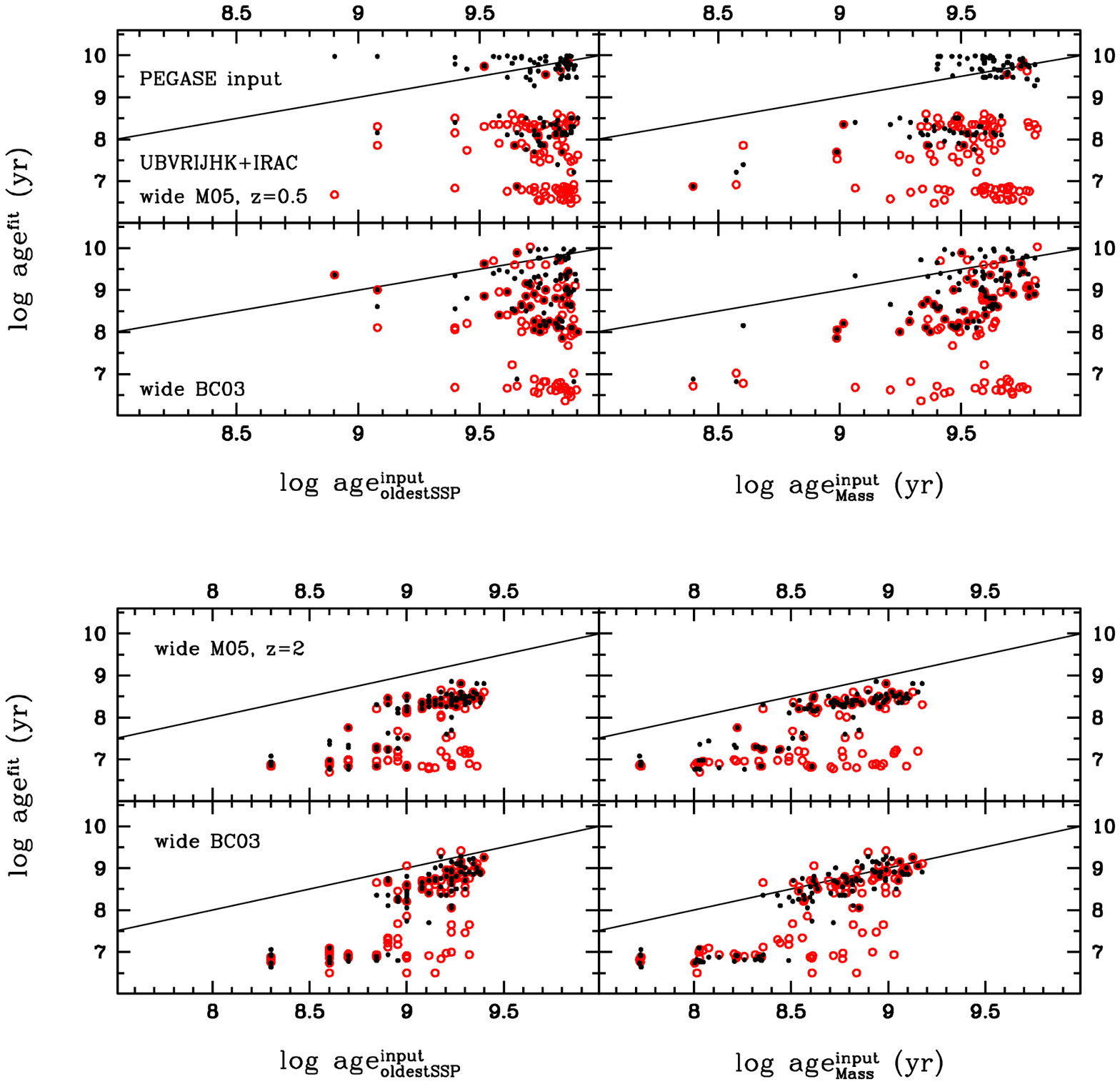}
\caption[Ages derived for BC03-type mock star-forming galaxies as a function of stellar population model when redshift is free]{\label{Pagezf} Age recovery with M05- and BC03-based templates (top and bottom in each panel) for mock star-forming galaxies based on PEGASE photometry when redshift is a free parameter in the fit. A wide setup and wavelength coverage was used. Black refers to the unreddened case, red to the case including reddening. \textit{Left:} Comparison to the age of the oldest SSP present. \textit{Right:} mass-weighted ages.}
\end{figure*}

\begin{figure*}
\centering\includegraphics[width=144mm]{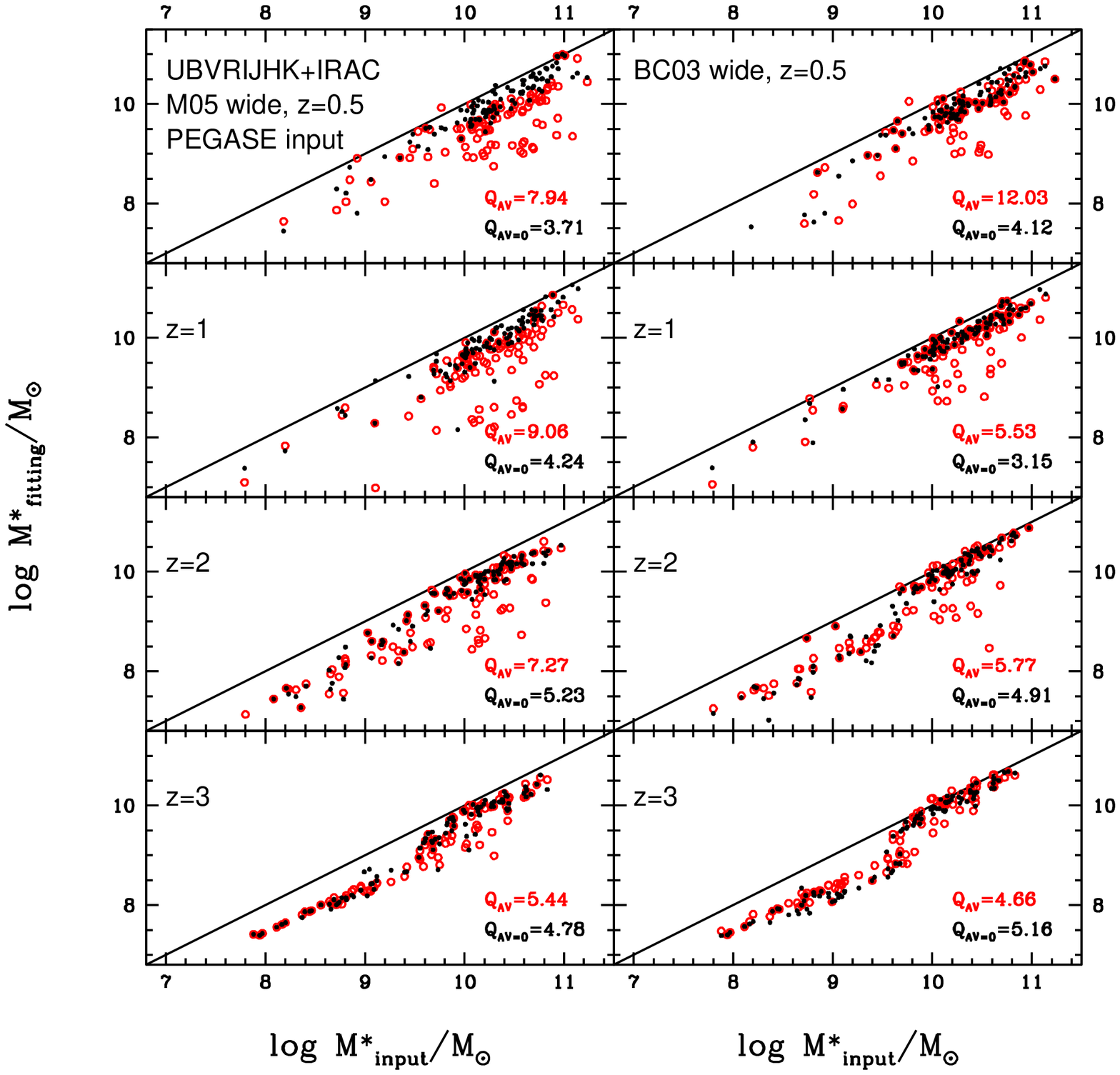}
\caption[Masses derived for BC03-type mock star-forming galaxies as a function of stellar population model when redshift is free]{\label{Pmasszf} Stellar mass recovery as a function of stellar population model (M05 and BC03 from left to right) or mock star-forming galaxies based on PEGASE photometry when redshift is left free. Red symbols refer to the reddened case, black to the unreddened case. A wide setup and a UBVRIJHK+IRAC wavelength coverage was used. Redshift increases from top to bottom. Quality factors refer to the entire mass range. }
\end{figure*}

\begin{figure*}
\centering\includegraphics[width=144mm]{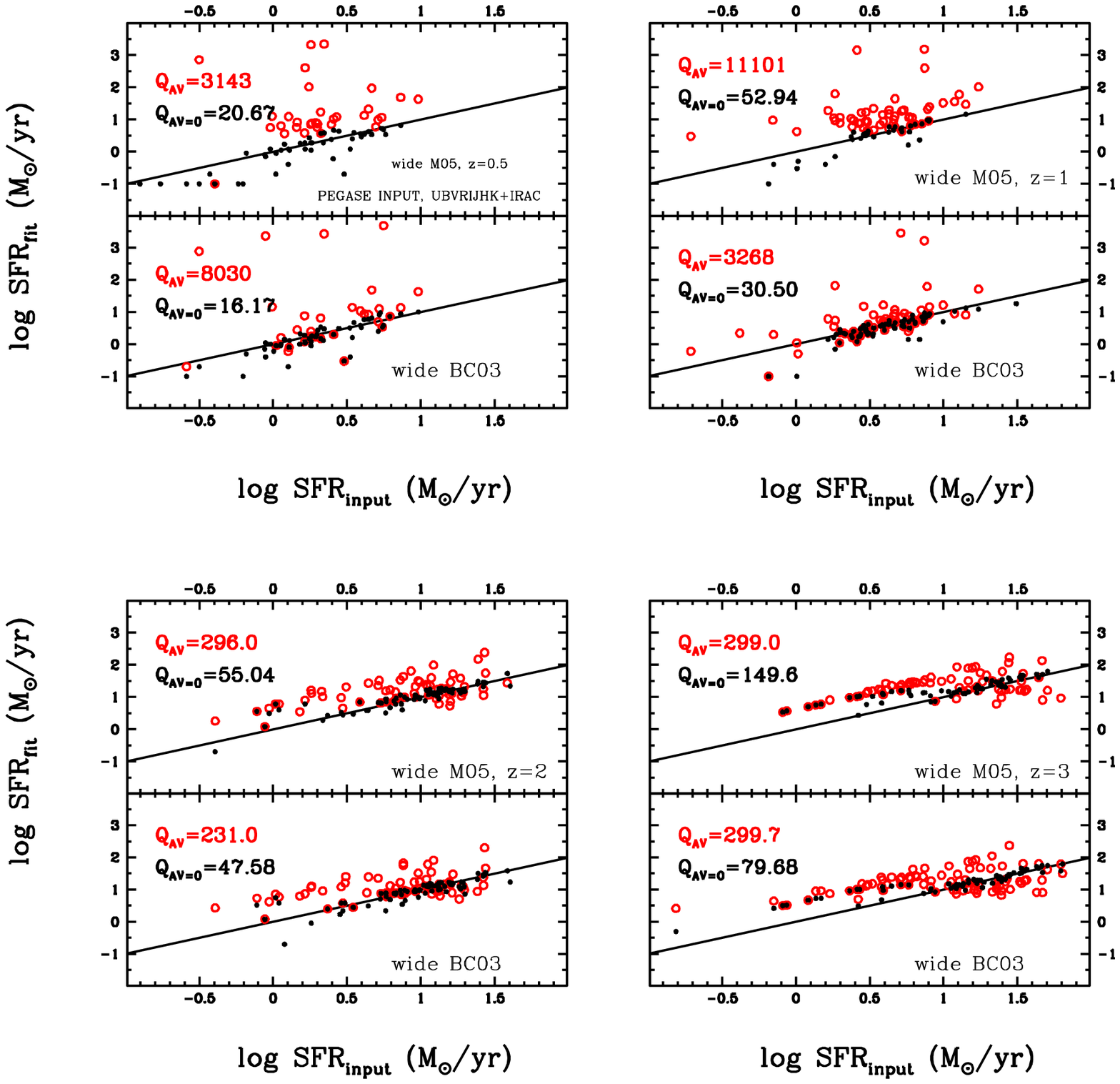}
\caption[SFRs derived for BC03-type mock star-forming galaxies as a function of stellar population model when redshift is free]{\label{Psfrzf} Star formation rate recovery as a function of stellar population model (M05 and BC03, top and bottom in each panel) for mock star-forming galaxies based on PEGASE photometry when redshift is a free parameter in the fit. Red symbols refer to the reddened case, black to the unreddened case. A wide setup and a UBVRIJHK+IRAC wavelength coverage was used. Redshift increases from top left to bottom right. Quality factors refer to the entire SFR range.}
\end{figure*}

\noindent As in Appendix \ref{bc03results} we also studied the case in which redshift is left free in the fit when we fit M05 or BC03 templates to semi-analytic galaxies with photometry based on PEGASE templates \citep{Pegase}. Note that PEGASE and BC03 stellar population models are very similar due to  similar input physics. Figs. \ref{Pzhist1} - \ref{Psfrzf} show the results. Overall, we find the same trends as in Paper I, ages and stellar masses are underestimated, reddening and SFRs are overestimated.

\noindent First, we study the redshift recovery in Fig. \ref{Pzhist1}. Apart from the highest redshift galaxies, the redshift recovery shows larger scatter on  average than when mock star-forming galaxies are based on M05 models, independently of the stellar population model of the templates that are used in the fit. Particularly, for $z\leq2$ redshift recoveries show a much wider distribution because the chosen filter set comprises the 4000\AA \space break but not the Lyman break. However, despite the larger scatter, the median offset for $z=2$ objects is smaller in Fig. \ref{Pzhist1} than in Fig. \ref{zhistdtzf} for both stellar population model templates and reddened/unreddened cases. Interestingly, M05 models perform better on PEGASE galaxies than BC03 models on M05 galaxies (compare Appendix \ref{bc03results} and offsets in Fig. C1), especially at low redshift. Hence, it seems that using M05 templates in the fitting provides overall a better photometric redshift determination independently of the real stellar evolution.

\noindent Clearly, age is the compensating factor. Fig. \ref{Pagezf} shows that ages derived with M05 models in the unreddenend case at low redshift are either very old ($>3$ Gyr) or young ($<300$ Myr), i.e. the age range for the TP-AGB is avoided. This distinction is clearer than at fixed redshift where ages from M05 templates were mostly younger than those derived with  BC03 templates. When redshift is free older ages are paired with lower redshifts and vice versa to compensate. Ages inferred from BC03 templates are generally younger than at fixed redshift. When reddening is included, the age-dust degeneracy dominates, driving the fit towards very young ages for both template setups. For M05 nearly all ages are below $300$ Myr in this case. At high redshift derived ages are very similar to the fixed redshift case. Ages derived with M05 templates are clearly younger than those derived with BC03 templates and therefore show a clear offset to the oldest and mass-weighted ages. This is the same effect which causes overestimated ages from BC03 templates for M05 galaxies.

\noindent As shown in Section \ref{results} the stellar mass recovery profits from the additional degree of freedom in the fit introduced by leaving the redshift free. This is also the case when PEGASE mock star-forming galaxies are fit with M05 and BC03 templates (Fig. \ref{Pmasszf}), particularly at low redshift. While at fixed redshift masses derived with M05 templates were underestimated on average by $\sim0.5$ dex in the unreddened case they are underestimated on average only by $\sim0.25$ dex when redshift is left free. For masses derived with BC03 templates the effect is very small instead. Including reddening in the fit worsens the result because of the age-dust degeneracy. This is similar to the case at fixed redshift and true for both stellar population models. It is worth to note that when reddening is included masses derived with M05 still show the same systematic offset as in the fixed redshift case.

\noindent The reddening determination is comparable for M05 and BC03 templates, independently of the flavour of the input galaxies. This is similar to the fixed redshift case. However, at low redshift both stellar population models require somewhat less reddening when redshift is free because redshift compensates.

\noindent Finally, we show the SFR recovery of PEGASE galaxies with M05 and BC03 templates when redshift is free in Fig. \ref{Psfrzf}. When reddening is excluded from the fit SFRs are generally well recovered, although at low redshift many of them are underestimated. With reddening in the fit SFRs are even more overestimated at low redshift than at fixed low redshift. At high redshift SFRs are overestimated for SFR$<10\,M_{\odot}$/yr and recovered well for SFRs larger than $10\,M_{\odot}$. This is equivalent to our results at fixed redshift.

\section{Properties of passive galaxies derived with simple stellar population templates}\label{SSPsetups}
\noindent In Paper I, we also studied how well the properties of mock passive galaxies are determined when the available SFHs in the template setups are restricted to simple stellar population models (SSPs). In particular, we used 2 template setups, one that contains SSPs of all available metallicities ($1/5$ solar, half solar, solar and twice solar, called 'only SSP' setup) and one that contains only the solar metallicity SSP ('solar SSP' setup). Although, it is not necessarily know a priori for what type of galaxy (passive or star-forming) the redshift needs to be determined via the SED fit, surveys such as SDSSIII/BOSS \citep{Schlegel2009,SDSSBOSS,Dawson2013} have demonstrated that predominantly passive galaxies can be selected by colour and colour-magnitude cuts alone \citep[see e.g.][]{Masters2011}. Hence, here we provide the results for using only SSP setups in the fitting of mock passive galaxies when redshift is also a free parameter.  
\begin{figure}
\centering\includegraphics[width=84mm]{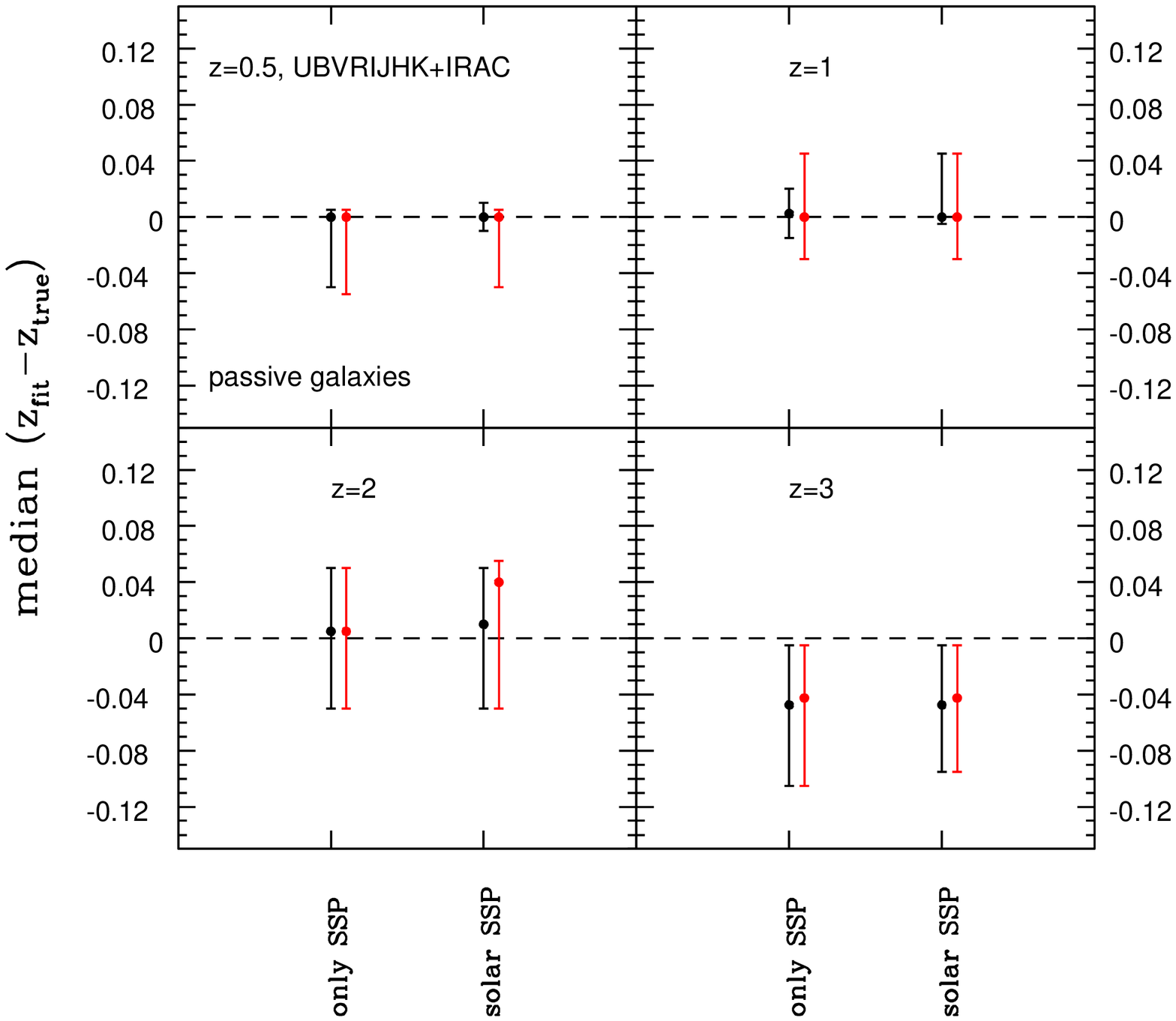}
\caption[Median redshift recovery for mock passive galaxies as a function of template setup]{\label{OLDzhistdtzf2}Median redshift recovery for mock passive galaxies as a function of only SSP template setups, namely only SSPs and solar SSP, using a broad wavelength coverage (UBVRIJHK+IRAC). Black symbols refer to the case without reddening, red ones to the reddened case. Errorbars are 68\% confidence levels.}
\end{figure}

\begin{figure}
\centering\includegraphics[width=84mm]{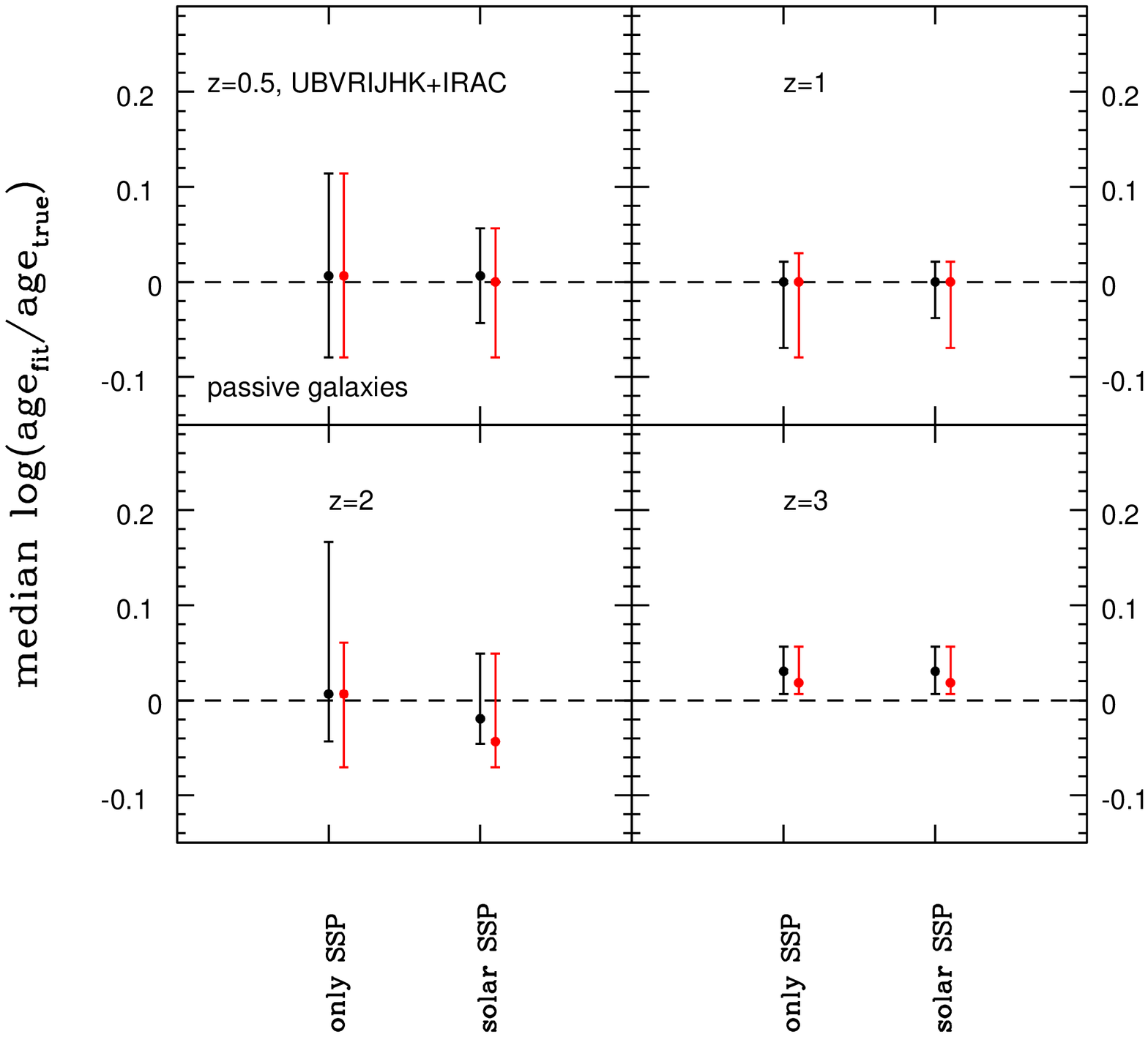}
\caption[Median age recovery for mock passive galaxies as a function of template setup]{\label{oldagedtzf2} Recovery of age (median) for mock passive galaxies for only SSP setups and redshifts. Symbols and errorbars are the same as in previous figures.}
\end{figure}

\noindent In Fig. E1 
 the median redshift recovery of mock passive galaxies is presented for the two only SSP setups. In principle, the solar SSP setup should provide the best parameter recovery overall, since this template matches the mock galaxies exactly in SFH, metallicity and age range. In the unreddened case redshifts are recovered very well for both setups at low redshift (small scatter and small or no offsets). At redshift $2$ scatter is larger and at redshift $3$ we also find a small offset ($\sim0.04$) that is larger than the offsets found with the wide and only-$\tau$ setup in Fig. \ref{OLDzhistdtzf}. We already explained that this is due to the slight mismatch between galaxy age and available age in the template age grid. While for the wide and only-$\tau$ setups the SFH can somewhat compensate for this, this is not possible for the only SSP setups. However, the scatter is smaller than for the wide and only-$\tau$ setups. When reddening is included, we find slightly larger scatter compared to the unreddened case. For redshift $2$ the increased offset in the reddened case for the solar SSP setup is due to the age dust degeneracy. While redshifts are overestimated, ages are underestimated (see also Fig. \ref{oldagedtzf2}).

\noindent As pointed out earlier, the fixed redshift (Paper I) and the free redshift case show similar results with regard to the effect of template setup on the derived ages of passive galaxies. We show the results for the two SSP setups in Fig. \ref{oldagedtzf2}. The median recovered ages show almost no offset at low redshift and very little offset at high redshift. In this respect, the age recovery is better with only SSP setups compared to the wide and only-$\tau$ setup. 
At $z=2$, the correct template underestimates the age because redshift is overestimated by a small amount ($\sim0.05$) in the majority of cases. At redshift $3$ offsets are comparable to those from template setups with more variety in SFH due to age mismatches and photometric uncertainties. We find the scatter to be similar to that in Fig. \ref{oldagedtzf} for the wide and only-$\tau$ setup, the scatter is smaller only at redshift $3$. Including reddening has a very small effect such that scatter is slightly increased at lower redshift. For a solar SSP setup at redshift $2$ the reddened case shows also a larger offset caused by the age-dust-redshift degeneracy. At redshift $3$ the offset is smaller for the same reason. 

\begin{figure}
\centering\includegraphics[width=84mm]{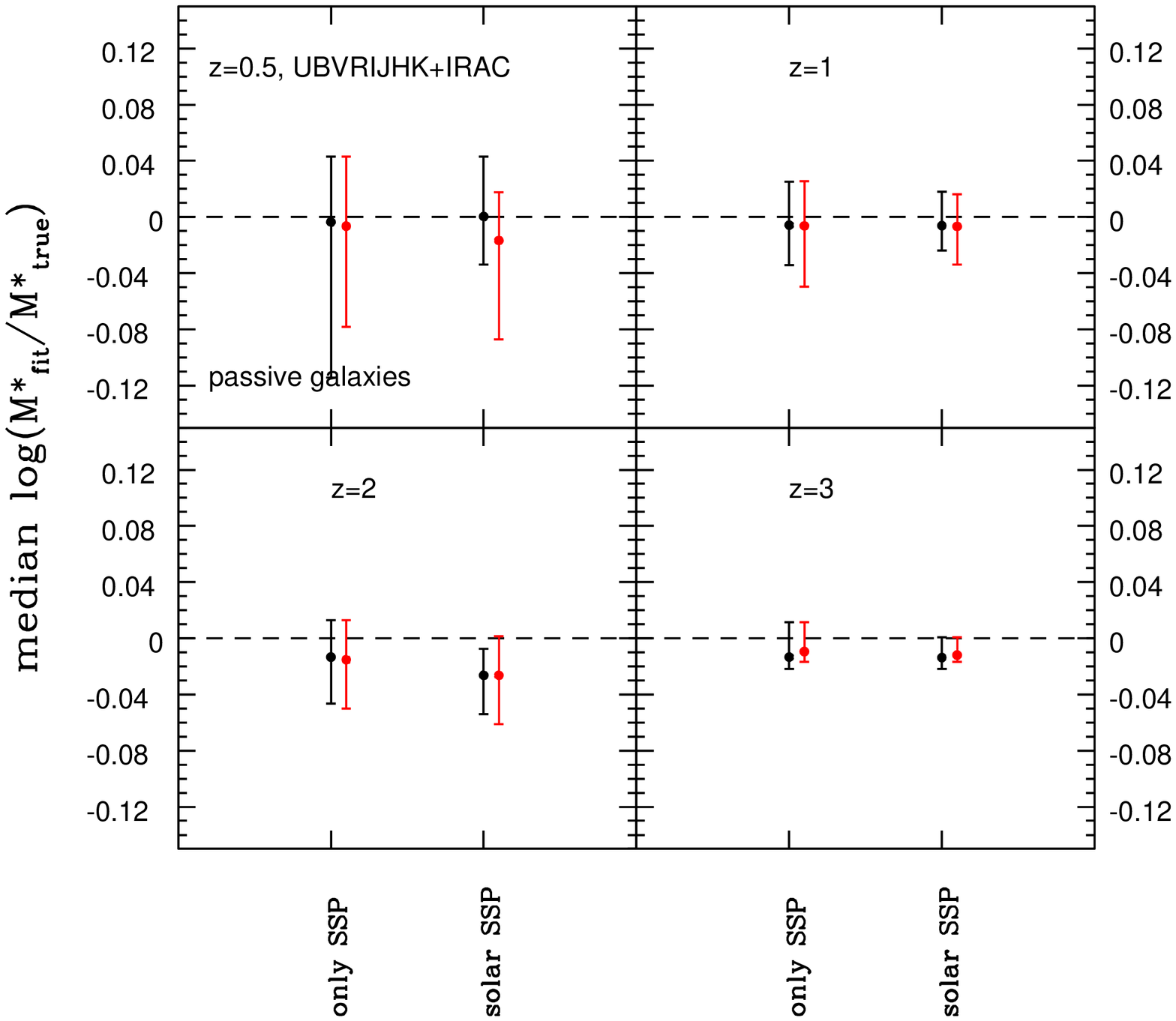}
\caption[Median mass recovery for mock passive galaxies as a function of template setup]{\label{oldmassdtzf2} Mass recovery (median) of mock passive galaxies achieved with only SSP setups as a function of redshift. Symbols and errorbars are the same as in previous figures.}
\end{figure}

\noindent Stellar masses of passive galaxies determined with only SSP setups are recovered similarly well than those recovered with a wide and only-$\tau$ setup as shown in Figs. \ref{oldmassdtzf} and \ref{oldmassdtzf2}. Median offsets are smaller then $\sim 0.02$ dex. Stellar masses tend to be slightly underestimated at higher redshift. This stems from the degeneracy between redshift and age because SFHs and metallicities match the input. The scatter is somewhat smaller at the lowest redshifts and the highest redshifts compared to that of the wide setup. The effect of reddening is largest at lower redshift where scatter is slightly increased in most cases but overall the effect of reddening is small.

\section{Comparison with Wuyts et al. (2009)}
\begin{figure*}
\centering\includegraphics[width=144mm]{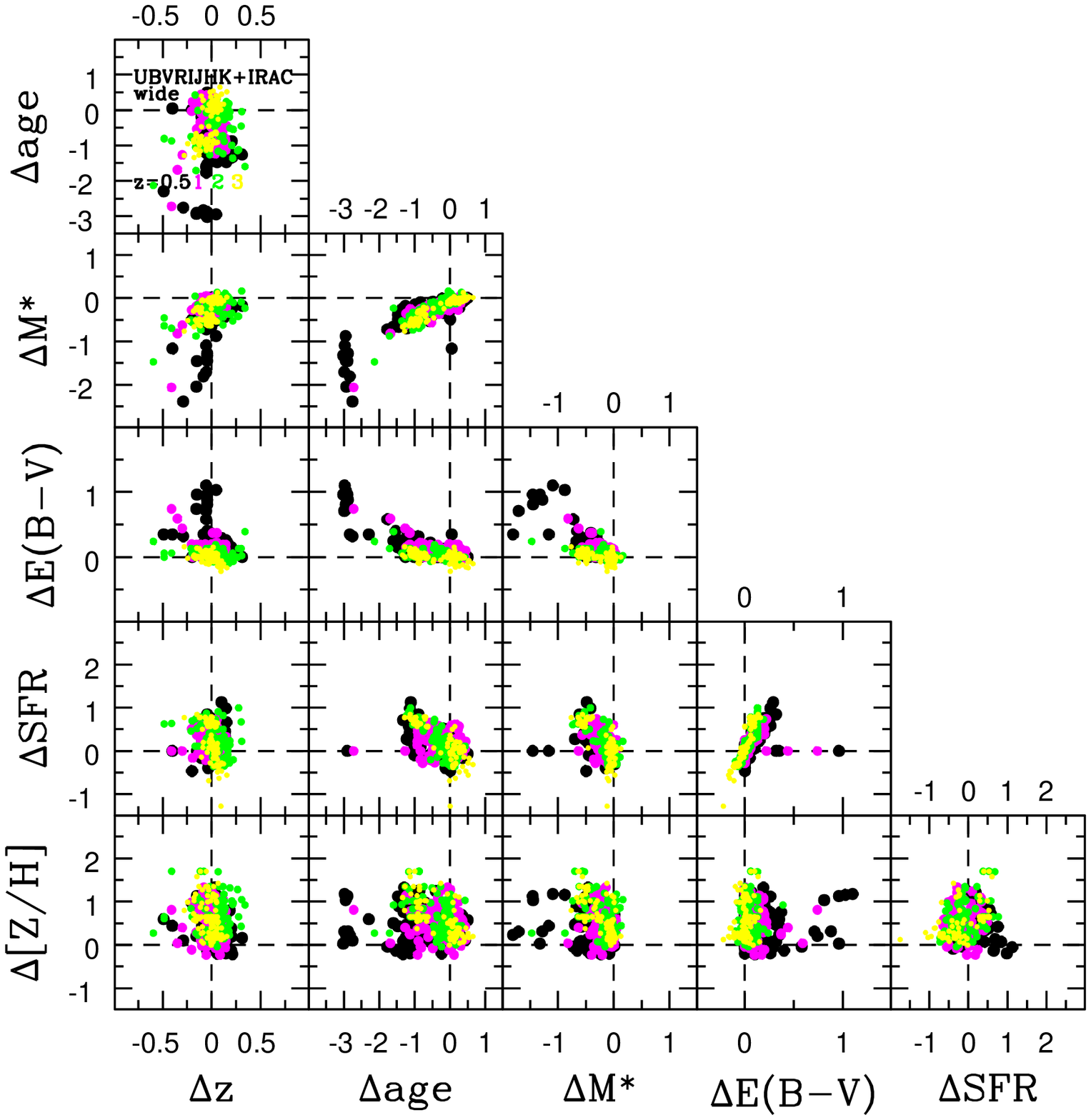}
\caption[Correlations and degeneracies between stellar population properties when redshift is a free parameter]{\label{trendszf} Basic correlations between redshift, age, stellar mass, dust reddening, and SFR derived from SED-fitting performed with a wide setup and a wide wavelength coverage (UBVRIJHK+IRAC) and reddening. $\Delta$ describes the difference between estimated and true quantity. Logarithmic values are used for stellar mass, age and SFR. The true age for the mock galaxies is represented by the mass-weighted age. Negative values of $\Delta$ mean underestimation, positive values stand for overestimation. We show objects at $z=0.5$ in black, at $z=1$ in magenta, $z=2$ in green and at $z=3$ in yellow.}
\end{figure*}

We reproduced Fig. 8 of \citet{Wuyts2009} also for the case in which redshift is left free in the fit. Accordingly we added a column showing the correlation of offsets in metallicity, SFR, $E(B-V)$, stellar mass and mass-weighted age with offsets in redshift. The results are shown in Fig. \ref{trendszf}.\\

\noindent Overall, we find a weak correlation between redshift and stellar mass such that when redshift is underestimated, stellar mass is underestimated, too. For all other properties correlations with redshift are even weaker because of the many degeneracies, e.g. redshifts and ages are underestimated when reddening is overestimated because they are degenerate. Median offsets for the stellar mass, age, reddening and SFR of mock galaxies are similar to the fixed redshift case.

\section{Scaling relations}

Tables \ref{scalezfree} and \ref{scalepasszfree} list the scaling relations for stellar masses between template setups when redshift is a free parameter in the fit. 
\begin{table*}
\caption[Scaling relations for mock star-forming galaxies when redshift is unknown]{Scaling relations to transform stellar masses for mock star-forming galaxies between different fitting setups $x$ as $\log M^*_{wide Salpeter}=a_1+a_2*x$. The coefficients $a_1$ and $a_2$ represent the unreddened case, $b_1$ and $b_2$ represent the reddened case. }
\begin{center}
\begin{tabular}{@{}clcccc}\hline
redshift & x & $a_1$ & $a_2$ & $b_1$ & $b_2$\\\hline
0.50  &  wide Chabrier   &    2.4652  &    0.7689  &    1.0713  &    0.9066 \\\hline
1.00  &  wide Chabrier   &    0.9667  &    0.9194  &    0.6539  &    0.9514 \\\hline
2.00  &  wide Chabrier   &    0.4726  &    0.9714  &    0.4078  &    0.9793 \\\hline
3.00  &  wide Chabrier   &    0.5029  &    0.9713  &    0.3796  &    0.9855 \\\hline
0.50  &  wide Kroupa   &    1.4633  &    0.8708  &    0.3393  &    0.9946 \\\hline
1.00  &  wide Kroupa   &    0.8264  &    0.9360  &    0.9359  &    0.9299 \\\hline
2.00  &  wide Kroupa   &    0.0834  &    1.0180  &   -0.0255  &    1.0289 \\\hline
3.00  &  wide Kroupa   &   -0.0968  &    1.0354  &   -0.1670  &    1.0440 \\\hline
0.50  &  wide $Z_{\odot}$   &    3.8500  &    0.6220  &    2.7551  &    0.7217 \\\hline
1.00  &  wide $Z_{\odot}$   &    1.2727  &    0.8723  &    1.1847  &    0.8779 \\\hline
2.00  &  wide $Z_{\odot}$   &    0.9155  &    0.9069  &    0.5792  &    0.9404 \\\hline
3.00  &  wide $Z_{\odot}$   &    0.7625  &    0.9258  &    0.5972  &    0.9421 \\\hline
0.50  &  only-$\tau$   &    1.8875  &    0.8135  &   -0.0098  &    0.9878 \\\hline
1.00  &  only-$\tau$   &    1.0996  &    0.8877  &    0.8531  &    0.9082 \\\hline
2.00  &  only-$\tau$   &    1.6317  &    0.8354  &    0.5326  &    0.9429 \\\hline
3.00  &  only-$\tau$   &    0.8059  &    0.9178  &    0.6248  &    0.9383 \\\hline
0.50  &  wide BC03   &    4.7388  &    0.5371  &    9.5668  &    0.0257 \\\hline
1.00  &  wide BC03   &    1.7411  &    0.8271  &    5.5878  &    0.4454 \\\hline
2.00  &  wide BC03   &    0.9954  &    0.8799  &    0.7234  &    0.9045 \\\hline
3.00  &  wide BC03   &    1.1207  &    0.8734  &    0.9429  &    0.8886 \\\hline
0.50  &  BC03 only-$\tau$   &    6.0939  &    0.4064  &    9.3264  &    0.0524 \\\hline
1.00  &  BC03 only-$\tau$   &    2.6001  &    0.7333  &    3.3156  &    0.6591 \\\hline
2.00  &  BC03 only-$\tau$   &    2.4519  &    0.7361  &    1.3094  &    0.8447 \\\hline
3.00  &  BC03 only-$\tau$   &    1.3444  &    0.8495  &    1.1954  &    0.8627 \\\hline
\end{tabular}
\end{center}
\label{scalezfree}
\end{table*}%

\begin{table*}
\caption[Scaling relations for mock passive galaxies when redshift is unknown]{Same as Table \ref{scalezfree} for passive galaxies}
\begin{center}
\begin{tabular}{@{}clcccc}\hline
redshift & x & $a_1$ & $a_2$ & $b_1$ & $b_2$\\\hline
0.50  &  wide Chabrier   &   -0.2011  &    1.0279  &   -0.1207  &    1.0214 \\\hline
1.00  &  wide Chabrier   &   -0.0626  &    1.0169  &    0.2429  &    0.9899 \\\hline
2.00  &  wide Chabrier   &    0.1971  &    0.9966  &    0.2053  &    0.9959 \\\hline
3.00  &  wide Chabrier   &    0.0941  &    1.0058  &    0.0792  &    1.0070 \\\hline
0.50  &  wide Kroupa   &   -0.1336  &    1.0206  &   -0.1749  &    1.0243 \\\hline
1.00  &  wide Kroupa   &    0.0118  &    1.0085  &    0.2174  &    0.9896 \\\hline
2.00  &  wide Kroupa   &    0.1517  &    0.9976  &    0.1451  &    0.9981 \\\hline
3.00  &  wide Kroupa   &    0.0728  &    1.0047  &    0.0566  &    1.0060 \\\hline
0.50  &  wide $Z_{\odot}$   &    0.0085  &    0.9954  &   -0.2862  &    1.0251 \\\hline
1.00  &  wide $Z_{\odot}$   &   -0.2216  &    1.0187  &   -0.4324  &    1.0365 \\\hline
2.00  &  wide $Z_{\odot}$   &   -0.0265  &    1.0026  &   -0.0251  &    1.0024 \\\hline
3.00  &  wide $Z_{\odot}$   &    0.1417  &    0.9881  &    0.1424  &    0.9881 \\\hline
0.50  &  only-$\tau$   &    0.0237  &    0.9940  &   -0.1676  &    1.0144 \\\hline
1.00  &  only-$\tau$   &   -0.2621  &    1.0224  &   -0.5312  &    1.0451 \\\hline
2.00  &  only-$\tau$   &    0.1464  &    0.9884  &    0.1460  &    0.9882 \\\hline
3.00  &  only-$\tau$   &    0.0489  &    0.9965  &    0.0354  &    0.9975 \\\hline
0.50  &  SSPs   &    0.0201  &    0.9973  &   -0.0340  &    1.0028 \\\hline
1.00  &  SSPs   &   -0.0973  &    1.0086  &   -0.0747  &    1.0065 \\\hline
2.00  &  SSPs   &    0.1210  &    0.9897  &    0.1676  &    0.9858 \\\hline
3.00  &  SSPs   &   -0.0371  &    1.0036  &   -0.0368  &    1.0036 \\\hline
0.50  &  wide BC03   &    0.5146  &    0.9411  &   11.0290  &    0.0195 \\\hline
1.00  &  wide BC03   &    0.3867  &    0.9504  &    2.0655  &    0.8387 \\\hline
2.00  &  wide BC03   &   -0.4636  &    0.9866  &    2.3464  &    0.7772 \\\hline
3.00  &  wide BC03   &    0.1364  &    0.9565  &    3.5024  &    0.6933 \\\hline
\end{tabular}
\end{center}
\label{scalepasszfree}
\end{table*}%

\end{document}